\documentclass{jfm}
\usepackage{graphicx}
\usepackage{epstopdf, epsfig}

\usepackage{array}
\usepackage{amsfonts,amssymb,amsbsy,amsmath}
\usepackage{subfigure}
\usepackage{textgreek}
\usepackage{upgreek}
\usepackage{bm}
\usepackage{url}
\usepackage{tikz}
\usetikzlibrary{shapes, shadows, arrows, positioning,patterns}
\usetikzlibrary{decorations.pathreplacing,calc}

\title{The effect of electrostatic charges on particle-laden duct flows}
\shorttitle{The effect of electrostatic charges on particle-laden duct flows}
\shortauthor{H. Grosshans et al.}

\author{Holger Grosshans\aff{1,2}\corresp{\email{holger.grosshans@ptb.de}},
 Claus Bissinger\aff{1},
 Mathieu Calero\aff{3},
 \and Miltiadis V. Papalexandris\aff{3}}

\affiliation{\aff{1}Physikalisch-Technische Bundesanstalt (PTB), Braunschweig, Germany
             \aff{2}Institute of Apparatus- and Environmental Technology, Otto von Guericke University of Magdeburg, Germany
             \aff{3}Institute of Mechanics, Materials and Civil Engineering, Universit\'{e} catholique de Louvain, Louvain-la-Neuve, Belgium}

\begin{document}

\maketitle

\begin{abstract}
We report on direct numerical simulations of the effect of electrostatic charges on particle-laden duct flows.
The corresponding electrostatic forces are known to affect particle dynamics at small scales and the associated turbophoretic drift.
Our simulations, however, predicted that electrostatic forces also dominate the vortical motion of the particles, induced by the secondary flows of Prandtl's second kind of the carrier fluid.
Herein we treated flows at two frictional Reynolds numbers ($Re_\mathrm{\tau}=$~300 and~600), two particle-to-gas density ratios ($\rho_\mathrm{p}/\rho=$~1000 and 7500), and three Coulombic-to-gravitational force ratios ($F_\mathrm{el}/F_\mathrm{g}=$~0, 0.004, and 0.026).
In flows with a high density ratio at $Re_\mathrm{\tau}=$~600 and $F_\mathrm{el}/F_\mathrm{g}=$~0.004, the particles tend to accumulate at the walls.
On the other hand, at a lower density ratio, respectively a higher $F_\mathrm{el}/F_\mathrm{g}$ of 0.026, the charged particles still follow the secondary flow structures that are developed in the duct.
However, even in this case, the electrostatic forces counteract the particles' inward flux from the wall and, as a result, their vortical motion in these secondary structures is significantly attenuated.
This change in the flow pattern results in an increase of the particle number density at the bisectors of the walls by a factor of five compared to the corresponding flow with uncharged particles.
Finally, at $Re_\mathrm{\tau}=$~300, $\rho_\mathrm{p}/\rho=$~1000, and $F_\mathrm{el}/F_\mathrm{g}=$~0.026 the electrostatic forces dominate over the aerodynamic forces and gravity and, consequently the particles no longer follow the streamlines of the carrier gas.
\end{abstract}




\section{Introduction}

The preferential concentration of solid particles in wall-bounded turbulent flows is of primary importance in a wide range of technological applications.
For this reason, it has been the subject of numerous experimental and numerical studies over the years.
In general, particle accumulation results from the complex interplay between the forces acting on the particles: aerodynamic, inertial, collisional, and gravitational forces.
In practice, however, in confined flows particles usually acquire a certain amount of electrostatic charge through triboelectric effects, i.e. via collisions of particles with the walls or with other pre-charged particles.
In some applications, such as powder coating~\citep{Yang16}, electrostatic precipitation~\citep{Hore14}, triboelectric separation~\citep{Lab17}, and spray painting~\citep{Boet02}, particle charging takes place deliberately to control the particles' trajectories. However, in some other powder-handling facilities, such as fluidized beds, it occurs incidentally and leads to operational problems~\citep{Foto17,Foto18}.
For example, due to particle-wall adhesion the charged particles can coat vessel walls which requires frequent cleaning~\citep{Sip18}.
Also, the dynamics of fluidized beds is greatly influenced by electrostatic interactions~\citep{Jal12,Jal15}.
In pharmaceutical devices, e.g., dry powder inhalers, electrostatic effects deteriorate the effectiveness of the final product~\citep{Wong15}.
During pneumatic conveying, where the emerging charge levels are particularly high~\citep{Kli18}, high concentrations of charged particles may lead to spark discharges with devastating consequences.
In the past, spark discharges have caused numerous dust explosions in chemical and process industries \citep{Eck03}.
Thus far, the effect of triboelectric charging and the ensuing electrostatic forces on particle concentrations has not been investigated in detail.
Therefore, many questions on this topic still remain open.

Uncharged solid particles are well-known to distribute non-uniformly in wall-bounded turbulent flows.
This is related to their turbophoretic drift, i.e. the tendency of particles to migrate towards regions of reduced turbulent kinetic energy \citep{Cap75,Ree83}.
The design of advanced flow solvers and the ever-increasing power of modern computer has made possible to study the phenomenon of turbophoresis via Direct Numerical Simulations~(DNS).
For example, one of the earliest DNS on this topic was that of \cite{Mcl89}.
Turbophoresis was further explored by \citet{Marc02} who performed DNS of particle-laden flow over a smooth flat plate.
Therein they reported that as the characteristic particle-response time-scale becomes shorter, the particle congregate faster and closer to the flat wall.
In the same study, the authors also identified the coherent structures that are responsible for the entrapment of particles in the near-wall region.
More recently, the study of \citet{Wang10} on turbophoresis in channel flows elucidated the preferential location for particles of different inertia and number density.

\begin{figure}
\begin{center}
 \def\SL{
 \draw [] plot [smooth cycle] coordinates {(.3,.7) (1.1,1.5) (.3,1.5)};
 \draw [] plot [smooth cycle] coordinates {(.4,.9) (.9,1.4) (.4,1.4)};
 \draw [] plot [smooth cycle] coordinates {(.5,1.1) (.7,1.3) (.5,1.3)};}
\begin{tikzpicture}
 \SL
 \begin{scope}[yscale=-1,xscale=1] \SL \end{scope}
 \begin{scope}[yscale=1,xscale=-1] \SL \end{scope}
 \begin{scope}[yscale=-1,xscale=-1] \SL \end{scope}
 \begin{scope}[yscale=1,xscale=1,rotate=90] \SL \end{scope}
 \begin{scope}[yscale=-1,xscale=1,rotate=90] \SL \end{scope}
 \begin{scope}[yscale=1,xscale=-1,rotate=90] \SL \end{scope}
 \begin{scope}[yscale=-1,xscale=-1,rotate=90] \SL \end{scope}
 \draw [thick] (-2,-2) rectangle (2,2);
 \draw [dash dot] (-2.5,0) -- (2.5,0);
 \draw [dash dot] (0,-2.5) -- (0,2.5);
 \draw [dash dot] (-2.3,-2.3) -- (2.3,2.3);
 \draw [dash dot] (2.3,-2.3) -- (-2.3,2.3);
 \draw [->, ultra thick, red] (-.3,-.3) -- (-1.5,-1.5) node [above,yshift=0,xshift=-4] {A};
 \draw [->, ultra thick, gray] (.3,.3) -- (1.5,1.5);
 \draw [->, ultra thick, gray] (-.3,.3) -- (-1.5,1.5);
 \draw [->, ultra thick, gray] (.3,-.3) -- (1.5,-1.5);
 \draw [->, ultra thick, blue] (-1.6,-1.8) -- (-.3,-1.8) node [above,xshift=4,yshift=-2] {B};
 \draw [->, ultra thick, gray] (1.6,-1.8) -- (.3,-1.8);
 \draw [->, ultra thick, gray] (-1.6,1.8) -- (-.3,1.8);
 \draw [->, ultra thick, gray] (1.6,1.8) -- (.3,1.8);
 \draw [->, ultra thick, gray] (1.8,1.6) -- (1.8,.3);
 \draw [->, ultra thick, gray] (-1.8,1.6) -- (-1.8,.3);
 \draw [->, ultra thick, gray] (1.8,-1.6) -- (1.8,-.3);
 \draw [->, ultra thick, gray] (-1.8,-1.6) -- (-1.8,-.3);
 \draw [->, ultra thick, green] (0,-1.7) -- (0,-.5) node [right,yshift=-1] {C};
 \draw [->, ultra thick, gray] (0,1.7) -- (0,.5);
 \draw [->, ultra thick, gray] (-1.7,0) -- (-.5,0);
 \draw [->, ultra thick, gray] (1.7,0) -- (.5,0);
 \draw [->,thick](-4,-2) -- (-3,-2) node[anchor=south,below]{$z^+$} ;
 \draw [->,thick](-4,-2) -- (-4,-1) node[anchor=south,left]{$y^+$} ;
\end{tikzpicture}
\end{center}
\caption{Symmetry planes of the duct and streamlines of the induced Prandtl's secondary flow of the second kind.
The arrows indicate the direction of the possible particle motion in the cross-section following the gas phase.}
\label{fig:secondary}
\end{figure}
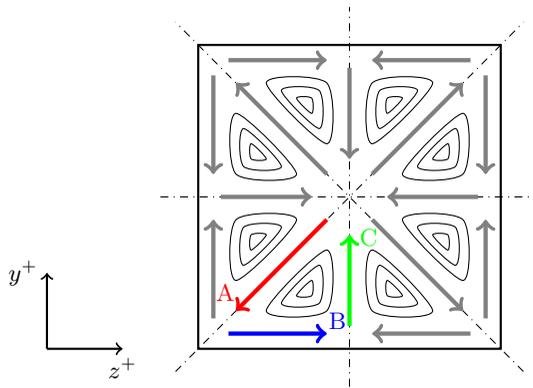

On the other hand, the properties of duct flows are quite different to those of flows over a flat plate or even channel flows.
This is due to the absence of a homogeneous direction which results in the development of secondary flow structures.
More specifically, as shown schematically in figure~\ref{fig:secondary}, duct flows are characterized by the formation of Prandtl's secondary flows of the second kind \citep{Pra27,Bra87,Nezu05}.
These secondary flows are formed due to turbulence-driven gradients of the Reynolds stress and are related to the variations of the probability distribution of the coherent flow structures \citep{Uhl07,Pin10,Kaw12}.

Although these secondary flows involve only small-amplitude fluid motion, they have a significant effect on the overall momentum and mass transport.
For example, in a duct with square cross-section, the mean secondary flow follows the well-known 8-vortex pattern, as can be observed in figure~\ref{fig:secondary}.
The intensity of the spanwise motion inside these vortices is between 1\% and 2\% of the bulk fluid velocity \citep{Pir18}.
Nonetheless, the role of these vortices is important because they transport high-momentum fluid from the duct core towards the corners.

It is also important to mention that the turbulence modelling of such secondary flows is particularly challenging because
the eddy viscosity can no longer be assumed to be isotropic.
Standard turbulence models, however, are based on the  assumption of isotropy and, therefore, cannot satisfactorily handle secondary flows induced by Reynolds-stress gradients \citep{Spe82,Mani13}.

In the case of particle-laden flows, these secondary motions of the second type have a strong affect on the local concentration of particles.
Actually, particle-laden flows in ducts have been the subject of several numerical studies that appeared in the literature, albeit they are not as much investigated as single-phase duct flows.
For example, \citet{Wang92} studied an open duct flow confined by three walls of the Reynolds number $Re=$~83\,000.
Therein, it was demonstrated that relatively light particles, of Stokes number $St$ between 3.2 and 96.6, tend to move diagonally in the direction indicated by the arrow~A in figure~\ref{fig:secondary} and to accumulate in the corners.
\citet{Wang92} also showed that on the other hand, heavy particles ($St>$~323) are much less affected by secondary flows and are more likely to concentrate in the bulk of the duct.
Further, \citet{Yao2010} performed Large-Eddy Simulations (LES) to study particle resuspension in a duct flow at $Re=$250\,000.
In that study, the authors demonstrated that secondary flows enable the displacement of particles to the wall region back into the bulk of the duct, which has a significant impact in the local particle concentration.
In the same study, it was also explained that lift forces counteract gravitation and support the transport of particles parallel to the horizontal walls and in the direction pointed by the arrow~B in figure~\ref{fig:secondary}.

In another study, \citet{Sha06} highlighted the importance of secondary flows for particle dispersion in ducts with square cross-sections.
They revealed that passive tracers and low-inertia particles are subject to a lateral advective transport that is absent in pipe and channel flows.
In this manner, the particles tend to circulate between the core and the boundary of the duct, i.e.~following the arrows A, B, and C that are shown in figure~\ref{fig:secondary}.
On the other hand, high-inertia particles follow mainly follow the directions of the arrows A and B and tend accumulate close to the wall.

With regard to electrostatics, as mentioned above, particles usually carry a certain amount of electric charge.
The emerging electric field significantly modifies the flow patterns of the particulate phase due to electrostatic forces between particles \citep{Dho91}.
In some cases, these forces have a counterintuitive effect; for example, they may cause particles to move upstream and against the fluid flow  \citep{Myl87}.
At present, numerical studies on triboelectric charging in particle-laden duct flows are particularly scarce.
This may be attributed to the complexity of the system of governing equations which consists of the Navier-Stokes equations for the fluid, the electric field equation, plus the equations of motion of the particulate phase suitably modified so as to take into account the particle-fluid and particle-electric field interactions.
As a result, our understanding of this type of flows remains rudimentary.

Recently, \citet{Yao20} studied via LES the effect of the electrostatic field on a particle-laden pipe flow at $Re=$~44\,000, by assuming one-way coupling between particles and carrier fluid.
According to these simulations, the electrostatic forces on particles dominate over aerodynamic forces and gravity in the near-wall region.
This resulted in higher particle concentrations close to the wall and lower ones in the bulk of the pipe, the latter one been defined by $y/R<$~0.95.
DNS of triboelectric powder charging were performed by \citet{Gro17a}; however, that study treated channel flows and was limited to the initial states of charging.
For this reason, the electrostatic charges had a minor impact on the flow patterns.

To our knowledge, the only available studies on the effect of electrostatic forces on particle-laden duct flows are the LES reported by \citet{Gro18d,Gro19d}.
Those studies, which considered flows at $Re=$~10\,000 and high-inertia particles, revealed that increasing the electrostatic charge results in i) higher particle concentration in the corners of the duct, and ii) more uniform particle concentrations away from them.
However, the underlying physical mechanism of this behavior was not explored in those simulations.
It is also worth mentioning that reciprocal investigations, i.e. studies on the influence of different flow parameters (such as $Re$ or $St$) on the electrostatic field in duct flows are also currently unavailable.

In summary, the two-way interaction between electrostatic field and particle-laden flows in ducts remains largely unexplored.
The present work aimed at providing new insight on this interaction by means of direct numerical simulations.
To this end, we employed our newly developed computational tool \citet{pafiX}; its name stands for \textit{particle flow simulation in explosion protection}.
This tool is available as freeware and is developed specifically for the study of
powder and fluid flows under conditions involving a hazard to operational safety.
In our study, we considered particle-laden flows at two different density ratios, $\rho_\mathrm{p}/\rho=$~1000 and 7500, and two different
frictional Reynolds numbers, $Re_\mathrm{\tau}=$~300 and 600.

The paper is structured as follows.
Section~2 provides an overview of the governing equations and the algorithms that are implemented in pafiX for their numerical treatment.
In Section~3 we first provide the validation tests of these algorithms and subsequently we present and discuss the results of our DNS study.
Finally, Section~4 concludes.

\section{Mathematical model and numerical method}
\label{sec:math}

In this section we outline the mathematical model and the numerical method for its solution.
The system of governing equations consists of three strongly coupled parts, namely (i)~the Navier-Stokes equations that describe the flow of the carrier gas, (ii)~Gauss's law for the electrostatic field, and (iii)~Newton's law of motion of the solid particles.

For a constant-density, particle-laden flow, the Navier-Stokes equations read,
\begin{subequations}
\begin{equation}
\label{eq:mass}
\nabla \cdot {\bm u} \;=\;0\,,
\end{equation}
\begin{equation}
\label{eq:mom}
\frac{\partial {\bm u}}{\partial t} + ({\bm u} \cdot \nabla) {\bm u}
\;=\; - \frac{1}{\rho} \nabla p  + \nu \nabla^2 {\bm u} + {\bm F}_{\mathrm s} + {\bm F}_{\mathrm f} \,.
\end{equation}
\end{subequations}
Using standard notation, $\bm u$, $\rho$, and $p$ stand for the fluid velocity, density, and dynamic pressure respectively.
Also, $\nu$ is the kinematic viscosity of the fluid.
Further,
${\bm F}_{\textrm s}$ is the source term accounting for the momentum transfer between the solid particles and the carrier gas.
More specifically, its integral over a control volume, e.g. a computational cell, is equal to the opposite of the
sum of the aerodynamic forces that act on the particles that are located inside the control volume.
Finally, ${\bm F}_{\textrm f}$ represents an externally applied forcing that balances the momentum loss of the fluid due to wall friction.

As is well known, the dynamic pressure enters the constant-density Navier-Stokes equations via its gradient and, therefore, its role is to constrain the fluid motion so that mass-continuity \eqref{eq:mass} is satisfied.
In other words, the role of the dynamic pressure is to limit the fluid to isochoric motions.
In the proposed algorithm, this constrain is satisfied by coupling the continuity and momentum equations through the distributed Gauss-Seidel scheme originally proposed by \citet{Bra78}.
Herein this algorithm is extended to three-dimensional domains and non-uniform grids.
The underlying idea of this scheme is to diminish the error in the continuity equation by iteratively adjusting the velocity field.
Afterward, the pressure field is modified accordingly so that the residuals of the momentum equation at all points remain unchanged.
Distributive relaxation represents an efficient and intuitive alternative to popular velocity-pressure schemes such as SIMPLE or PISO \citep{Ferz02}.

As mentioned above, the electric field strength ${\bm E}$ follows Gauss's law.
In the electrostatic approximation,  ${\bm E}$ is defined in terms of the electric potential $\varphi_{\rm el}$,
\begin{equation}
\label{eq:elecpot}
{\bm E} \;=\; -\nabla \varphi_{\rm el}\,,
\end{equation}
so that Gauss's law reduces to,
\begin{equation}
\label{eq:gauss}
\nabla^2 \varphi_{\rm el} \;=\; -\dfrac{\rho_{\mathrm{el}}}{\varepsilon} \,,
\end{equation}
which is a Poisson equation.
In this equation, ${\rho_{\mathrm{el}}}$ is the electric charge density.
If no external electric field is present, the electric charge density results directly from the positions of the individual particles and their individual charge.
The electric permittivity of the solid-gaseous mixture, ${\varepsilon}$, can be approximated by the value of the free space either if the permittivity of the particles is low or if the solid volume fraction is very small~\citep{Rokk10,Rivas07} which is the case in our simulations.
Thus, we apply a value of $\varepsilon=8.85 \times 10^{-12}$ F/m.

The numerical solution of the above equations requires the discretization of the spatial derivatives of $\bm u$, $p$, and $\varphi_{\rm el}$.
The convective term of the Navier-Stokes equations is approximated by a fifth-order accurate Weighted Essentially Non-Oscillatory, WENO, schemes~\citep{Jang96}.
The pressure gradient and viscous terms in \eqref{eq:mom}, as well as the velocity derivatives of \eqref{eq:mass}, are discretized via fourth-order central differences.
Further, the left-hand side of Gauss's law \eqref{eq:gauss} is discretized via second-order central differences.
Then the discretization of the Poisson equation results in a linear system that can be solved by standard linear solvers.

In order to keep the high-order discretization schemes simple, the spatial derivatives of the convective terms are transformed via stretching of the spatial coordinates.
For a generic flow quantity $\phi$, this transformation is performed with the application of the chain rule as follows,
\begin{equation}
\label{eq:mapping}
\dfrac{d \phi}{d x} \;=\; \dfrac{1}{x'(\xi)} \dfrac{d \phi}{d \xi} \,,
\end{equation}
where the prime symbol denotes the derivative of the $x$ variable with respect to the stretched $\xi$ variable.
Thus, the non-uniform grid in the physical $x$ direction is mapped to the uniform grid in the $\xi$ direction on which the derivatives are then discretized.

Further, in order to reduce the required memory and computing time, the deferred-correction method is applied.
According to it, the numerical approximation of a quantity $\phi$ is expressed by,
\begin{equation}
\label{eq:dc}
\phi \;=\; \phi^{\mathrm{l}} + \left( \phi^{\mathrm{h}} - \phi^{\mathrm{l}} \right)^{\mathrm{old}} \,,
\end{equation}
where the superscript~`l' denotes an approximation by a low-order scheme and `h' an approximation by a high-order scheme.
The terms indicated by `old' are computed explicitly using values from the previous outer iteration.
For the computational cells in the vicinity of a solid boundary there are not sufficient number of points available to form a five-point stencil for the high-order discretization of the spatial derivatives.
In those computational cells, only the low-order schemes are retained.

Time integration of equation~(\ref{eq:mom}) is performed via  an implicit second-order scheme using a variable time-step, i.e.
\begin{equation}
\label{eq:dt}
\dfrac{\partial {\bm u}}{\partial t} \;\approx\; \dfrac{
  \left( 1+\tau^{n+\frac{1}{2}} \right) {\bm u}^{n+1}
- \left( 1+\tau^{n+\frac{1}{2}}+\tau^{n-\frac{1}{2}} \right) {\bm u}^{n}
+ \left( \tau^{n-\frac{1}{2}} \right) {\bm u}^{n-1}
}
 {\tau^{n+\frac{1}{2}} \Delta t^{n+1}}
\end{equation}
with
\begin{equation}
\tau^{n+\frac{1}{2}} = \dfrac{\Delta t^{n+1}}{\Delta t^{n+1} + \Delta t^{n}}
\quad \mathrm{and} \quad
\tau^{n-\frac{1}{2}} = \dfrac{\Delta t^{n}}{\Delta t^{n} + \Delta t^{n-1}} \, .
\end{equation}
In the above equations, the superscript $n$ is a label for the current time instance, $t^n$, i.e. the most advanced time at which the solution has been computed.
Also, $\Delta t^{n+1}$ is the time-step used to advance the solution from the $n$-th time level to the next one.
The time-step is determined via the CFL condition.
For example, in the simulations presented herein, the maximum Courant number was set equal to 0.2.

The above algorithm has been implemented in staggered grids which are not prone to the well-known odd-even decoupling between the pressure and velocity fields.
More specifically, the variables $p$, ${\bm E}$ and ${\rho_{\mathrm{el}}}$ are computed at the cell centers whereas the velocity vector ${\bm u}$ and the source term ${\bm F}_\mathrm{s}$ are calculated at the center of the cell faces.
The resulting linear system of equations is solved using the Jacobi method which facilitates the straightforward parallelization of the code.
At each time-step, equations~(\ref{eq:mass}), (\ref{eq:mom}), and~(\ref{eq:gauss}) are relaxed until the $L_2$ norm of the error of each variable drops by three orders of magnitude from its initial value.

The position of the particles is calculated in the Lagrangian frame of reference.
 More specifically, Newton's second law of motion is solved separately for each particle,
\begin{equation}
\label{eq:newton}
\dfrac{\mathrm{d} \bm u_{\textrm p}}{\mathrm{d} t} \;=\; {\bm f}_{\mathrm{fl}} + {\bm f}_{\mathrm{coll}} + {\bm f}_{\mathrm{el}} + {\bm f}_{\mathrm{g}}\, .
\end{equation}
In this equation, ${\bm f}_{\mathrm{fl}}$ represents the acceleration due to the force exerted by the surrounding fluid on the particle (aerodynamic force), ${\bm f}_{\mathrm{coll}}$ represents the acceleration due to particle-particle or wall-particle collisions, and ${\bm f}_{\mathrm{el}}$ is the acceleration due to electrostatic forces.
Finally, ${\bm f}_{\mathrm{g}}$ stands for the gravitational acceleration.

The acceleration due to the aerodynamic force is computed by the following expression, \citep{Som12},
\begin{equation}
{\bm f}_{\rm fl} \;=\; - \frac{3 \rho}{8 \rho_{\rm p} r} C_{\rm d} \vert {\bm u}_{\rm rel} \vert \,{\bm u}_{\rm rel}\,,
\end{equation}
where $\rho_\mathrm{p}$ is the particle density, $C_{\rm d}$ is the particle drag coefficient and ${\bm u}_{\rm rel}$ the particle velocity relative to the fluid,  ${\bm u}_{\rm rel} = {\bm u}_{\rm p} - {\bm u}$.
The particle drag coefficient, $C_{\rm d}$, is computed as a function of the particle Reynolds number, $Re_{\rm p} = 2 \vert {\bm u}_{\rm rel} \vert \, r / \nu$, according to the relation provided by \citet{Sch33},
\begin{align}
\label{eq:drag}
C_{\rm d} =
\begin{cases}
\quad \dfrac{4}{Re_{\rm p}} \left(6 + Re_{\rm p}^{2/3} \right)\ & \text{for}~Re_{\rm p} \leq 1000\,,\\ \nonumber \\
\quad 0.424 \,, &\text{for}~Re_{\rm p} > 1000 \,.
\end{cases}
\end{align}

Also, the acceleration due to the net effect of gravity on the particle reads
\begin{equation}
{\bm f}_{\rm g} = \left( 1- \dfrac{\rho}{\rho_{\rm p}} \right) \bf{g} \, ,
\end{equation}
where ${\bf g}$ is the gravitational acceleration.

As regards the modeling of particle-particle collisions, common methodologies include the hard-sphere approach of \citet{Cam85} and the soft-sphere approach of \citet{Cun79}.
According to the hard-sphere approach, each particle is propagated to the next collision, so that the time increment is adaptive and independent of the time-step used to integrate the Navier-Stokes equations, cf.~equation(\ref{eq:dt}).
Consequently, if the particulate phase is dense and the collision frequency between particles is high, then the time increment is reduced significantly and the computation of the particle trajectories becomes computationally expensive.

On the other hand, the soft-sphere approach allows the volumes of different particles to occupy at the same time the same space.
At each instance, the resulting collisional particle acceleration is computed and deduced from the amount of overlapping of their volumes.
The advantage of this approach is that it allows for a constant time-step.
However, to obtain physically realistic results, the particle overlapping has to be as limited as possible, which in turn implies that the computational time-step has to be kept small.

\begin{figure}
\begin{center}
\begin{tikzpicture}[scale=2.5]

%
\newcommand\radp{0.4}

\coordinate (A) at (0.2,.5);
\coordinate (B) at (2,.5);

\shade[ball color = gray,opacity=.8]  (A) circle (\radp cm);
\shade[ball color = gray,opacity=.8]  (B) circle (\radp cm);

\draw[arrows=-latex,thick,dotted] (A) -- +(1.27,.6) coordinate (C);
\draw[arrows=-latex,thick,blue!100] (A) -- +(.805,0.38) node [above,pos=0.6] {$\bm v^*_1$} coordinate (v1);

\draw[arrows=-latex,semithick] (B) --  +(-100:\radp cm) node[left,pos=.8] {$r_2$};
\draw[arrows=-latex,semithick] (A) --  +(-100:\radp cm) node[left,pos=.8] {$r_1$};

\shade[ball color = gray,opacity=.2]  (C) circle (\radp cm);

\draw[arrows=-latex,thick,blue!100,dotted] (C) -- +(.8,0.38) node [above,pos=0.6] {$\bm v^*_1$} coordinate (D);
\draw[arrows=-latex,thick] (A) -- (B) node (z1) [below,pos=0.6]{$\bm z_{12}$};
\draw[arrows=-latex,thick,dotted] (C) -- (B) node [right,pos=.7] {$\bm n$};
\draw[arrows=-latex,thick,blue!100,shorten >=1.3cm] (C) -- (B) node [left,pos=.35] {$\bm v^*_{n,1}$};
\draw[arrows=-latex,thick,blue!100] (1.65,.9) -- (D) node [right,pos=.5] {$\bm v^*_{t,1}$};


\draw[fill] (A) circle (.5pt) node[anchor=east] {$\bm x_1$};
\draw[fill] (B) circle (.5pt) node[anchor=west] {$\bm x_2$};
\draw[fill] (C) circle (.5pt) node[anchor=south] {$\bm x_\mathrm{c,1}$};

\end{tikzpicture}
\vspace{-5pt}
\caption{Sketch of the parameters used in the ray casting collision detection approach in the rest frame of particle~2.}
\label{fig:raycast}
\end{center}
\end{figure}
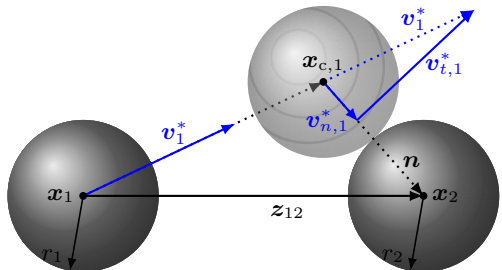
In our computational tool pafiX, we implemented a variant of the hard-sphere approach, namely, the ray casting method.
It was proposed by \citet{Roth82} in the field of computer graphics to solve a variety of intersection problems and was subsequently extended to collision detection by \citet{Schr01}.
For our purposes, we adapted this approach to detect collisions between spherical particles which allows the use of larger time-steps.
More specifically, the time-step can be the same as the time-step $\Delta t$ for the Navier-Stokes equations.

Due to the size of the time-step, two particles may not be in contact with each other in the next two time-instances, say at $t^{n+1}$ and $t^{n+2}$, even if they collide at some intermediate time instance (between $t^{n}$ and $t^{n+2}$).
According to our implementation of the ray casting method, collisions between two particles (see figure~\ref{fig:raycast}) are anticipated when a number of criteria are fulfilled.
These criteria are the following, sorted in order of ascending computational cost.
\begin{enumerate}
\item Check whether the particles are in the same or a neighboring computational cell.
This check involves the  assumption that particles propagate with a velocity of the order of the fluid phase and therefore do not transverse more than one computational cell per time step.
\item Check if the vector connecting both particle centroids, ${\bm z}_{12}$, and the relative velocity $\bm{v}^*_1 = \bm{v}_1 - \bm{v}_2$ form an acute angle.
This test examines if the particles are propagating towards each other, that is, if they are on a collision course.
\item Determine whether the particles will collide with each other if they propagate with their current velocity, i.e.~if they would collide at a later time instance.
\item Control if the particle collision anticipated by the previous condition takes place during the following time increment, $\Delta t^{n+1}$.
\end{enumerate}
If any of these conditions is not fulfilled, then the subsequent conditions are not further checked and, instead, the particles are propagated with ${\bm f}_\mathrm{coll}=0$.

If all criteria are fulfilled, a collision is anticipated to take place.
In general, this will be an oblique collision.
The collision parameters are computed in the rest frame of one of the colliding particles.
For  the collision depicted in figure~\ref{fig:raycast}, the particle velocities in the rest frame of  particle~2 are $\bm{v}^*_1$ and $\bm{v}^*_2$, with $\bm{v}^*_2 =0$.
Then, particle~1 is propagated to the fictitious point $\bm{x}_{\mathrm{c},1}$ so that it gets in contact with particle~2.
The unit vector along the line that connects the centers of the two particles is $\bm{n} = \bm{x}_{\mathrm{c},1} - \bm{x}_2$.
Further, the particle velocity can be decomposed into a component in the direction of $\bm{n}$,
\begin{equation}
\bm{v}^*_{\mathrm{n},1}\;=\; \dfrac{\bm{n} \cdot \bm{v}^*_1}{|\bm{n}|^2} \bm{n}\,,
\end{equation}
and a component that lies on the tangential plane of contact,
\begin{equation}
\bm{v}_{\mathrm{t},1}^* \;=\; \bm{v}^*_1 - \bm{v}^*_{\mathrm{n},1} \,.
\end{equation}
The latter component, $\bm{v}_{\mathrm{t},1}^*$, does not change during collision.

The velocity of a particle after collision is given as the sum of the rest-frame velocity and the post-collision velocity component in the direction of $\bm{n}$.
Accordingly, the post-collision velocity of particle~1 reads,
\begin{equation}
\label{eq:post1}
\bm{v}'_1 \;=\; \bm{v}_2 + \bm{v}_{\mathrm{t},1}^* + \dfrac{r_1^3 - e r_2^3}{r_1^3  + r_2^3} \bm{v}^*_{\mathrm{n},1} \,,
\end{equation}
while, the post-collision velocity of particle~2 reads,
\begin{equation}
\label{eq:post2}
\bm{v}'_2 \;=\; \bm{v}_2 + (1+e)\dfrac{r_1^3}{r_1^3  + r_2^3} \bm{v}^*_{\mathrm{n},1} \, .
\end{equation}
where $r_1$ and $r_2$ are the radii of the particles and $e$ is the restitution ratio.
For the derivation of  equations~(\ref{eq:post1}) and~(\ref{eq:post2}), it was assumed that the particles are made of the same material and, therefore, the particle densities are equal.
Also, for the derivation of (\ref{eq:post2}) we took into account the fact that in the rest frame of particle~2 the velocity of this particle is identically zero.

With regard to wall-particle collisions, we remark that when a particle impacts a wall, the wall-normal velocity component changes according to
\begin{equation}
\label{eq:udashn}
\bm{v}' \cdot {\bm n} \;=\; e \, \bm{v} \cdot {\bm n} \,,
\end{equation}
with $\bm{n}$ being in this case the unit vector normal to the wall.
On the other hand, the tangential velocity component of the particle remains constant.

Finally, the last term in equation~(\ref{eq:newton}) describes the electrostatic force acting on a given particle and is given by
\begin{equation}
\label{eq:fel}
{\bm f}_{\mathrm{el}} = \dfrac{Q \, {\bm E}}{m_{\mathrm p}}
\end{equation}
where $Q$ and $m_{\mathrm p}$ are, respectively, the charge and the mass of the particle.
This force is calculated by superposing the Coulomb interactions between the particle and its neighboring particles (say, the particles that at a given time instance reside in the same computational cell) according to the hybrid method proposed by \citet{Gro17e} which is similar to that of \citet{Kol16}.
This approach is more suitable for wall-bounded flows as compared to Ewald summation and the P$^3$M method \citep{Yao18}.

Finally, it is worth mentioning that the entire algorithm is implemented in pafiX in a Fortran~90 computer code parallelized via the Message Passing Interface protocol (MPI).
As described in Appendix~A, the algorithm scales excellently on up to 256 processors.

\section{Results and discussion}

\subsection{Numerical set-up}%
\label{sub:set_up}

\begin{figure}
\centering
\subfigure[]{
\begin{tikzpicture}[scale=.9]
 \draw [->, thick] (-.8,-.8) -- (0,0) node [below,xshift=5] {$x$};
 \draw [thick] (-1,-1) rectangle (1,1);
 \draw [thick] (1,3) -- (3,3) -- (3,1);
 \draw [thick] (-1,1) -- (1,3);
 \draw [thick] (1,1) -- (3,3);
 \draw [thick] (1,-1) -- (3,1);
 \draw [thick,|-|] (-1.5,-1) -- node [above,rotate=90] {$H$} (-1.5,1);
 \draw [thick,|-|] (-1.3,1.2) -- node [above,rotate=45] {$L$} (.7,3.2);
\end{tikzpicture}
\label{fig:grida}}
\qquad\qquad
\subfigure[]{
\begin{tikzpicture}
    \node[anchor=south west,inner sep=0] (Bild) at (-.03,-.13)
    {\includegraphics[trim=0cm 0cm 0cm 0cm,clip=true,width=0.25\textwidth]{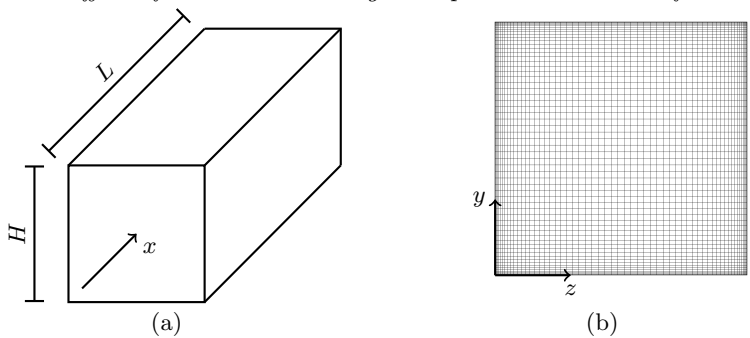}};
    \begin{scope}
	\draw [->,thick](0,0) -- (1,0) node[anchor=south,below]{$z$} ;
	\draw [->,thick](0,0) -- (0,1) node[anchor=south,left]{$y$} ;
    \end{scope}
\end{tikzpicture}
\label{fig:gridb}}
\caption{(a) Graphic representation and dimensions of the flow domain.
The arrow points to the direction of the flow.
(b) Computational mesh in the cross section, consisting  of 60 $\times$ 60 cells.}
\end{figure}

We consider gravity-driven,  particle-laden flows in a duct with square cross section, as depicted in figure~\ref{fig:grida}.
The direction of the flow is along the $x$-axis and aligned with the gravity vector.
In other words, the gravity vector is given by ${\bm g}=(9.81,0,0)$ m/s$^2$.
In order to mimic a very long duct, periodic boundary conditions are applied on the streamwise $x$-direction, whereas solid walls are assumed in the $y$- and $z$-directions.
Also, the duct is assumed to be grounded and fully conductive, i.e.~zero-Dirichlet boundary conditions are prescribed for the electric potential $\varphi_{\rm el}$ at the walls.

The grid is refined closed to the solid walls since the near-wall regions are the ones with the highest velocity gradients.
The distribution of grid points in both wall-normal directions ($y$- and $z$-directions) is performed in a fashion similar to that employed by \citet{Kim87} for grid refinement in the wall-normal direction in channel flows.
Thus, the grid points at the $y$-direction are located at $y_j=\cos \theta_j$ with $\theta_j = (j-1) \pi / (N-1)$ and $j=1,\ldots,N$ where $N$ is equal to the total number of points in the $y$-direction.
The number of points, $N$, and the grid-point distribution in the $z$-direction are the same as in the $y$-direction.
The resulting grid for $N=60$ is depicted in figure~\ref{fig:gridb}.
On the other hand, the mesh is uniform in the streamwise $x$-direction.
The DNS presented in this paper have been performed on the grid consisting of 180\,$\times$\,120\,$\times$\,120 cells.

The length of the domain, $L$, equals six times the duct height, $H$, which ensures the statistical independence of the flow quantities in the streamwise direction, i.e. the flow quantities at $L/2$ are not correlated to the flow quantities  at the periodic boundaries.
Therefore, our results are independent of the size of the chosen computational domain.
In the simulations presented herein, the dimensions of the duct were $H=4$~cm and $L=24$~cm.

\renewcommand{\arraystretch}{1.2}
\setlength{\tabcolsep}{8pt}
\begin{table}
 \centering
 \caption{Parameters of the numerical simulations.}
 \label{tab:param}
 \medskip
 {
 \begin{tabular}{ccrcccc}
 $Re_\mathrm{\tau}$  & $\rho_\mathrm{p} / \rho$  & $St$ ~ &   $Q$  &   $\rho_N$ &  $St_\mathrm{el}$  & $F_\mathrm{el}/F_\mathrm{g}$ \\
 \hline
 600   & 1000   &  31.5   & 0.0~pC   & 10$^8$/m$^3$ & 0.0 & 0\\
 600   & 1000   &  31.5   & 0.1~pC   & 10$^8$/m$^3$ & 0.1 & 0.026\\
 300   & 1000   &   7.9   & 0.0~pC   & 10$^8$/m$^3$ & 0.0 & 0\\
 300   & 1000   &   7.9   & 0.1~pC   & 10$^8$/m$^3$ & 0.1 & 0.026\\
 600   & 7500   & 197.1   & 0.0~pC   & 10$^8$/m$^3$ & 0.0 & 0\\
 600   & 7500   & 197.1   & 0.1~pC   & 10$^8$/m$^3$ & 0.3 & 0.004\\
 \end{tabular}}
\end{table}
\renewcommand{\arraystretch}{1.0}
\setlength{\tabcolsep}{1pt}

The flows parameters in our numerical simulations are presented in table~\ref{tab:param}.
The friction Reynolds number is defined as $Re_\mathrm{\tau} = u_\mathrm{\tau} H / \nu$, where $u_\mathrm{\tau} = \sqrt{\tau_\mathrm{w}/\rho}$ is the friction velocity and $\tau_\mathrm{w}$ is the stress at the wall.
The Stokes number is defined as the ratio of the particle response time-scale to the flow time-scale, i.e.,~$St=\tau_\mathrm{p}/\tau_\mathrm{f}$.
The particle response time-scale is given by $\tau_\mathrm{p}=2 \rho_\mathrm{p} r^2 / (9 \rho \nu)$ with $\rho_\mathrm{p}$ being the particles density.
The flow time-scale is based on the friction velocity, namely $\tau_\mathrm{f} = \nu / u^2_\mathrm{\tau}$.
Finally, $Q$ and $\rho_N$ stand, respectively, for the electrostatic charge and number density of the particles.
These quantities are expressed in a dimensionless form by the electrostatic Stokes number that we define similar to \citet{Bou20} as the ratio of the particle response time-scale to a time-scale characterizing the Coulombic particle interaction, i.e. $St_\mathrm{el}=\tau_\mathrm{p}/\tau_\mathrm{el}$.
Therein, $\tau_\mathrm{el}$ was derived as $2 \left(\pi\,\varepsilon\,m_\mathrm{p}/\left(\rho_N\,Q^2\right)\right)^{1/2}$.
Alternatively, $Q$ and $\rho_N$ can be non-dimensionalized as the ratio of the Coulombic to gravitational forces~\citep{Kol18b}, $F_\mathrm{el}/F_\mathrm{g}=Q^2\rho_N^{2/3}/\left(4\,\pi\,\varepsilon\,m_\mathrm{p}\,g\right)$.
In the definition of $St_\mathrm{el}$ and $F_\mathrm{el}/F_\mathrm{g}$ the reference length-scale is the distance between the centers of two particles assuming a uniform distribution in the domain.

According to table~\ref{tab:param}, we investigate the effect of electrical forces for three different flow conditions:
First, a flow of $Re_\mathrm{\tau}=600$ and $\rho_\mathrm{p} / \rho=1000$ is simulated.
Further, $Re_\mathrm{\tau}$ is reduced to 300 and, finally, $\rho_\mathrm{p} / \rho$ is increased to 7500.
For these three cases, the flow and particle dynamics are studied for uncharged particles.
Then, we investigate how these flows are affected if each particle carries an electrostatic charge of $Q=0.1$~pC.
More specifically, the particle number density in all cases is 10$^8/$m$^3$.
This implies that the particulate phase inside the computational domain consists of 38400 particles.
Also, the particles are monodisperse and spherical, with a radius $r=25~\upmu$m.
It is worth mentioning that we assume there is no charge exchange during particle-particle or wall-particle collisions.

The validation of the flow solver is presented in Appendix~B.
Additionally to the parameters in table~\ref{tab:param}, we explored the effect of higher charges on the particles.
These results are briefly discussed in Appendix~C.

\subsection{Particle concentrations and dynamics}

\begin{figure*}
\begin{center}
\subfigure[$Re_\mathrm{\tau}=$~600, $\rho_\mathrm{p}/\rho=$~1000, $F_\mathrm{el}/F_\mathrm{g}=$~0]{\includegraphics[trim=0cm 0cm 0cm 0cm,clip=true,width=0.47\textwidth]{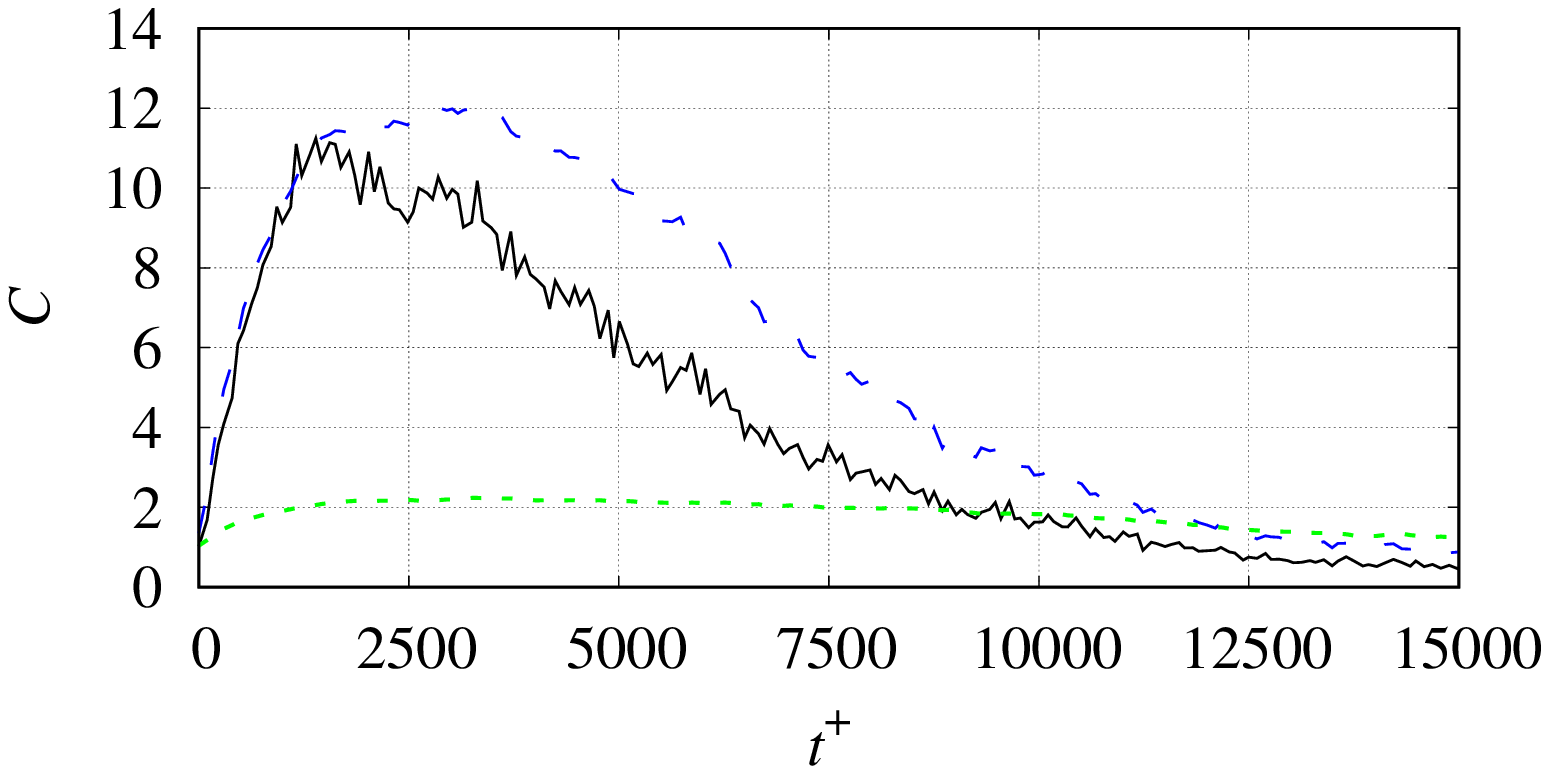}\label{fig:convergence01036}}\quad
\subfigure[$Re_\mathrm{\tau}=$~600, $\rho_\mathrm{p}/\rho=$~1000, $F_\mathrm{el}/F_\mathrm{g}=$~0.026]{\includegraphics[trim=0cm 0cm 0cm 0cm,clip=true,width=0.47\textwidth]{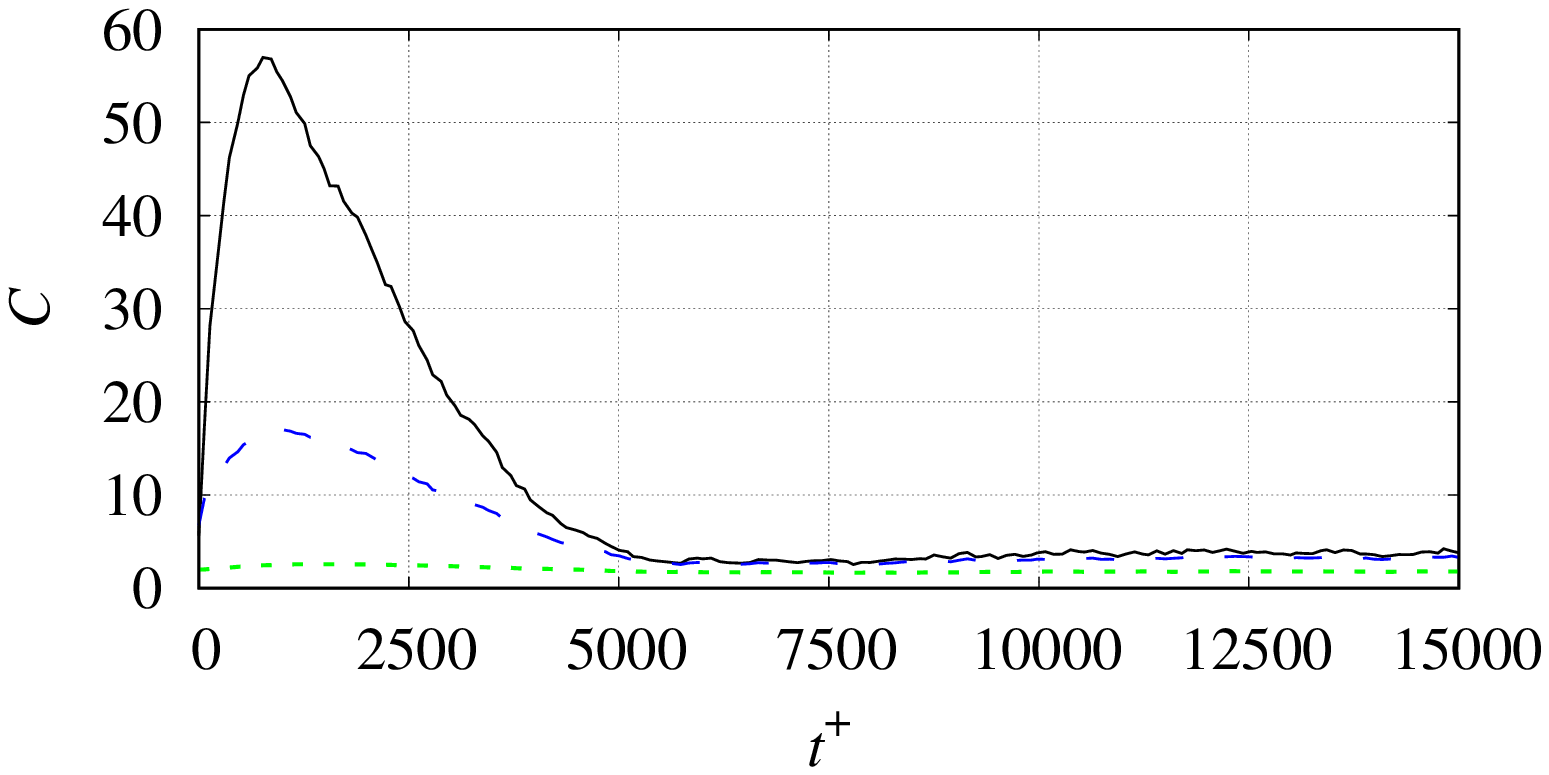}\label{fig:convergence01027}}\\
\subfigure[$Re_\mathrm{\tau}=$~300, $\rho_\mathrm{p}/\rho=$~1000, $F_\mathrm{el}/F_\mathrm{g}=$~0]{\includegraphics[trim=0cm 0cm 0cm 0cm,clip=true,width=0.47\textwidth]{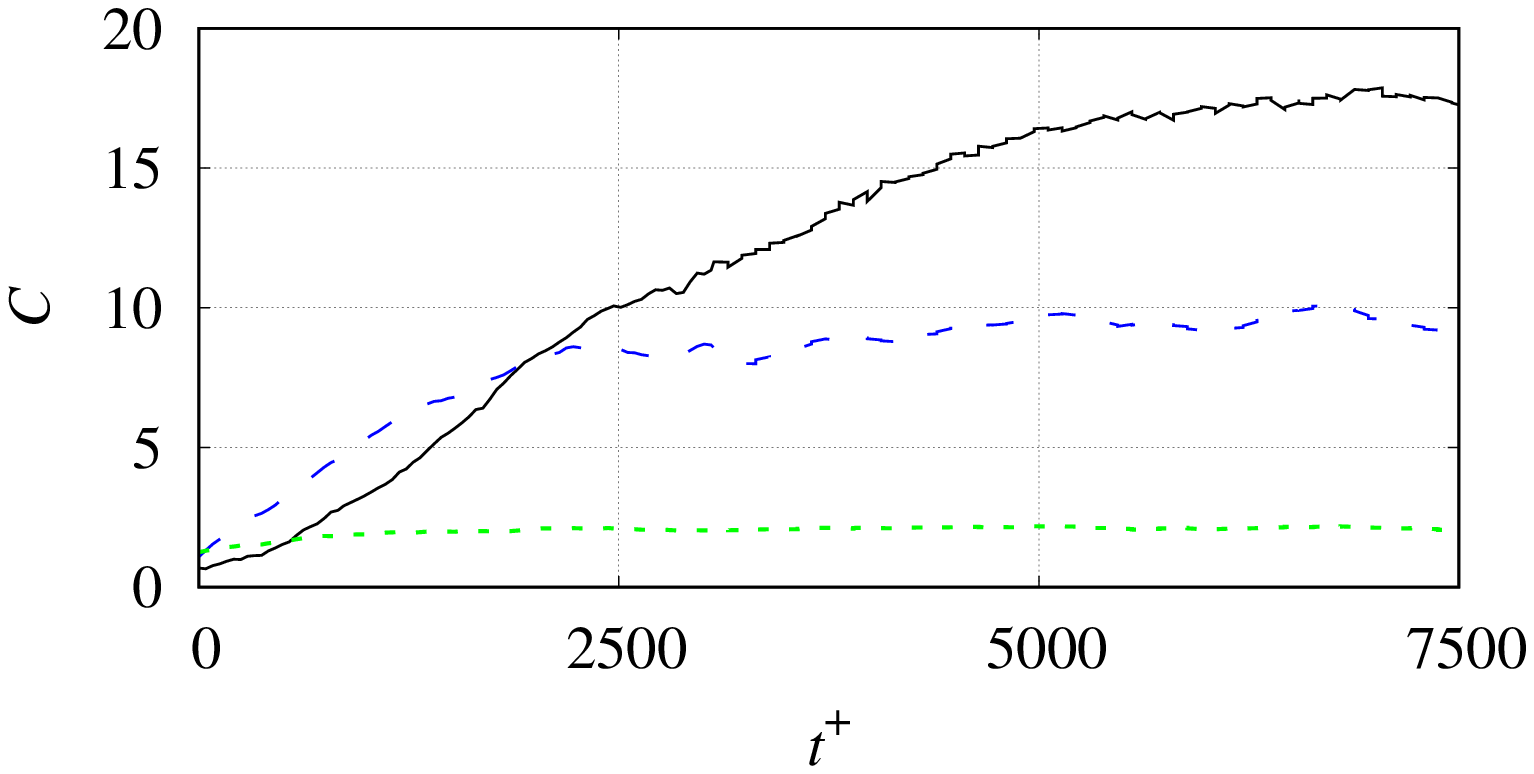}\label{fig:convergence01037}}\quad
\subfigure[$Re_\mathrm{\tau}=$~300, $\rho_\mathrm{p}/\rho=$~1000, $F_\mathrm{el}/F_\mathrm{g}=$~0.026]{\includegraphics[trim=0cm 0cm 0cm 0cm,clip=true,width=0.47\textwidth]{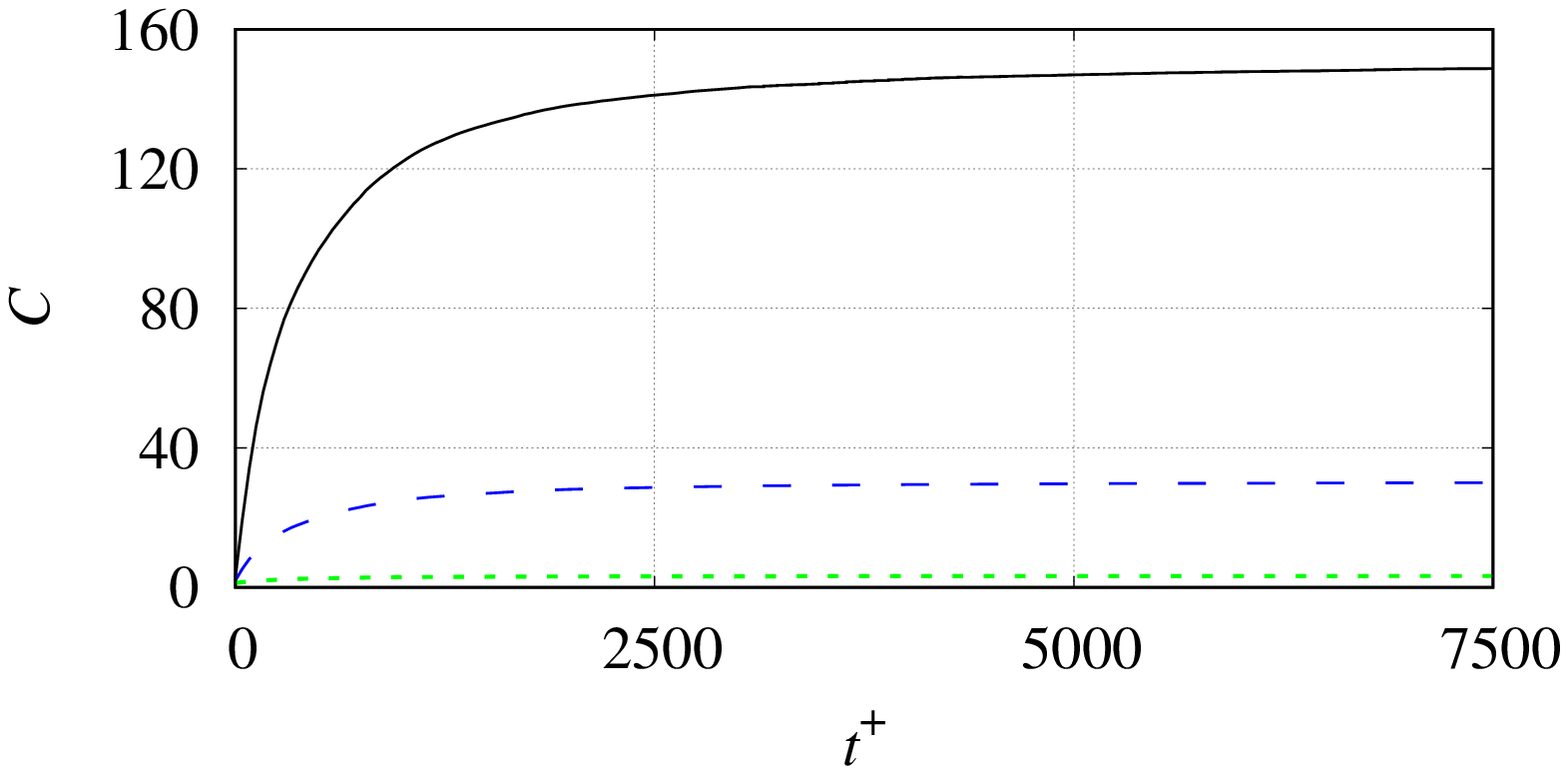}\label{fig:convergence01028}}\\
\subfigure[$Re_\mathrm{\tau}=$~600, $\rho_\mathrm{p}/\rho=$~7500, $F_\mathrm{el}/F_\mathrm{g}=$~0]{\includegraphics[trim=0cm 0cm 0cm 0cm,clip=true,width=0.47\textwidth]{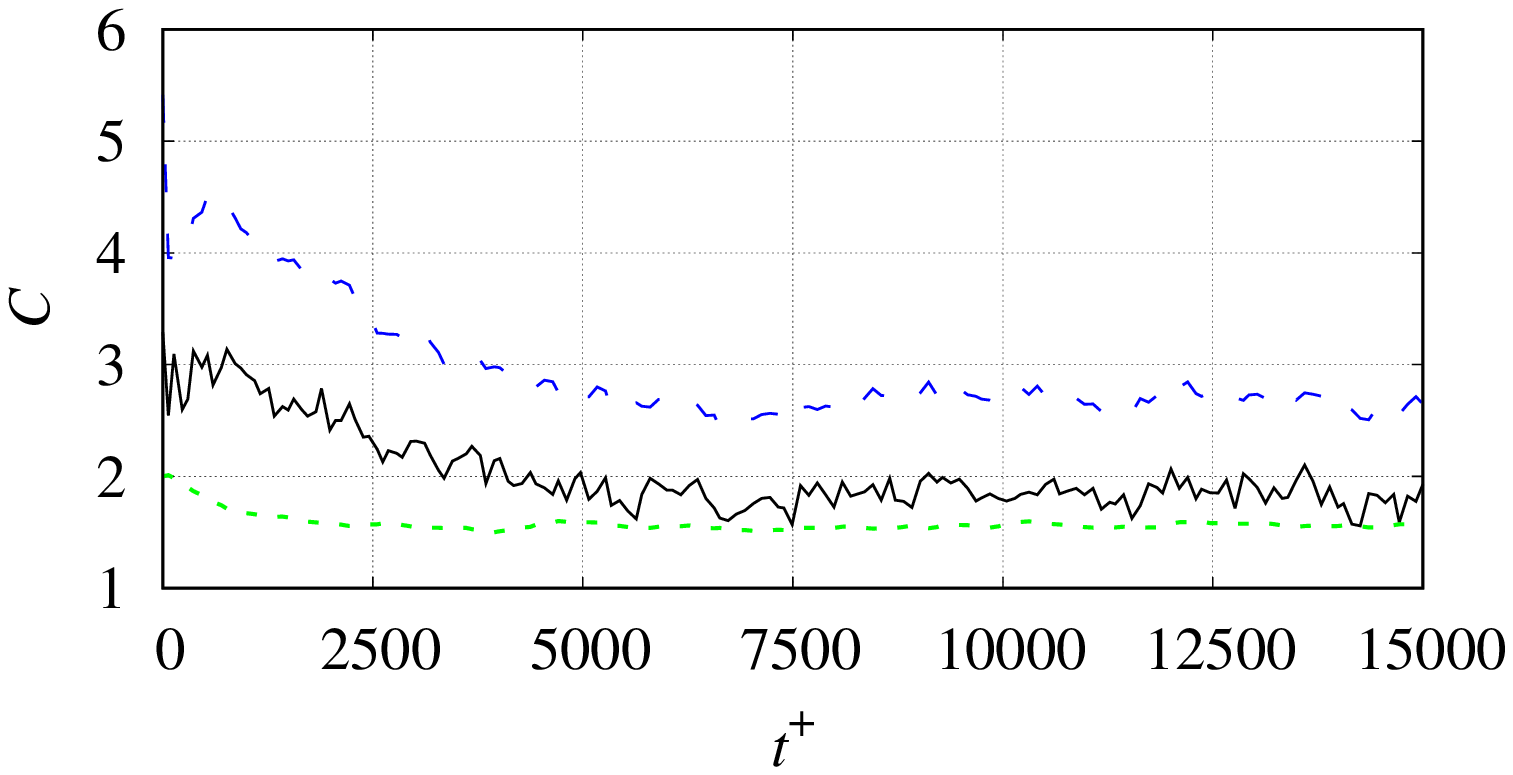}\label{fig:convergence01043}}\quad
\subfigure[$Re_\mathrm{\tau}=$~600, $\rho_\mathrm{p}/\rho=$~7500, $F_\mathrm{el}/F_\mathrm{g}=$~0.004]{\includegraphics[trim=0cm 0cm 0cm 0cm,clip=true,width=0.47\textwidth]{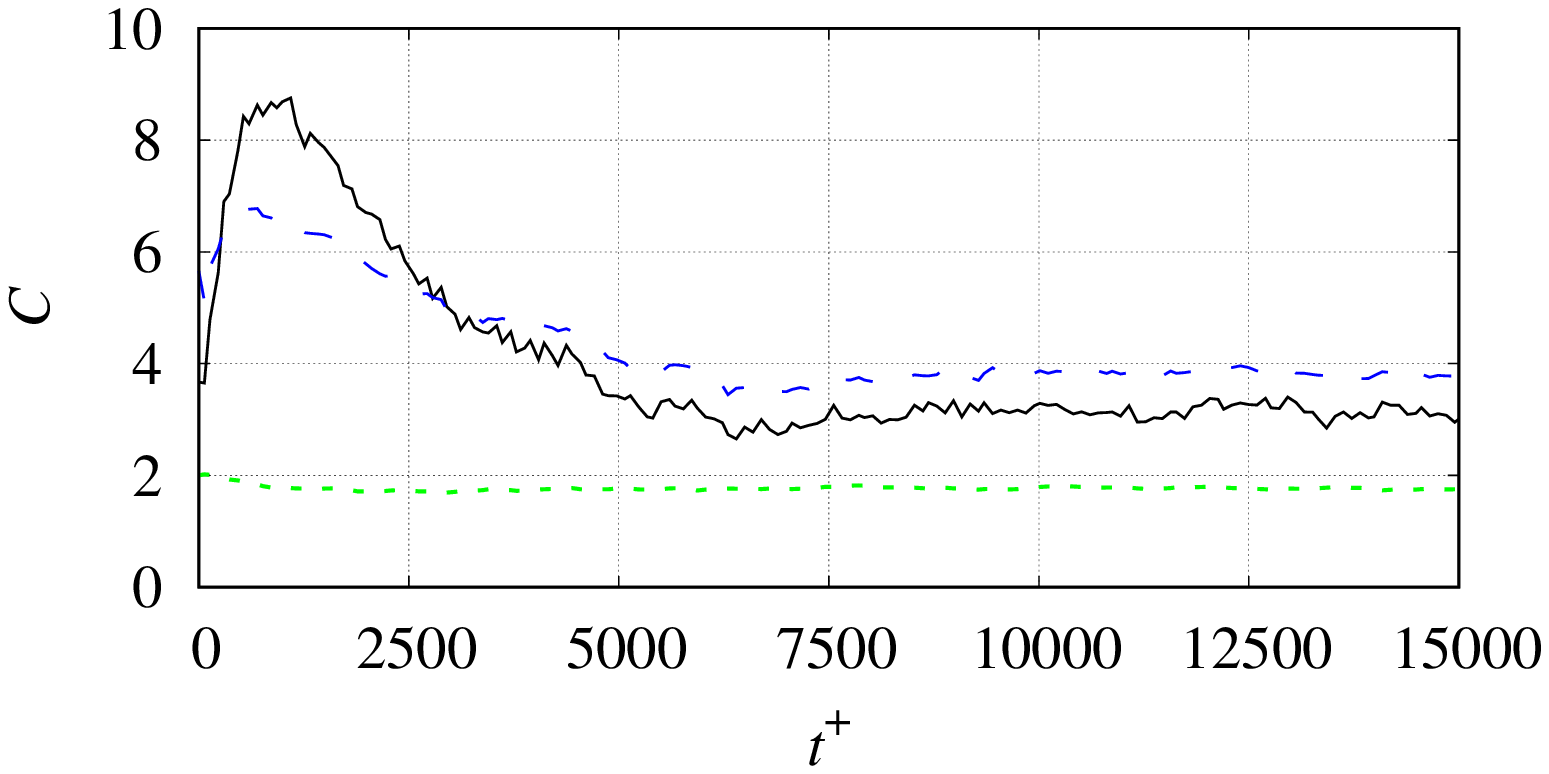}\label{fig:convergence01044}}
\end{center}
\caption[]{
Temporal evolution of the normalized particle number density $C$ near the walls. \\
\begin{tabular}{lllll}
(\raisebox{1mm}{\tikz{\draw[thick] (0,0)--(.5,0);}}) &  $y^+$ or $z^+$ $<1$  &for $Re_\mathrm{\tau}=$~600 & \quad and \quad $y^+$ or $z^+$ $<0.5$ &for $Re_\mathrm{\tau}=$~300 \\
(\raisebox{1mm}{\tikz{\draw[thick,blue,dashed] (0,0)--(.5,0);}}) &  $y^+$ or $z^+$ $<5$  &for $Re_\mathrm{\tau}=$~600 & \quad and \quad $y^+$ or $z^+$ $<2.5$  &for $Re_\mathrm{\tau}=$~300 \\
(\raisebox{1mm}{\tikz{\draw[thick,green,dotted] (0,0)--(.5,0);}}) &  $y^+$ or $z^+$ $<50$ &for $Re_\mathrm{\tau}=$~600
& \quad and \quad $y^+$ or $y^+$ $<50$ &for $Re_\mathrm{\tau}=$~300
\end{tabular}
}
\label{fig:convergence}
\end{figure*}

In this subsection, we present results for the particle concentration and particle dynamics for the six cases mentioned in table~\ref{tab:param}.
For each case, we started the simulation without particles, until the flow turbulence becomes fully developed.
Subsequently, we seeded particles at random locations and with velocities equal to those of the fluid in the same locations.
In order to save computing time and since particles tend to migrate from high to low turbulence-intensity regions due to turbophoretic drift, the injection was such that the initial particle concentrations were higher close to the walls of the duct.
It is noted however, that the predicted properties of statistically stationary particle-laden flows are totally independent of the initial location of particles.
The charge was assigned to the particles instantly when they were embedded in the flow. 
Accordingly, electrostatic forces between particles modify the dynamics of the particulate phase from the beginning.

First, we examined the temporal evolution of the particle concentrations in the near-wall regions.
The results for all six cases are presented in figure~\ref{fig:convergence}.
In these plots, the time is non-dimensionalized in terms of the frictional velocity, i.e., $t^+=t u^2_\mathrm{\tau}/\nu$.
Also, the distance from a wall is given in terms of wall units, i.e. $y^+$ and $z^+$.
For each flow case, the three curves shown in the figure represent the normalized number density $C$ of the particles residing within three different distances from the walls of the duct.
The number density $C$ is normalized with the overall number density of the particles in the duct $\rho_N$.
For example, for flows at $Re_\mathrm{\tau}=$~600, the black solid lines depict the normalized number density $C$ of the particles confined within a very thin layer from the walls.
The thickness of this layer is equal to one viscous length-scale.
Also, the blue dashed curves show the temporal evolution of the particles located within the viscous sublayer, i.e.~less than five viscous length-scales from the wall; inside this layer viscous stresses dominate over Reynolds stresses.
Further, the green dotted lines show the normalized number density of the particles located in the viscous wall-region which extends to 50 viscous length-scales, where both Reynolds and viscous stresses are significant.

In figure~\ref{fig:convergence} we readily observe that, in all cases, the concentrations in the viscous wall-region reach much faster their final values, i.e.
their values when the flow of the mixture has become statistically-stationary.
The particles are transported from the bulk of the flow towards this region by the large eddies of the flow.
Since these eddies have the highest amounts of kinetic energy, the transportation of the particles into this region occurs relatively fast.
Once inside the viscous wall-region, the particles adopt their preferential locations but are driven by small-scale fluctuations; this is, therefore, a slower process.
For the interpretation of the plots shown in figure~\ref{fig:convergence}, it is worth recalling that the frictional quantities of the flow of $Re_\mathrm{\tau}=$~300 are different from those of $Re_\mathrm{\tau}=$~600; this has an impact on the normalization of all spatial coordinates and time.

In some cases there is an overshoot of $C$ in the early stages of the evolution of the particulate phase, as can be seen in figures~\ref{fig:convergence01036}, \ref{fig:convergence01027}, and \ref{fig:convergence01044}.
More specifically, the particles are first accelerated from their initial locations towards a location very close to the wall.
Subsequently they move away from the wall and settle to their preferential location at a distance further from it.
As mentioned above, the initial particle positions are random and do not necessarily correspond to the actual dynamics of fully developed turbulence.
Therefore, this overshoot characterizes the transition of the particulate phase from its initial state to the physically meaningful and statistically stationary state.

Further, the curves in the right column of figure~\ref{fig:convergence} show that the effect of electrostatic forces takes place gradually and not abruptly.
Upon comparison to the left column it can be deduced when the particles are charged, they assume their stationary-state positions faster than when they do not carry any charge.
This is particularly noticeable in the comparison of figure~\ref{fig:convergence01036} with~\ref{fig:convergence01027} as well as in the comparison of figure~\ref{fig:convergence01037} with~\ref{fig:convergence01028}.
Further, this observation is in accordance with the finding of \citet{Gro18d} that charges of the same polarity dampen the particle velocity fluctuations.

Also, from figure~\ref{fig:convergence}, it is readily inferred that the particle concentrations at the statistically stationary state vary significantly from one case to the other, depending on the underlying turbulence dynamics and the charge they carry.
In the following paragraphs, we explore this phenomenon and examine the properties of the particle concentrations when the flow has become statistically stationary.
Statistics have been collected for a period of $t^+ =$~17\,500.
This time period is equivalent to 100 flow-through times, i.e.
the ratio of the length of the computational domain to the bulk velocity of the fluid.

\begin{figure*}
\begin{center}
\subfigure[$Re_\mathrm{\tau}$\,=\,600, $\rho_\mathrm{p}/\rho$\,=\,1000, $F_\mathrm{el}/F_\mathrm{g}=$~0]{\includegraphics[trim=6cm 0cm 3cm 0cm,clip=true,width=0.47\textwidth]{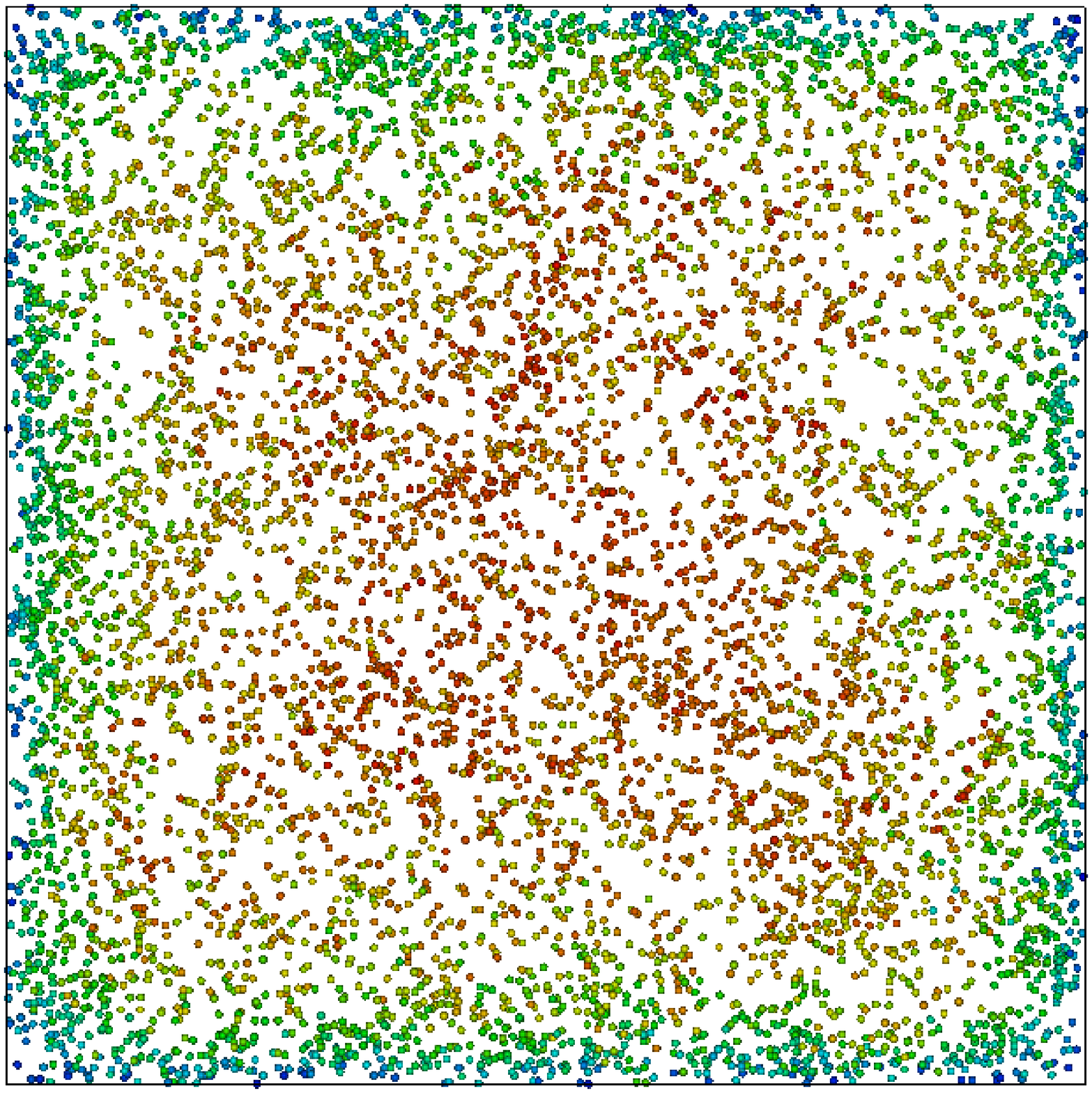}\label{fig:inst_01036}}\quad 
\subfigure[$Re_\mathrm{\tau}$\,=\,600, $\rho_\mathrm{p}/\rho$\,=\,1000, $F_\mathrm{el}/F_\mathrm{g}=$~0.026]{\includegraphics[trim=6cm 0cm 3cm 0cm,clip=true,width=0.47\textwidth]{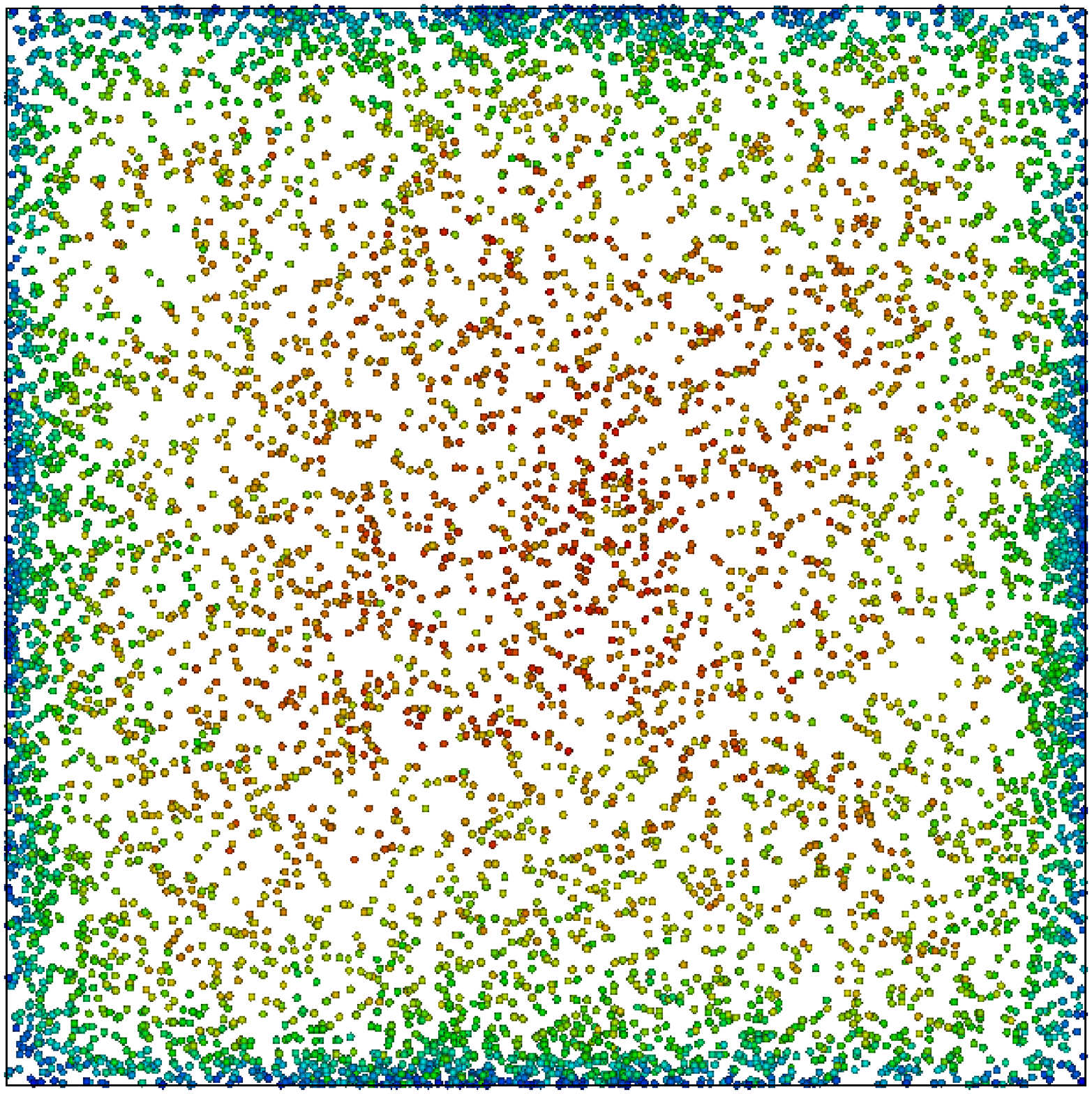}\label{fig:inst_01027}}\\ 
\subfigure[$Re_\mathrm{\tau}$\,=\,300, $\rho_\mathrm{p}/\rho$\,=\,1000, $F_\mathrm{el}/F_\mathrm{g}=$~0]{\includegraphics[trim=6cm 0cm 3cm 0cm,clip=true,width=0.47\textwidth]{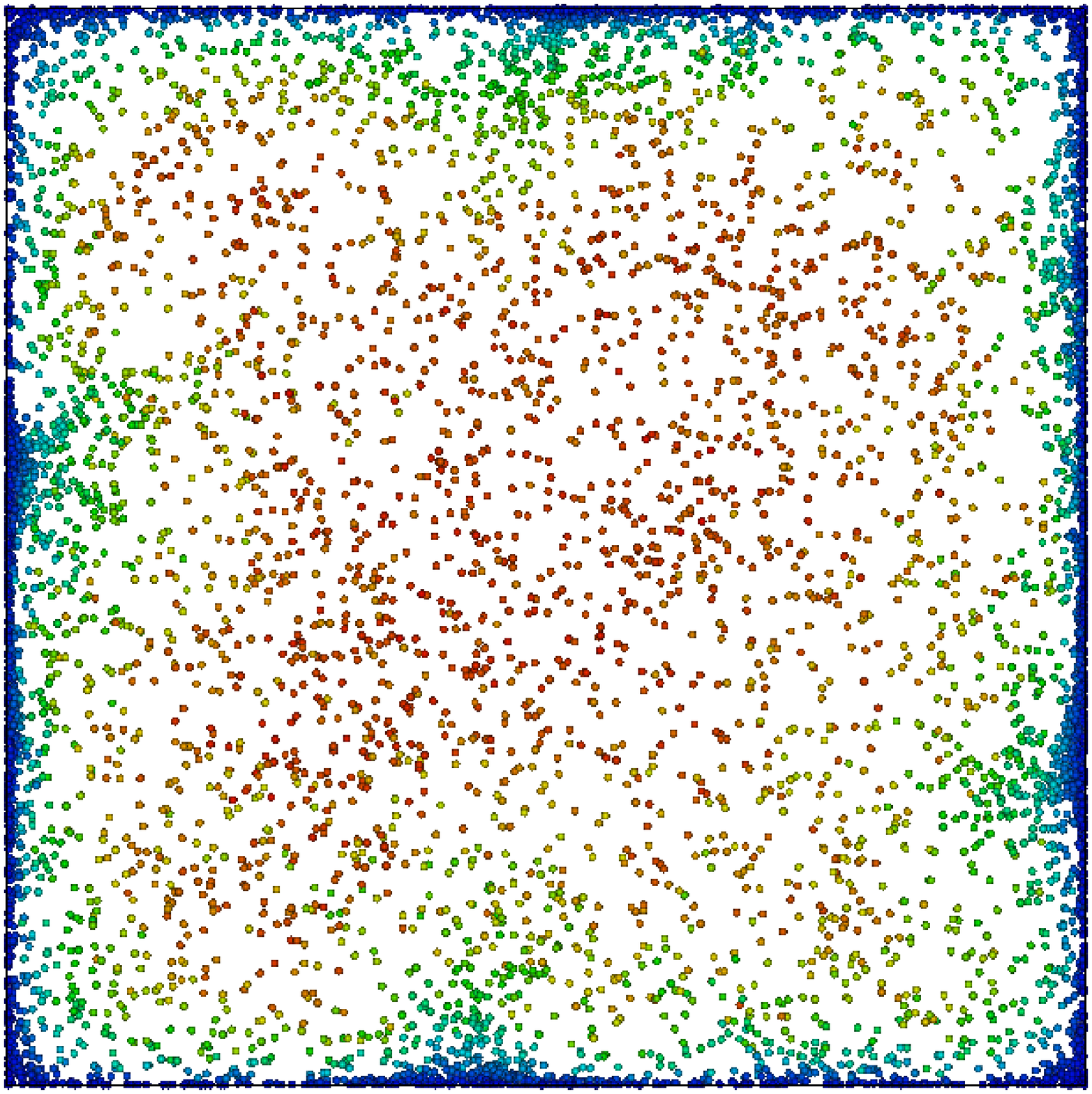}\label{fig:inst_01037}}\quad 
\subfigure[$Re_\mathrm{\tau}$\,=\,300, $\rho_\mathrm{p}/\rho$\,=\,1000, $F_\mathrm{el}/F_\mathrm{g}=$~0.026]{\includegraphics[trim=6cm 0cm 3cm 0cm,clip=true,width=0.47\textwidth]{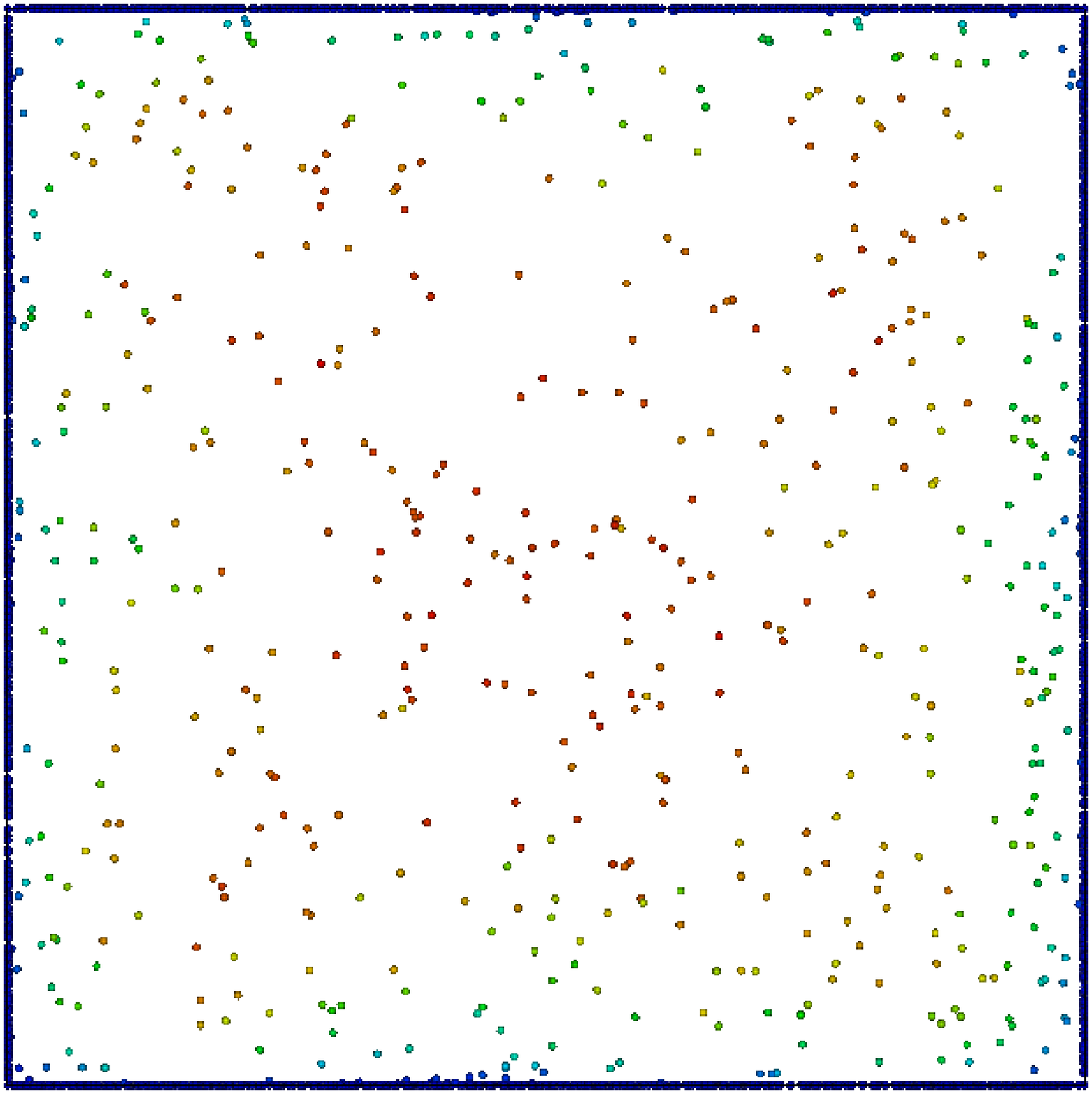}\label{fig:inst_01028}}\\ 
\subfigure[$Re_\mathrm{\tau}$\,=\,600, $\rho_\mathrm{p}/\rho$\,=\,7500, $F_\mathrm{el}/F_\mathrm{g}=$~0]{\includegraphics[trim=6cm 0cm 3cm 0cm,clip=true,width=0.47\textwidth]{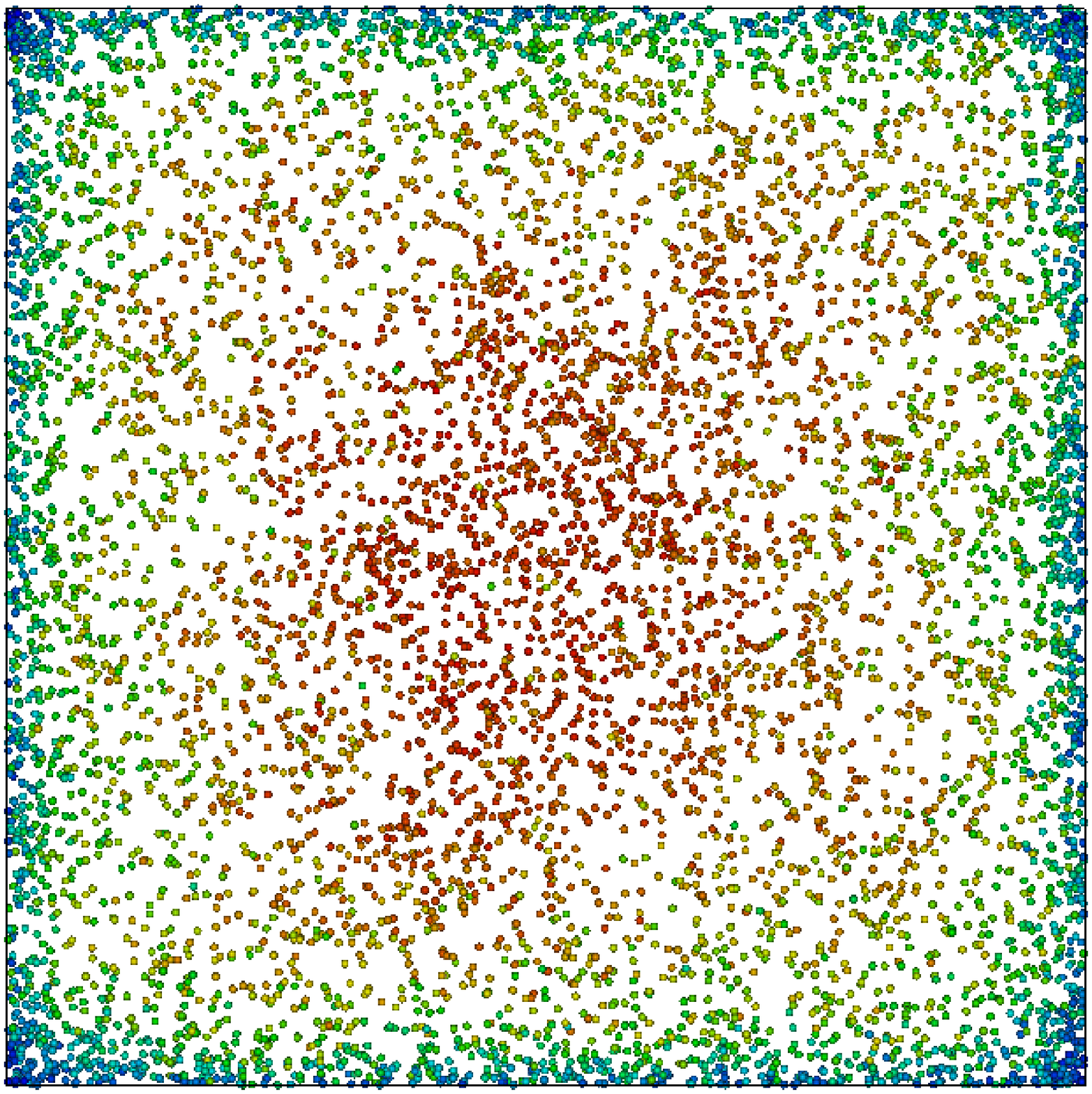}\label{fig:inst_01043}}\quad 
\subfigure[$Re_\mathrm{\tau}$\,=\,600, $\rho_\mathrm{p}/\rho$\,=\,7500, $F_\mathrm{el}/F_\mathrm{g}=$\,0.004]{\includegraphics[trim=6cm 0cm 3cm 0cm,clip=true,width=0.47\textwidth]{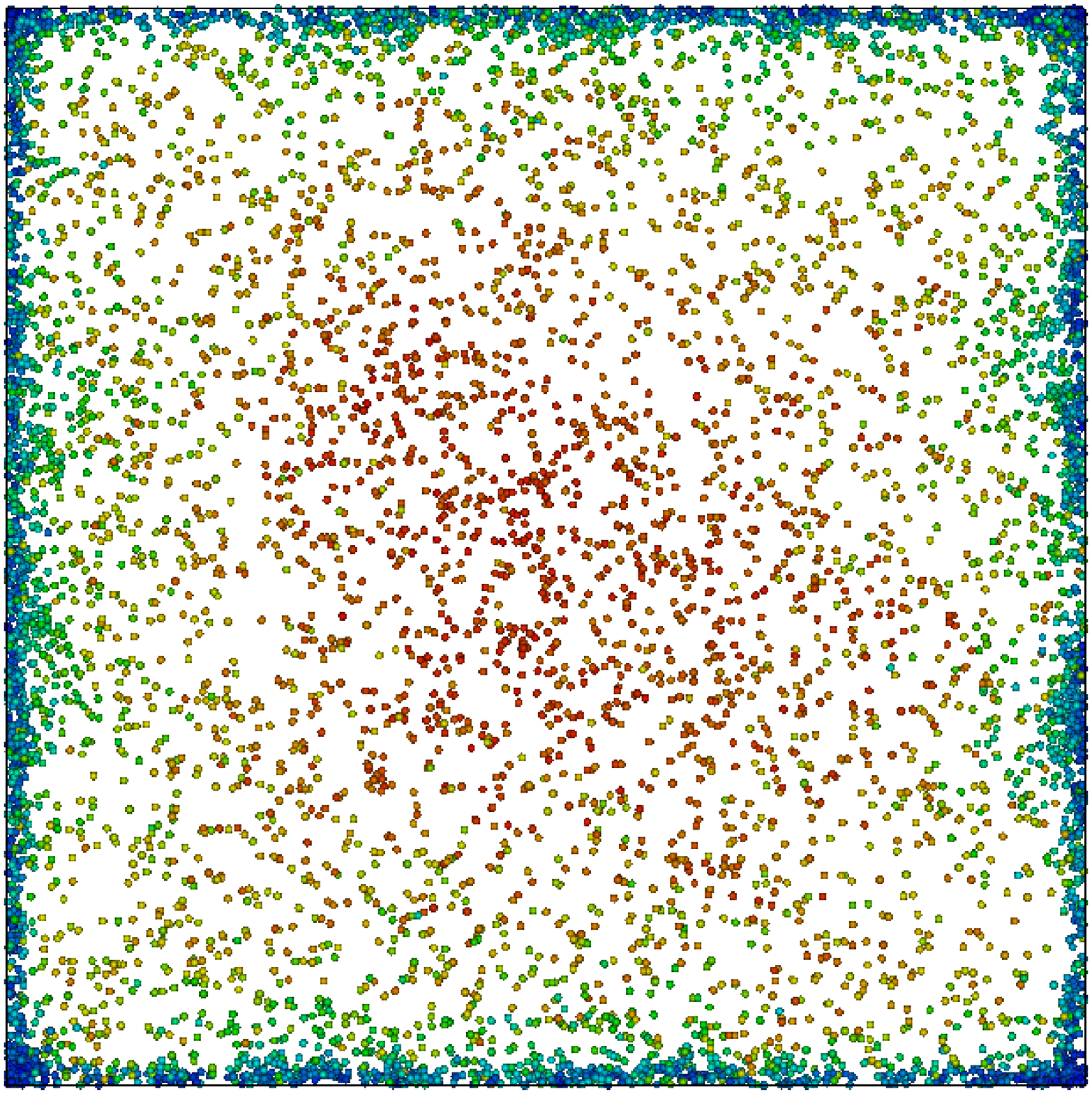}\label{fig:inst_01044}}\\ 
\end{center}
\caption{Instantaneous particle positions in the cross-section of the du ct ($y$-$z$ plane), recorded once the flow has  become statistically stationary.
The colors indicate the streamwise velocity of the particles; for each case, the red colour corresponds to the fastest and the blue colour to the slowest particle.
For visualization purposes the particle size is scaled up and only one in every five particles is shown.}
\label{fig:inst}
\end{figure*}

\begin{figure*}
\begin{center}
\subfigure[$Re_\mathrm{\tau}$~=~600, $\rho_\mathrm{p}/\rho$~=~1000, $F_\mathrm{el}/F_\mathrm{g}=$~0]{
\begin{tikzpicture}
    \node[anchor=south west,inner sep=0] (Bild) at (0,0)
    {\includegraphics[trim=3cm 7cm 2cm 7cm,clip=true,width=0.43\textwidth]{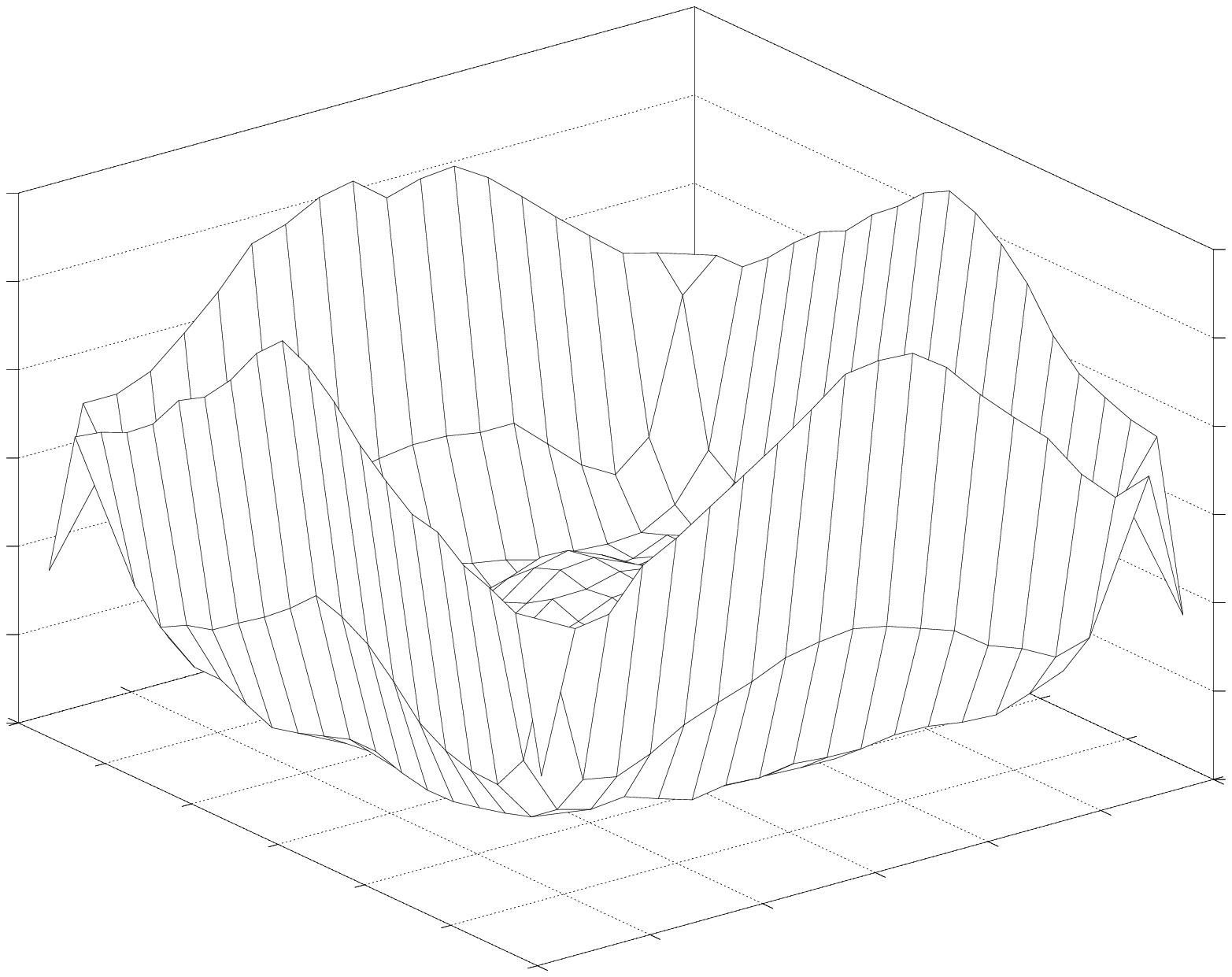}};
    \begin{scope}[x=(Bild.south east),y=(Bild.north west)]
	\draw [](0.17,0.02) node[anchor=south]{$z^+$} ;
	\draw [](0.8,0.0) node[anchor=south]{$y^+$} ;
	\draw [](0.1,0.85) node[anchor=south]{$\langle C \rangle$} ;
	\draw [](0.45,0.02) node[anchor=center]{\scriptsize 0} ;
	\draw [](0.03,0.23) node[anchor=center]{\scriptsize 600} ;
	\draw [](0.92,0.2) node[anchor=center]{\scriptsize 600} ;
    \end{scope}
\end{tikzpicture}
\label{fig:01036}}
\qquad
\subfigure[$Re_\mathrm{\tau}$~=~600, $\rho_\mathrm{p}/\rho$~=~1000, $F_\mathrm{el}/F_\mathrm{g}=$~0.026]{
\begin{tikzpicture}
    \node[anchor=south west,inner sep=0] (Bild) at (0,0)
    {\includegraphics[trim=3cm 7cm 2cm 7cm,clip=true,width=0.43\textwidth]{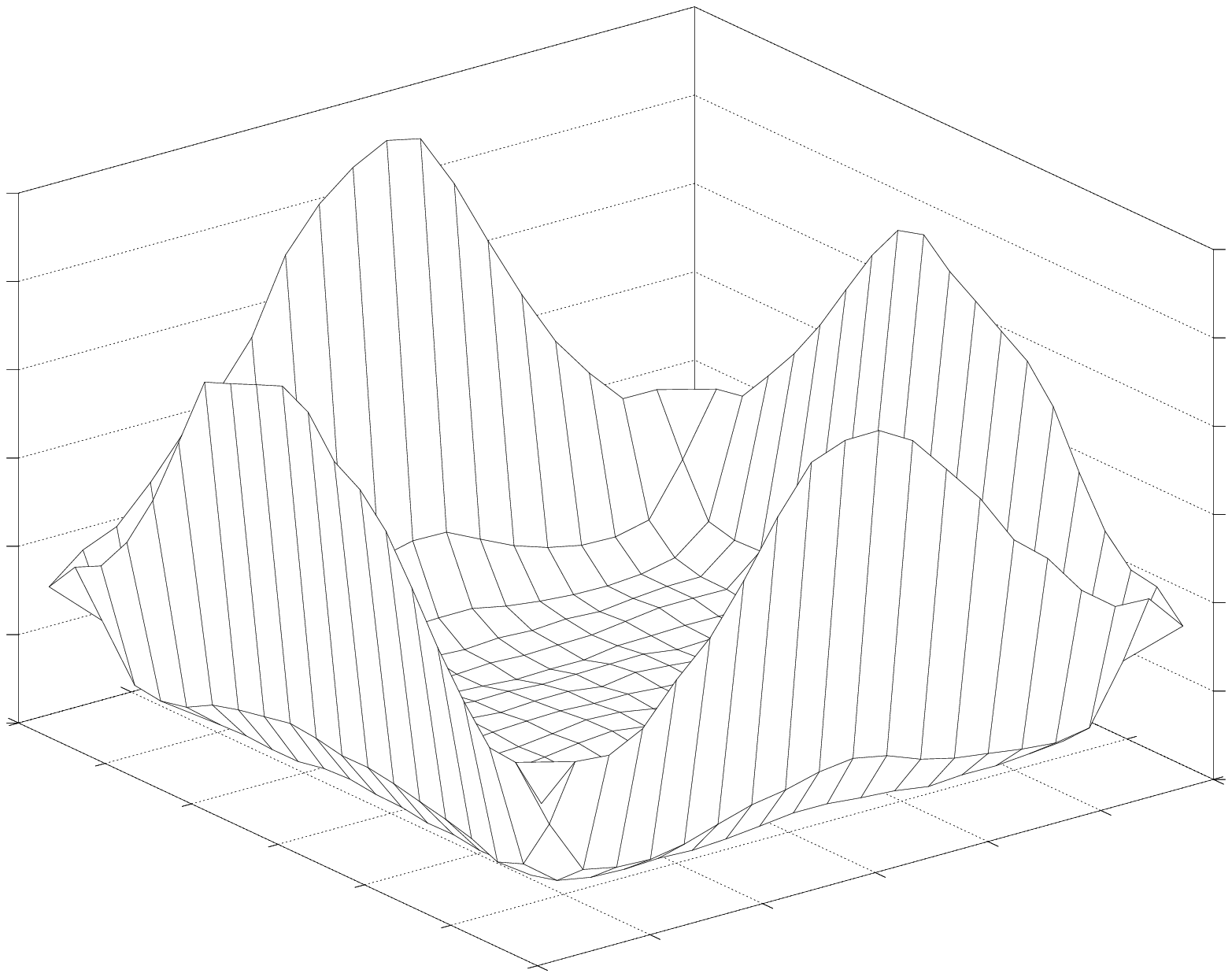}};
    \begin{scope}[x=(Bild.south east),y=(Bild.north west)]
	\draw [](0.17,0.02) node[anchor=south]{$z^+$} ;
	\draw [](0.8,0.0) node[anchor=south]{$y^+$} ;
	\draw [](0.1,0.85) node[anchor=south]{$\langle C \rangle$} ;
	\draw [](0.45,0.02) node[anchor=center]{\scriptsize 0} ;
	\draw [](0.03,0.23) node[anchor=center]{\scriptsize 600} ;
	\draw [](0.92,0.2) node[anchor=center]{\scriptsize 600} ;
    \end{scope}
\end{tikzpicture}
\label{fig:01027}}
\\
\subfigure[$Re_\mathrm{\tau}$~=~300, $\rho_\mathrm{p}/\rho$~=~1000, $F_\mathrm{el}/F_\mathrm{g}=$~0]{
\begin{tikzpicture}
    \node[anchor=south west,inner sep=0] (Bild) at (0,0)
    {\includegraphics[trim=3cm 7cm 2cm 7cm,clip=true,width=0.43\textwidth]{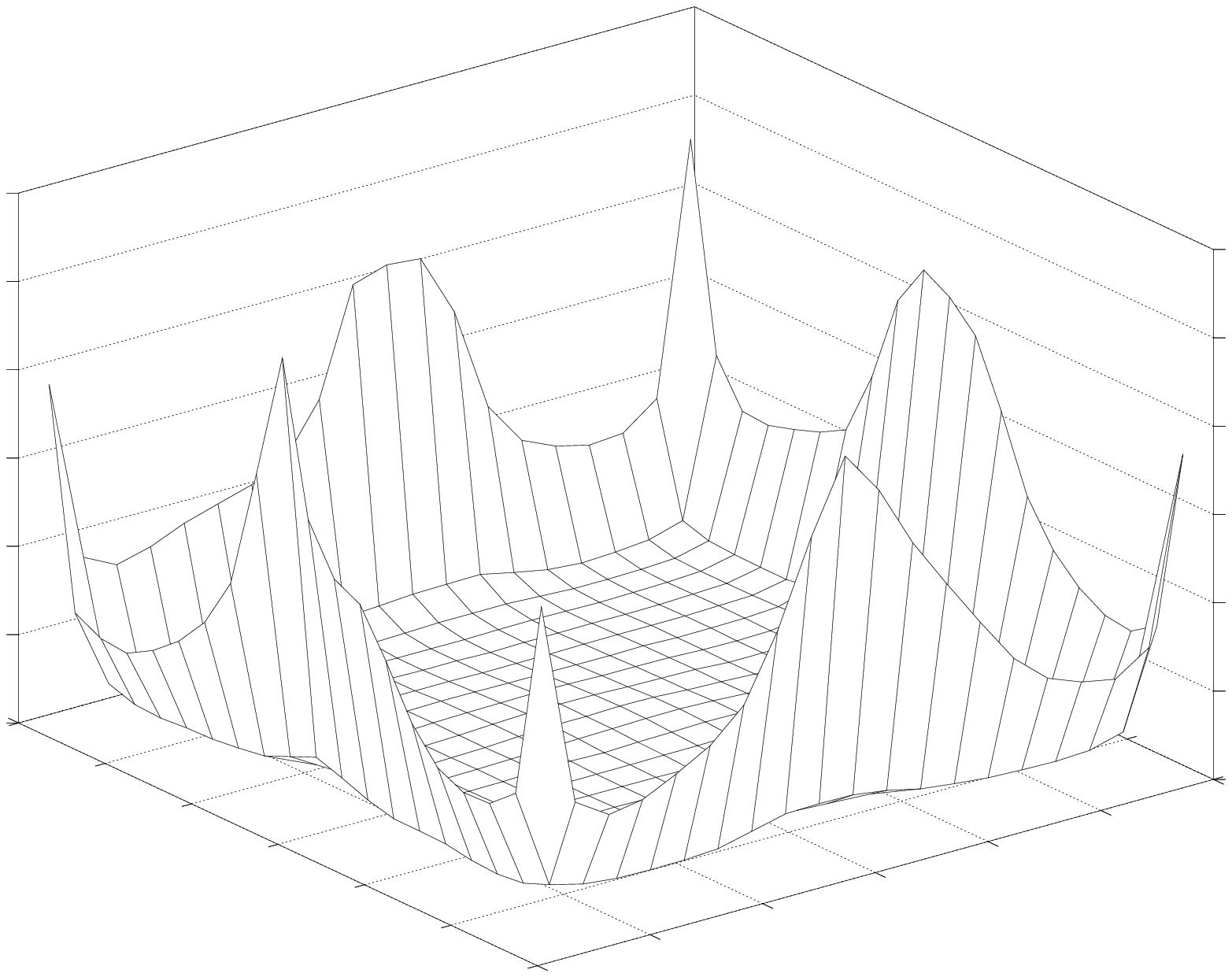}};
    \begin{scope}[x=(Bild.south east),y=(Bild.north west)]
	\draw [](0.17,0.02) node[anchor=south]{$z^+$} ;
	\draw [](0.8,0.0) node[anchor=south]{$y^+$} ;
	\draw [](0.1,0.85) node[anchor=south]{$\langle C \rangle$} ;
	\draw [](0.45,0.02) node[anchor=center]{\scriptsize 0} ;
	\draw [](0.03,0.23) node[anchor=center]{\scriptsize 300} ;
	\draw [](0.92,0.2) node[anchor=center]{\scriptsize 300} ;
    \end{scope}
\end{tikzpicture}
\label{fig:01037}}
\qquad
\subfigure[$Re_\mathrm{\tau}$~=~300, $\rho_\mathrm{p}/\rho$~=~1000, $F_\mathrm{el}/F_\mathrm{g}=$~0.026]{
\begin{tikzpicture}
    \node[anchor=south west,inner sep=0] (Bild) at (0,0)
    {\includegraphics[trim=3cm 7cm 2cm 7cm,clip=true,width=0.43\textwidth]{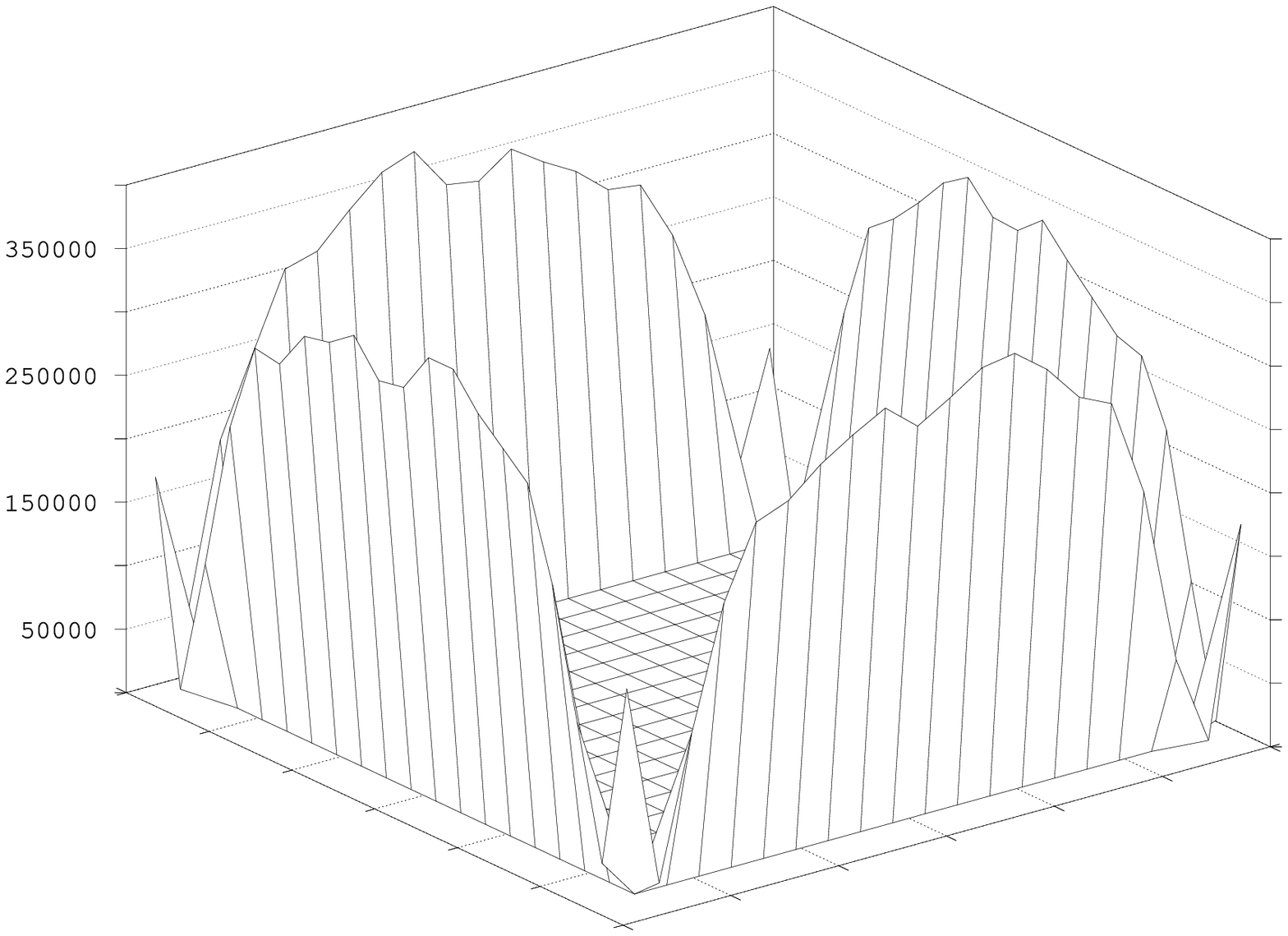}};
    \begin{scope}[x=(Bild.south east),y=(Bild.north west)]
	\draw [](0.17,0.02) node[anchor=south]{$z^+$} ;
	\draw [](0.8,0.0) node[anchor=south]{$y^+$} ;
	\draw [](0.1,0.85) node[anchor=south]{$\langle C \rangle$} ;
	\draw [](0.45,0.02) node[anchor=center]{\scriptsize 0} ;
	\draw [](0.03,0.23) node[anchor=center]{\scriptsize 300} ;
	\draw [](0.92,0.2) node[anchor=center]{\scriptsize 300} ;
    \end{scope}
\end{tikzpicture}
\label{fig:01028}}
\\
\subfigure[$Re_\mathrm{\tau}$~=~600, $\rho_\mathrm{p}/\rho$~=~7500, $F_\mathrm{el}/F_\mathrm{g}=$~0]{
\begin{tikzpicture}
    \node[anchor=south west,inner sep=0] (Bild) at (0,0)
    {\includegraphics[trim=3cm 7cm 2cm 7cm,clip=true,width=0.43\textwidth]{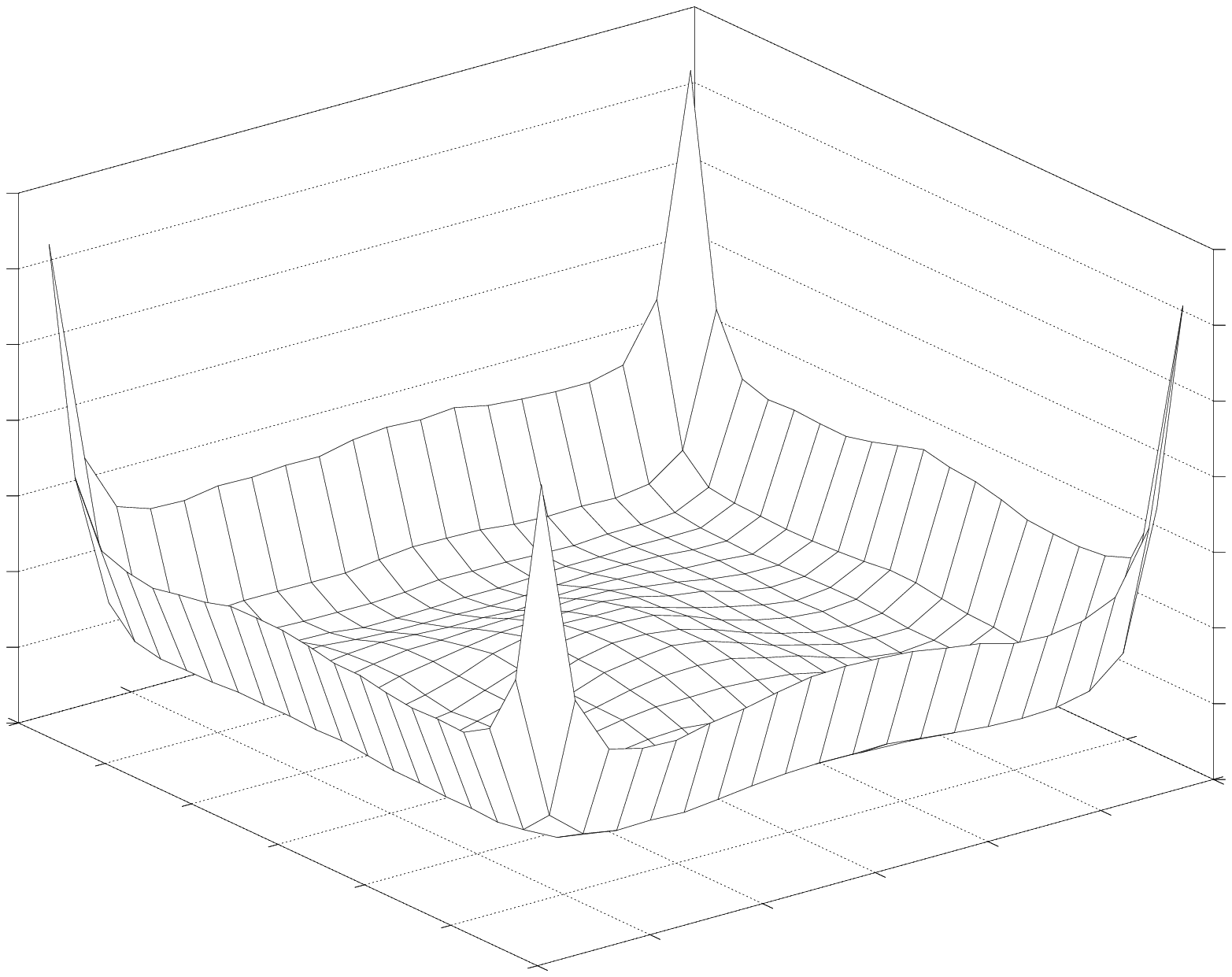}};
    \begin{scope}[x=(Bild.south east),y=(Bild.north west)]
	\draw [](0.17,0.02) node[anchor=south]{$z^+$} ;
	\draw [](0.8,0.0) node[anchor=south]{$y^+$} ;
	\draw [](0.1,0.85) node[anchor=south]{$\langle C \rangle$} ;
	\draw [](0.45,0.02) node[anchor=center]{\scriptsize 0} ;
	\draw [](0.03,0.23) node[anchor=center]{\scriptsize 600} ;
	\draw [](0.92,0.2) node[anchor=center]{\scriptsize 600} ;
    \end{scope}
\end{tikzpicture}
\label{fig:01043}}
\qquad
\subfigure[$Re_\mathrm{\tau}$~=~600, $\rho_\mathrm{p}/\rho$~=~7500, $F_\mathrm{el}/F_\mathrm{g}=$~0.004]{
\begin{tikzpicture}
    \node[anchor=south west,inner sep=0] (Bild) at (0,0)
    {\includegraphics[trim=3cm 7cm 2cm 7cm,clip=true,width=0.43\textwidth]{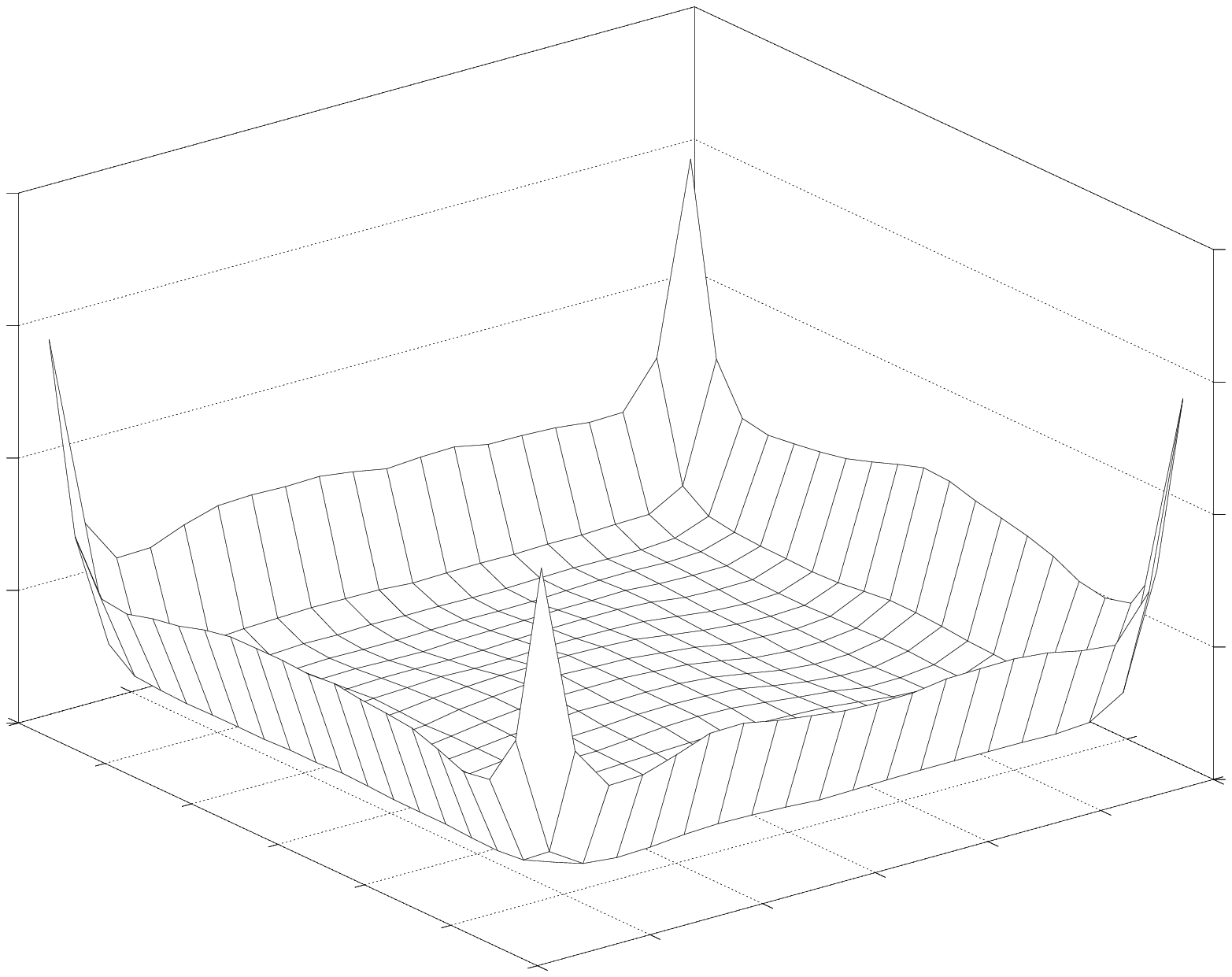}};
    \begin{scope}[x=(Bild.south east),y=(Bild.north west)]
	\draw [](0.17,0.02) node[anchor=south]{$z^+$} ;
	\draw [](0.8,0.0) node[anchor=south]{$y^+$} ;
	\draw [](0.1,0.85) node[anchor=south]{$\langle C \rangle$} ;
	\draw [](0.45,0.02) node[anchor=center]{\scriptsize 0} ;
	\draw [](0.03,0.23) node[anchor=center]{\scriptsize 600} ;
	\draw [](0.92,0.2) node[anchor=center]{\scriptsize 600} ;
    \end{scope}
\end{tikzpicture}
\label{fig:01044}}
\end{center}
\caption{Mean particle concentration in arbitrary units, $\langle C \rangle$, in the cross-section of the duct.
Note that $\langle C \rangle$ is normalized by the maximum local concentration of each case, i.e.~the vertical axes of the figures are scaled differently.
}
\label{fig:part_conc}
\end{figure*}

Instantaneous snapshots of the particle positions in the cross-section of the duct ($y$-$z$ plane) are presented in figure~\ref{fig:inst} for each case examined herein.
These snapshots were taken after the flow became  statistically stationary.
Also, for visualization  purposes, only one in every five particles is shown.
These images are complemented with the snapshots of the mean particle concentration $\langle C \rangle$ in figure~\ref{fig:part_conc}.
Therein, $\langle C \rangle$ is normalized by the maximum local concentration for each case.
The bin size for establishing these histograms is 30\,$\times$\,30 viscous length-scales for flows at $Re_\mathrm{\tau}$=~600, and 15\,$\times$\,15 for the flows at $Re_\mathrm{\tau}$=~300.

As a first observation, these figures confirm that particles tend to accumulate close to the walls, as mentioned in the Introduction.
Nonetheless, despite this common trend, the resulting particle distributions vary both quantitatively and qualitatively from one case to another.
According to the figures, concentration peaks appear at two different locations, namely in the corners and in the bisectors (the symmetry lines of each wall) of the duct.
For the case of $Re_\mathrm{\tau}$~=~600 and $\rho_\mathrm{p}/\rho$~=~1000, particles accumulate preferably at the bisectors whereas relatively few particles reside near the corners of the duct, as can be seen in figures~\ref{fig:inst_01036} and~\ref{fig:01036}.
Furthermore, when the particles are charged, figures~\ref{fig:inst_01027} and \ref{fig:01027}, the particle density close to the walls increase significantly, especially the peaks at the bisectors.

In the cases of low frictional Reynolds number, $Re_\mathrm{\tau}$=~300, the particle concentration peaks at both locations, i.e. corners and wall bisectors.
In particular, the snapshot in figure~\ref{fig:inst_01037} clearly illustrates the impact of secondary flows of the second kind on the particle dynamics.
Therein, one can identify a high-concentration region at each wall that is stretching towards the centreline of the duct.
These noticeable structures correspond to the arrow~C sketched in figure~\ref{fig:secondary} and are induced by the vortical motion of the gas.
Thus, at these regions, the particles are ejected from the wall back towards the bulk flow.
As can be inferred from figure~\ref{fig:inst_01037}, the location of these ejection points is not exactly at the bisectors.
Instead, our simulations predicted that these ejection points move with a low temporal frequency back and forth along the wall.
When the particles carry electrostatic charge, they are pushed against the wall and no longer follow the turbulent eddies to the bulk of the flow.
Consequently, as can be seen in figure~\ref{fig:inst_01028}, only a few particles are airborne.

Finally, it is worth noting that for the case with a high density ratio, figure~\ref{fig:01043}, the mean concentration
$\langle C \rangle$ has peaks only in the corners and not at the walls.
Furthermore, the electrostatic charge carried by the particles, figures~\ref{fig:inst_01044} and~\ref{fig:01044}, does not appear to alter this general flow pattern.

Further, with regard to the mean particle concentration, comparison of figures~\ref{fig:01036}, \ref{fig:01037}, and~\ref{fig:01043} shows that the flow parameters employed in our study result in qualitatively different concentration profiles.
Further, the effect of electrostatic charge is distinctly different for each case and depends on the underlying dynamics of the fluid flow.
It is conjectured that this difference is linked to the strength of the vortical structures of the carrier gas (i.e. the turbulence intensity) and illustrates the sensitivity of the particles to these structures.

\begin{figure*}
\begin{center}
\subfigure[Slice: $2.5<z^+<5$ for $Re_\mathrm{\tau}=$~300 and $5<z^+<10$  for  $Re_\mathrm{\tau}=$~600.]{\includegraphics[trim=0cm 0cm 0cm 0cm,clip=true,width=0.47\textwidth]{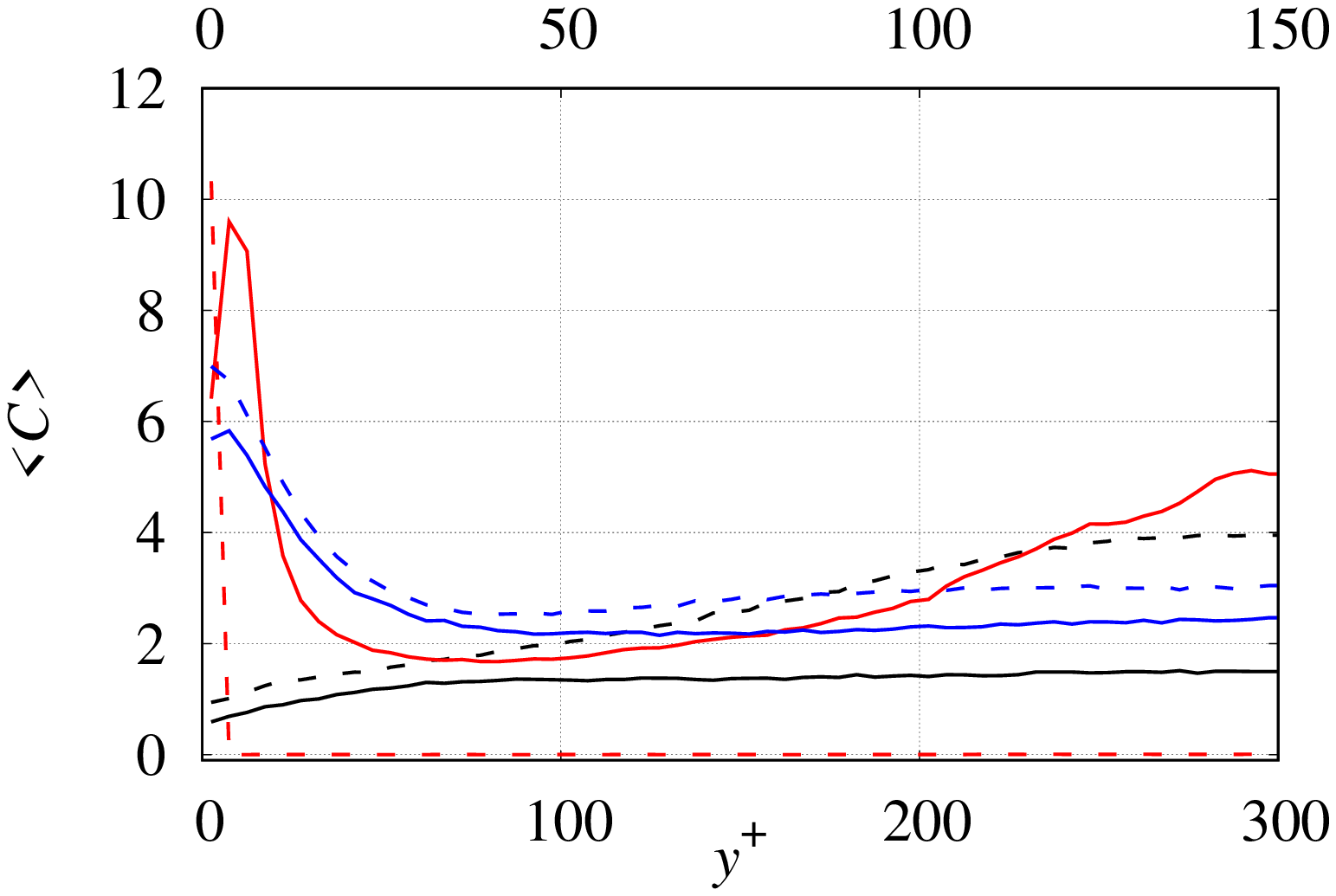}\label{fig:1d_1}}\quad
\subfigure[Slice: $10<z^+<12.5$ for $Re_\mathrm{\tau}=$~300 and $20<z^+<25$ for $Re_\mathrm{\tau}=$~600.]{\includegraphics[trim=0cm 0cm 0cm 0cm,clip=true,width=0.47\textwidth]{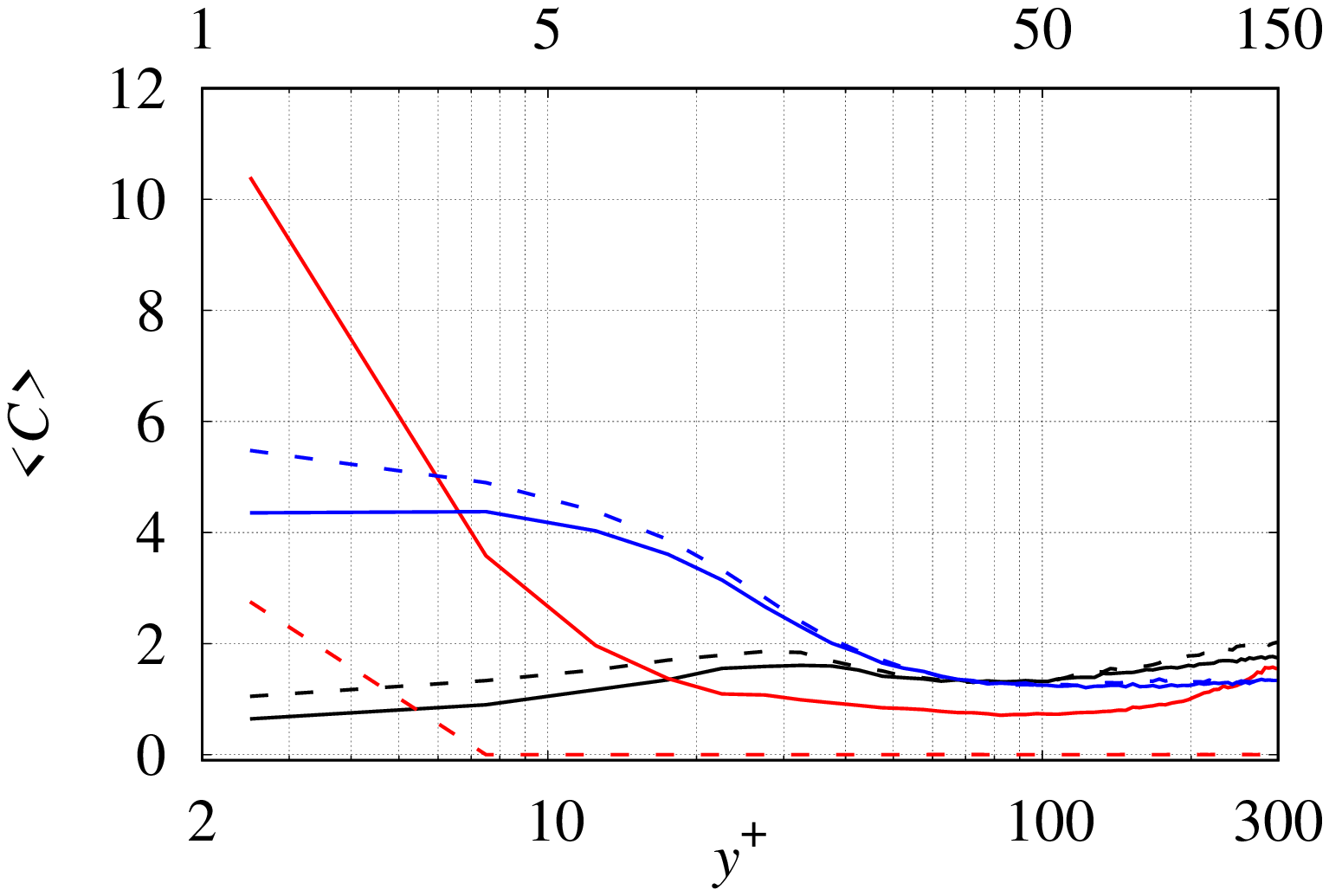}\label{fig:1d_2}}\\
\subfigure[Slice: $27.5<z^+<30$ $Re_\mathrm{\tau}=$~300 and $55<z^+<60$ for  $Re_\mathrm{\tau}=$~600.]{\includegraphics[trim=0cm 0cm 0cm 0cm,clip=true,width=0.47\textwidth]{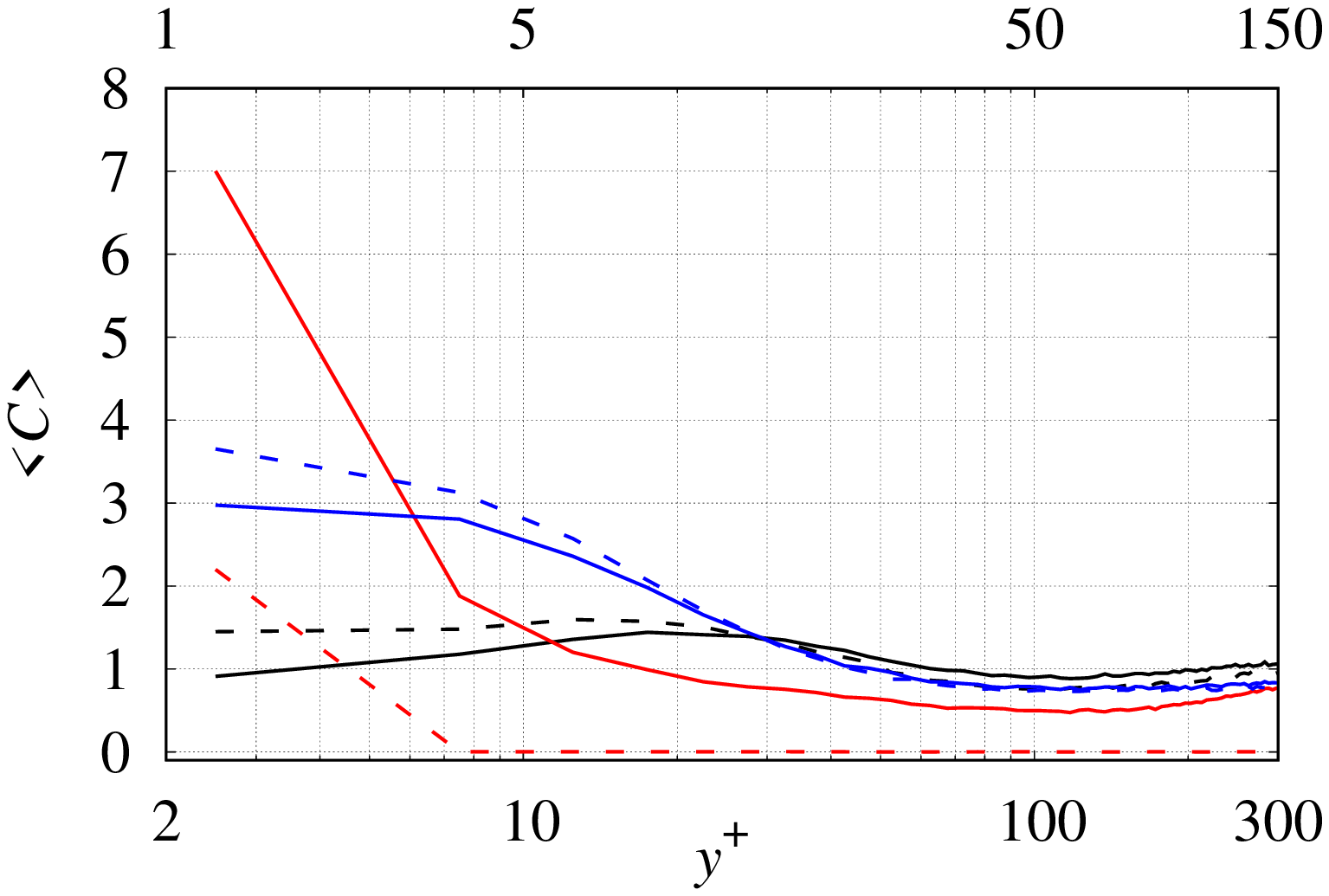}\label{fig:1d_3}}\quad
\subfigure[Slice $147.5<z^+<150$ for $Re_\mathrm{\tau}=$~300 and $295<z^+<300$ for  $Re_\mathrm{\tau}=$~600.]{\includegraphics[trim=0cm 0cm 0cm 0cm,clip=true,width=0.47\textwidth]{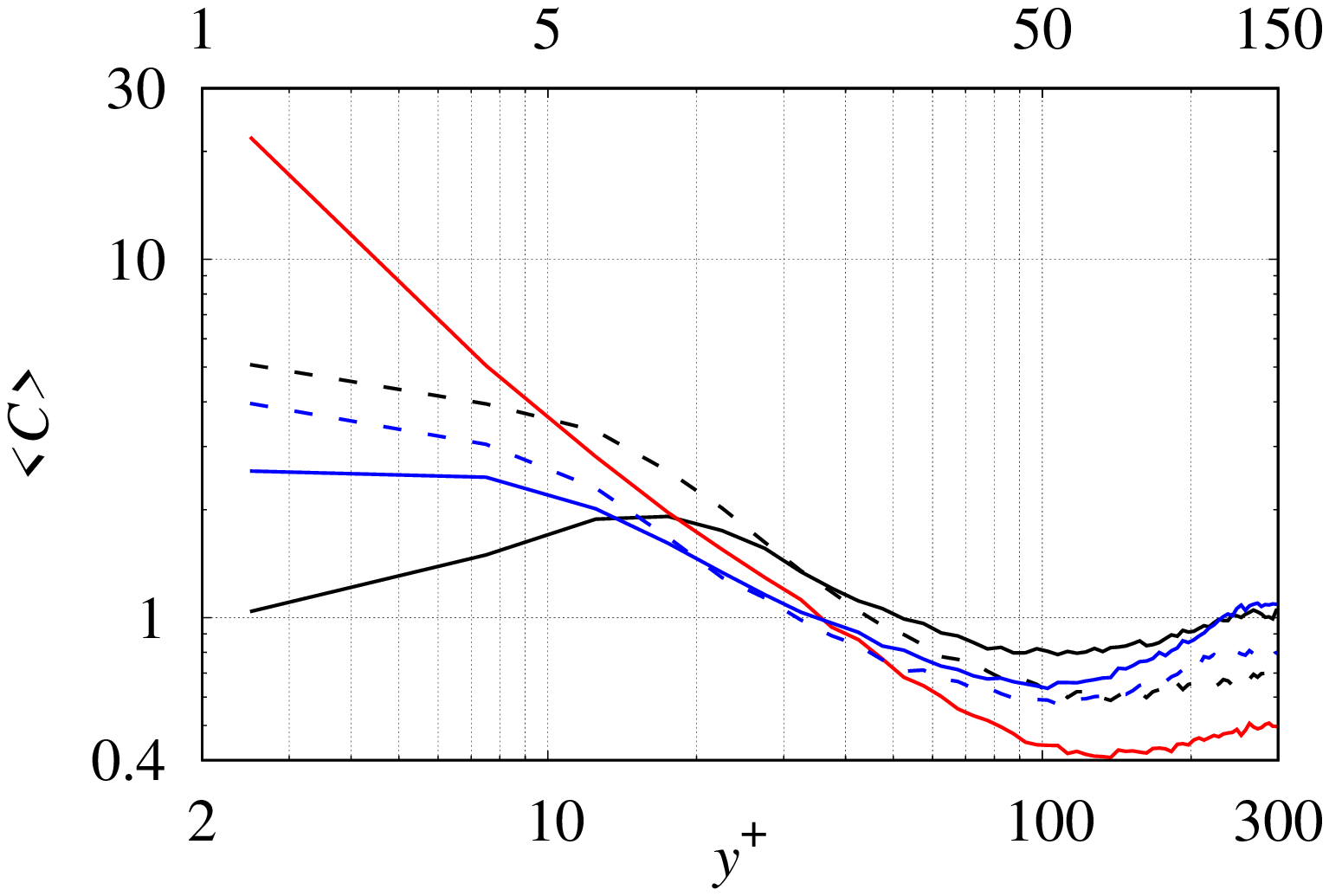}\label{fig:1d_4}}\quad
\end{center}
\caption[]{
Mean normalized particle number density, $\langle C\rangle$, in four different slices.
For visibility purposes, in (a) the horizontal axes are linear, whereas in (b), (c), and (d) they are logarithmic.
In each figure, the top horizontal axis corresponds to the flows at $Re_\mathrm{\tau}=$~300 and the lower one to flows $Re_\mathrm{\tau}=$~600.
\begin{tabular}{llll}
(\raisebox{1mm}{\tikz{\draw[thick] (0,0)--(.5,0);}}) &
$Re_\mathrm{\tau}=$~600, $\rho_\mathrm{p} / \rho=$~1000, $F_\mathrm{el}/F_\mathrm{g}=$~0 &
\qquad  (\raisebox{1mm}{\tikz{\draw[thick,dashed] (0,0)--(.5,0);}}) &
$Re_\mathrm{\tau}=$~600, $\rho_\mathrm{p} / \rho=$~1000, $F_\mathrm{el}/F_\mathrm{g}=$~0.026 \\
(\raisebox{1mm}{\tikz{\draw[thick,red] (0,0)--(.5,0);}}) &
$Re_\mathrm{\tau}=$~300, $\rho_\mathrm{p} / \rho=$~1000, $F_\mathrm{el}/F_\mathrm{g}=$~0 &
\qquad  (\raisebox{1mm}{\tikz{\draw[red,thick,dashed] (0,0)--(.5,0);}}) &
$Re_\mathrm{\tau}=$~300, $\rho_\mathrm{p} / \rho=$~1000, $F_\mathrm{el}/F_\mathrm{g}=$~0.026 \\
(\raisebox{1mm}{\tikz{\draw[thick,blue] (0,0)--(.5,0);}}) &
$Re_\mathrm{\tau}=$~600, $\rho_\mathrm{p} / \rho=$~7500, $F_\mathrm{el}/F_\mathrm{g}=$~0 &
\qquad (\raisebox{1mm}{\tikz{\draw[blue,thick,dashed] (0,0)--(.5,0);}}) &
$Re_\mathrm{\tau}=$~600, $\rho_\mathrm{p} / \rho=$~7500, $F_\mathrm{el}/F_\mathrm{g}=$~0.004
\end{tabular}
}
\label{fig:part_conc1d}
\end{figure*}

In order to examine in more detail the influence of the electrostatic charge, in figure~\ref{fig:part_conc1d} we present plots of the particle concentrations in four different stations.
As indicated in the sketch of figure~\ref{fig:secondary}, a duct with square cross-section possesses eight symmetry planes.
Due to this symmetry, we have plotted results for only one quadrant of the duct in figure~\ref{fig:part_conc1d}.
The horizontal axes on these plots are scaled differently for $Re_\mathrm{\tau}=$~300 and for $Re_\mathrm{\tau}=$~600 and in such a way that the physical coordinates of all cases are consistent with each other.
In this figure, the solid lines correspond to flows with uncharged particles, whereas the dashed lines correspond to the equivalent flows with charged particles.

Figure~\ref{fig:1d_1} shows the profile of the mean normalized number density $\langle C\rangle$ in a slice very close to the wall of the duct.
As noted earlier, for the case with $Re_\mathrm{\tau}=$~600 and $\rho_\mathrm{p}/\rho=$~1000, $\langle C\rangle$ does not exhibit a peak in the corners of the duct.
Instead, according to figure~\ref{fig:1d_1}, $\langle C\rangle$ peaks at the bisectors of the wall.
Moreover, the peak value is significantly increased, by a factor of 2.6, when the particles are electrostatically charged.
By contrast, the particle concentration in the corners is not significantly affected by the electrostatic charge.

The case of $Re_\mathrm{\tau}=$~300 and $\rho_\mathrm{p}/\rho=$~1000 is characterized by higher concentration peaks in the corners and the walls compared to the other cases.
Also, for this specific case, the particle concentration across the other slices, shown in figures~\ref{fig:1d_2}--\ref{fig:1d_4}, exhibits the highest gradients among all cases  examined herein.
Thus, in this case, the particle concentration is most influenced by the turbulence dynamics.
This implies that the particles are most sensitive to variations of the attacking forces.
Accordingly, the presence of electrostatic forces can  dramatically influence the particle concentrations.
More precisely, the repelling forces push the particles towards the walls of the duct where they accumulate.

By comparison, in the flow at $Re_\mathrm{\tau}=$~600 and $\rho_\mathrm{p}/\rho=$~7500, the particle concentration is only slightly affected, as can be seen in figure~\ref{fig:1d_1}.
The importance of inertial forces is also indicated by the relatively high $F_\mathrm{el}/F_\mathrm{g}$ of 0.026.
The electrostatic charge results in an increase in the local concentration of approximately 30\% along the wall.
The same can be inferred from figures~\ref{fig:1d_2}--\ref{fig:1d_3} where the concentration in the bulk of the flow is not affected by the electrostatic charges.
In fact, in this region the normalized particle number density is close to unity.
However, the peak close to the wall, $y^+<30$, increases by as much as 30\%.
The effect of electrostatic charge in the case $Re_\mathrm{\tau}=$~600 and $\rho_\mathrm{p}/\rho=$~1000 and in the region of the slices examined in figures~\ref{fig:1d_2}--\ref{fig:1d_3} is similar.
When the particles are charged, the particle concentration increases close to the walls but is only slightly modified in the bulk of the flow.

The most drastic effect of the electrostatic forces can be observed in the slice which characterizes the centre plane of the duct, as shown in figure~\ref{fig:1d_4}.
The highest concentration of particles occurs in the case with $Re_\mathrm{\tau}=$~300 and $\rho_\mathrm{p}/\rho=$~1000 and uncharged particles.
The particle concentration decrease with the distance from the wall until it reaches a minimum  at $y^+\approx 75$.
 In  the corresponding case with electrostatic charging, the particles migrate towards the wall (due to the scaling of the axis, this case is not visible in figures~\ref{fig:1d_4}).
For the flow at $Re_\mathrm{\tau}=$~600 and $\rho_\mathrm{p}/\rho=$~7500 the effect of the electrostatic charges is still significant albeit less pronounced.
The particle concentration at the walls is increased by 54\% and the peak in the bulk of the duct is reduced by 27\%.
The influence of electrostatic charges in the case with $Re_\mathrm{\tau}=$~600 and $\rho_\mathrm{p}/\rho=$~1000 is  also very significant.
More specifically, without electrostatic charge the concentration peak is located at $y^+=$~18, whereas when the particles are charged, the peak is located right next to the wall.
Additionally, this peak is increased by a factor of five.
Further, the concentration at the center is reduced by 31~\% and the corresponding peak nearly completely diminishes.

\begin{figure*}
\begin{center}
 \subfigure[Slice: $2.5<z^+<5$ for  $Re_\mathrm{\tau}=$~300 and $5<z^+<10$ for $Re_\mathrm{\tau}=$~600.]{\includegraphics[trim=0cm 0cm 0cm 0cm,clip=true,width=0.47\textwidth]{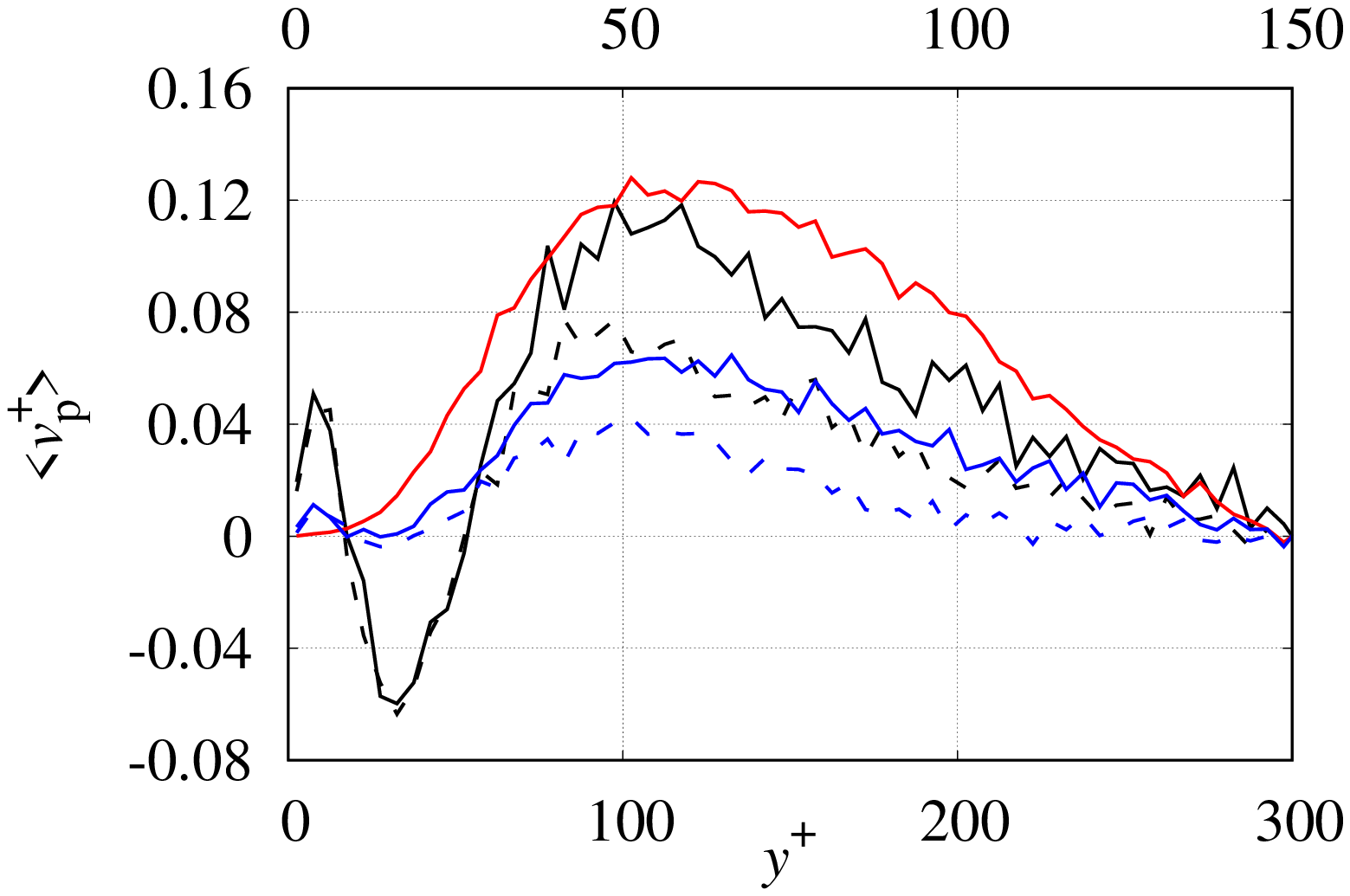}\label{fig:1dv_1}}\quad
 \subfigure[Slice: $22.5<z^+<25$ for  $Re_\mathrm{\tau}=$~300 and $45<z^+<50$ for $Re_\mathrm{\tau}=$~600.]{\includegraphics[trim=0cm 0cm 0cm 0cm,clip=true,width=0.47\textwidth]{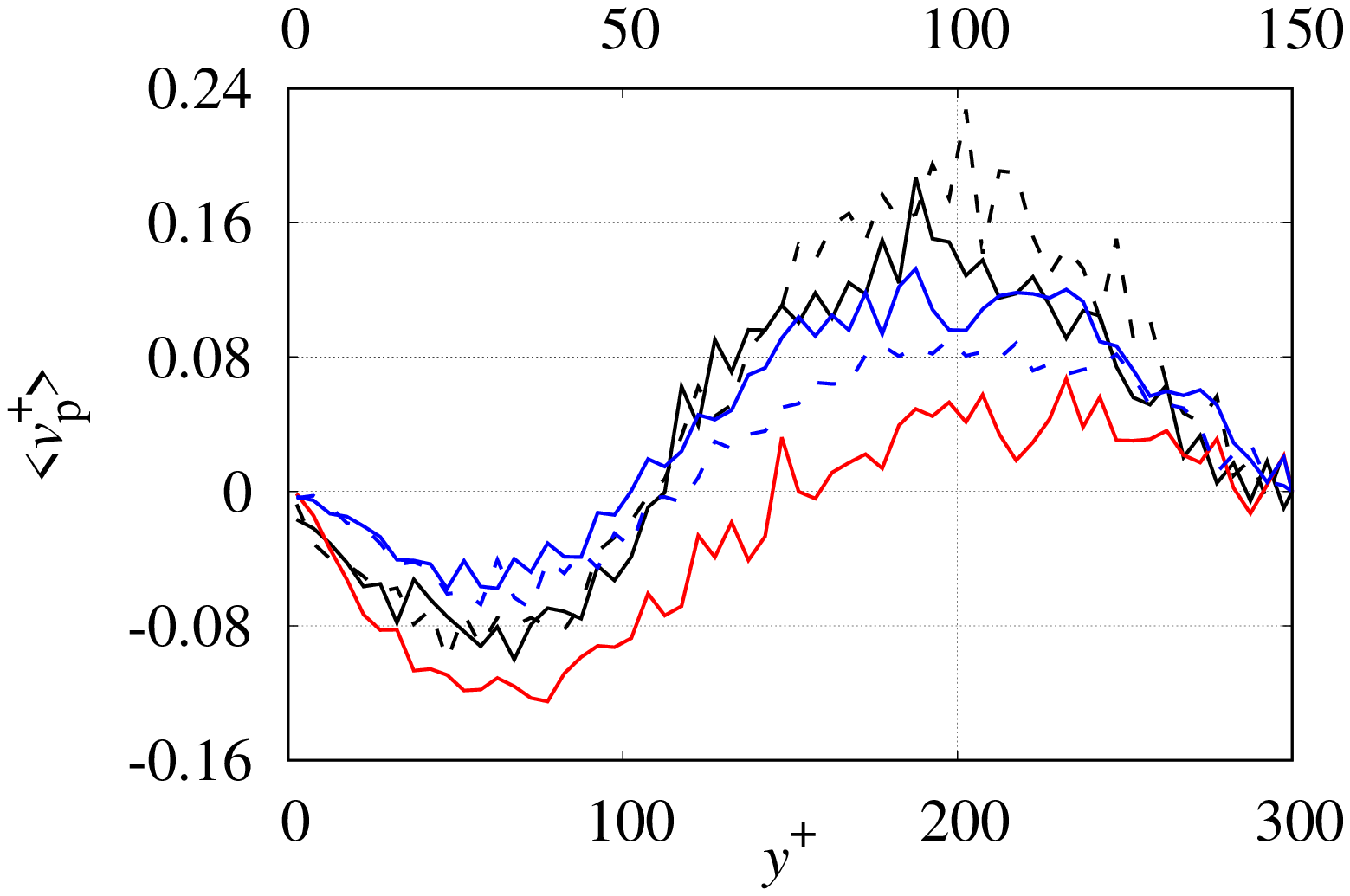}\label{fig:1dv_2}}\\
 \subfigure[Slice: $35<z^+<37.5$ for $Re_\mathrm{\tau}=$~300 and $70<z^+<75$ for $Re_\mathrm{\tau}=$~600.]{\includegraphics[trim=0cm 0cm 0cm 0cm,clip=true,width=0.47\textwidth]{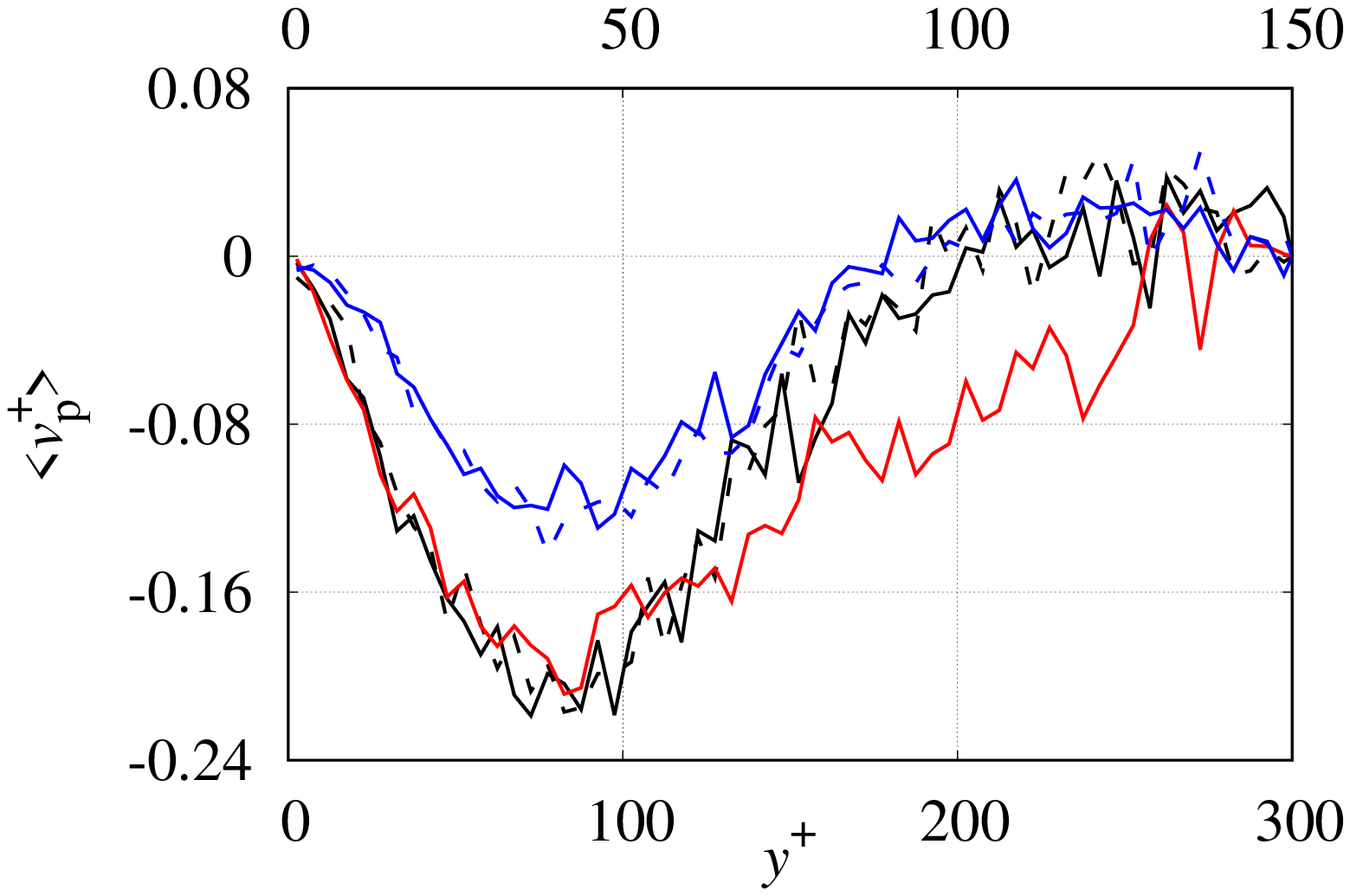}\label{fig:1dv_3}}\quad
 \subfigure[Slice: $147.5<z^+<150$ for$Re_\mathrm{\tau}=$~300 and $295<z^+<300$ for $Re_\mathrm{\tau}=$~600.]{\includegraphics[trim=0cm 0cm 0cm 0cm,clip=true,width=0.47\textwidth]{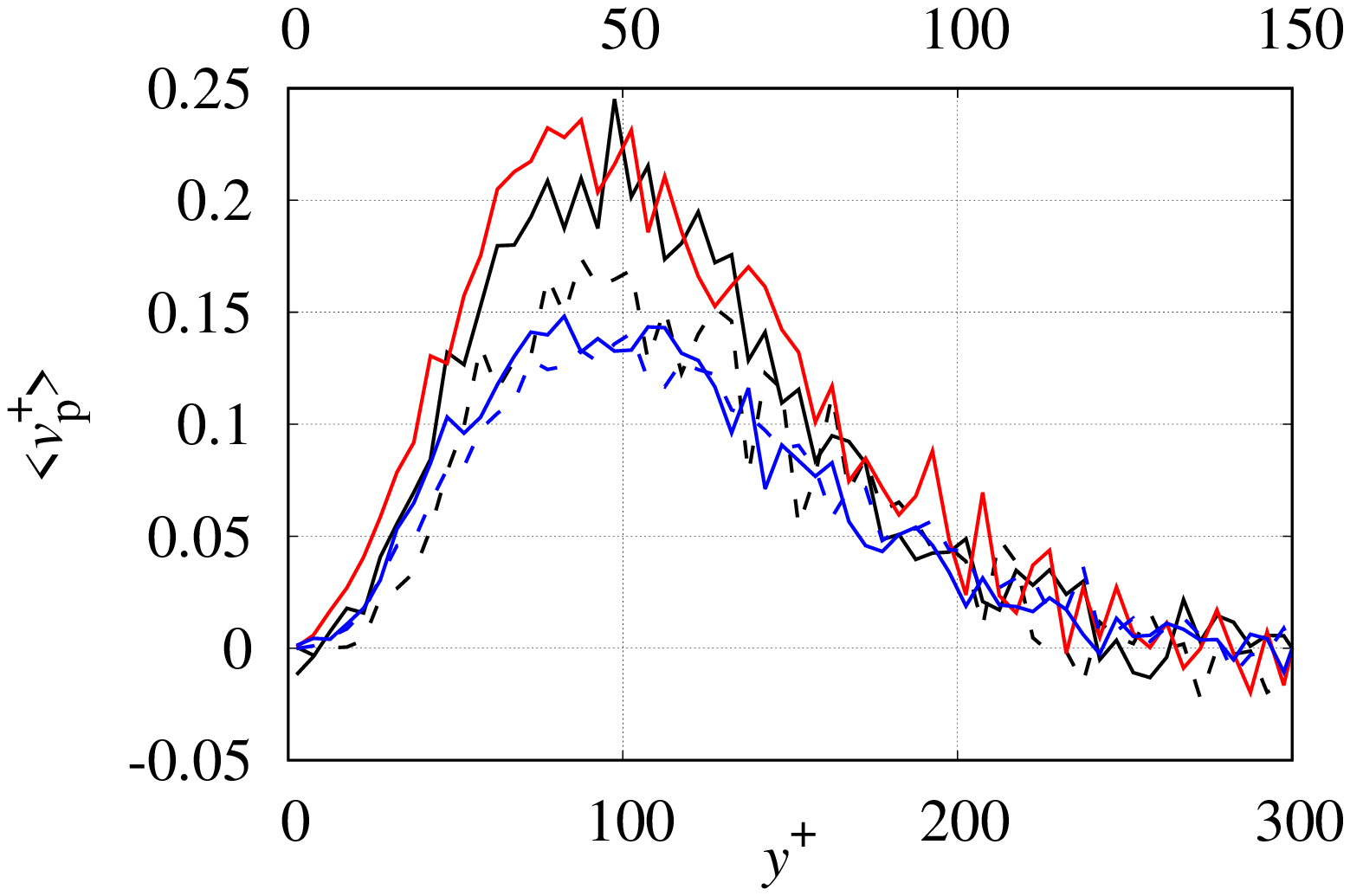}\label{fig:1dv_4}}\quad
\end{center}
\caption[]{
Normalized mean particle velocity in a wall-normal direction in four different slices.
In each figure, the top horizontal axis corresponds to the flows at $Re_\mathrm{\tau}=$~300 and the lower one to the flows at $Re_\mathrm{\tau}=$~600.
\begin{tabular}{llll}
(\raisebox{1mm}{\tikz{\draw[thick] (0,0)--(.5,0);}}) &
$Re_\mathrm{\tau}=$~600, $\rho_\mathrm{p} / \rho=$~1000, $F_\mathrm{el}/F_\mathrm{g}=$~0 &
\qquad  (\raisebox{1mm}{\tikz{\draw[thick,dashed] (0,0)--(.5,0);}}) &
$Re_\mathrm{\tau}=$~600, $\rho_\mathrm{p} / \rho=$~1000, $F_\mathrm{el}/F_\mathrm{g}=$~0.026 \\
(\raisebox{1mm}{\tikz{\draw[thick,red] (0,0)--(.5,0);}}) &
$Re_\mathrm{\tau}=$~300, $\rho_\mathrm{p} / \rho=$~1000, $F_\mathrm{el}/F_\mathrm{g}=$~0 &
\qquad  (\raisebox{1mm}{\tikz{\draw[red,thick,dashed] (0,0)--(.5,0);}}) &
$Re_\mathrm{\tau}=$~300, $\rho_\mathrm{p} / \rho=$~1000, $F_\mathrm{el}/F_\mathrm{g}=$~0.026 \\
(\raisebox{1mm}{\tikz{\draw[thick,blue] (0,0)--(.5,0);}}) &
$Re_\mathrm{\tau}=$~600, $\rho_\mathrm{p} / \rho=$~7500, $F_\mathrm{el}/F_\mathrm{g}=$~0 &
\qquad (\raisebox{1mm}{\tikz{\draw[blue,thick,dashed] (0,0)--(.5,0);}}) &
$Re_\mathrm{\tau}=$~600, $\rho_\mathrm{p} / \rho=$~7500, $F_\mathrm{el}/F_\mathrm{g}=$~0.004
\end{tabular}
}
\label{fig:1dv}
\end{figure*}

In figure~\ref{fig:1dv} we present plots of the mean particle velocities.
These plots clearly substantiate the effect of secondary flows on the particle dynamics.
As before, by virtue of the symmetries of the flow domain, we present results for one wall-normal velocity component and in one quadrant of the duct.
In general, the flows at low frictional Reynolds number, $Re_\mathrm{\tau}=$~300, exhibit the highest wall-normal velocities (in absolute terms).
On the other hand, the flows at high density ratio, $\rho_\mathrm{p} / \rho=$~7500, exhibit the lowest ones.
These trends are directly related to the difference in the particles' Stokes number which measures the sensitivity of the particles to the velocity fluctuations of the surrounding fluid.

The positive velocities in figure~\ref{fig:1dv_1} at distances larger than $y^+\approx 50$ are due to particles moving in the direction of the arrow~B of figure~\ref{fig:secondary}.
In other words, the particles move away from the corner of the duct and towards the bisectors of the walls.
As can be inferred from the plots shown in figure~\ref{fig:1dv_1}, this particle motion is substantially subdued when the particles carry electrostatic charge.
As a matter of fact, for the case at $Re_\mathrm{\tau}=$~300 and $\rho_\mathrm{p} / \rho=$~1000, the dominance of the electrostatic forces over the aerodynamic ones completely inhibits the vortical motion of particles.
For this reason, the velocity profiles for this case have not been included in figure~\ref{fig:1dv_1}.

The particle velocities in the region $y^+<50$ relate to the tail of the structure of particles that are transported from the duct's centreline to the corners in the direction of the arrow A of figure~\ref{fig:secondary}.
A more detailed perspective of this outward flux is provided in figures~\ref{fig:1dv_2} and~\ref{fig:1dv_3}; therein this outward flux is represented by the peaks of the wall-normal velocity amplitudes at $y^+\approx 60$ and $y^+\approx 90$, respectively.
Interestingly, this part of the vortical motion of the particles is not affected by the electrostatic field; instead, the motion of charged particles is very similar to that of uncharged ones.

The inward particle flux from the wall bisectors towards the centreline of the duct follows the direction of the arrow C in figure~\ref{fig:secondary} and can be observed in figure~\ref{fig:1dv_4}.
Similarly to the flux along the wall, the particle velocity in this direction is strongly reduced in the case of charged particles.
For both types of motions, the extent of the influence of electrostatic forces on the particle velocity is stronger for the cases with $\rho_\mathrm{p} / \rho=$~1000 than for the cases with $\rho_\mathrm{p} / \rho=$~7500.

In summary, both the particle flux along the walls and the inward motion from the wall towards the bulk of the duct are significantly reduced when the particles carry electrostatic charge.
On the other hand, the particle transport from the duct's centreline towards the corners is not affected by the electrostatic forces.
This finding explains the particle concentration profiles shown in figure~\ref{fig:part_conc1d}.
In particular, due to electrostatic charges, the particles are driven with roughly the same velocity towards the walls but are slowly ejected back to the centre.
As a result, the particle concentration increases at the walls and especially in the corners.
Consequently, the notable difference in the particles' wall-normal velocity between charged and uncharged particles when $Re_\mathrm{\tau}=$~600, $\rho_\mathrm{p} / \rho=$~1000 substantiates the strong increase of the particle concentration in the vicinity of walls.

\begin{figure*}
\begin{center}
 \subfigure[$Re_\mathrm{\tau}=$\,600, $\rho_\mathrm{p} / \rho=$\,1000]{\includegraphics[trim=7.2cm 2cm 6.5cm 1cm,clip=true,width=0.29\textwidth]{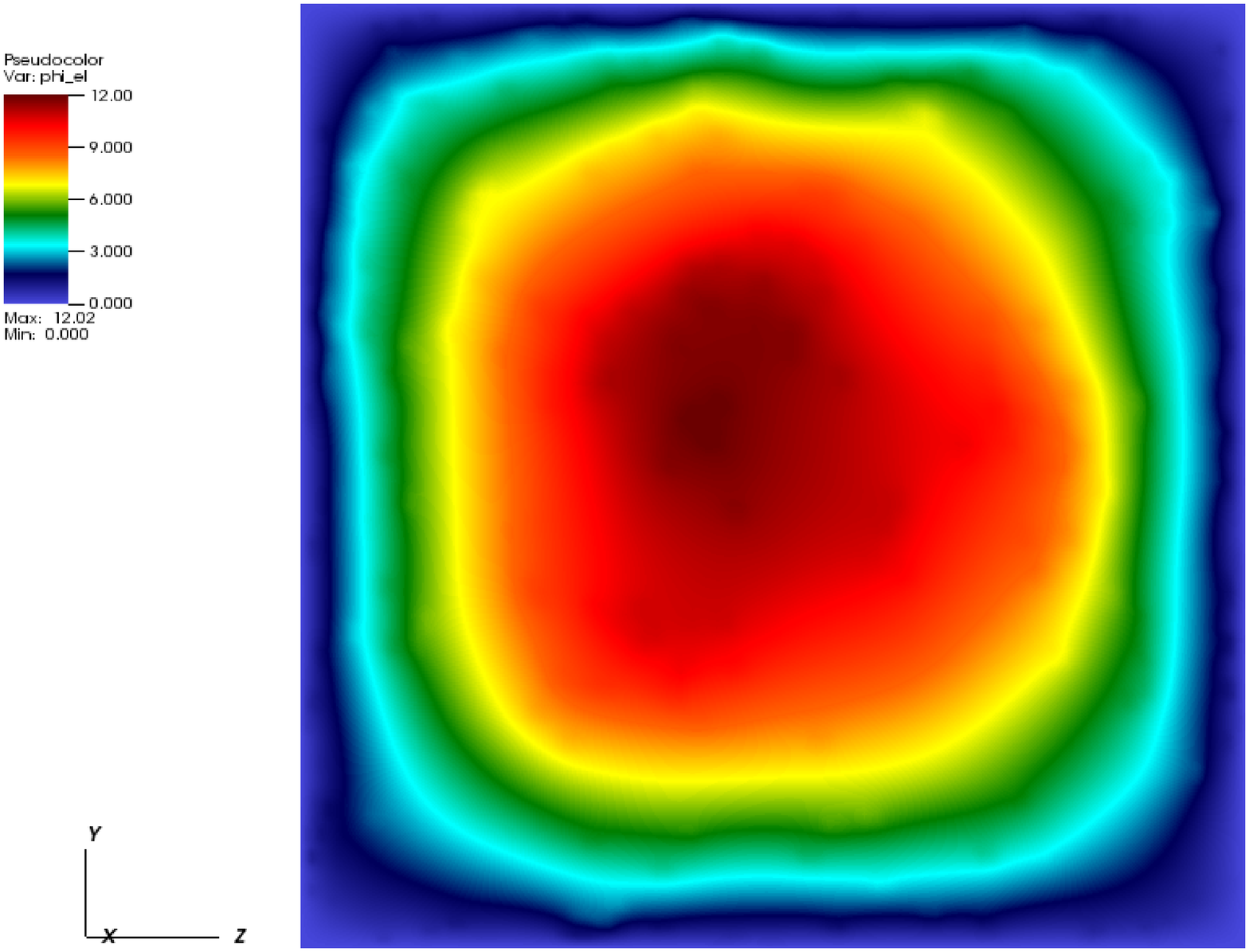}\label{fig:}}~
 \subfigure[$Re_\mathrm{\tau}=$\,300, $\rho_\mathrm{p} / \rho=$\,1000]{\includegraphics[trim=7.2cm 2cm 6.5cm 1cm,clip=true,width=0.29\textwidth]{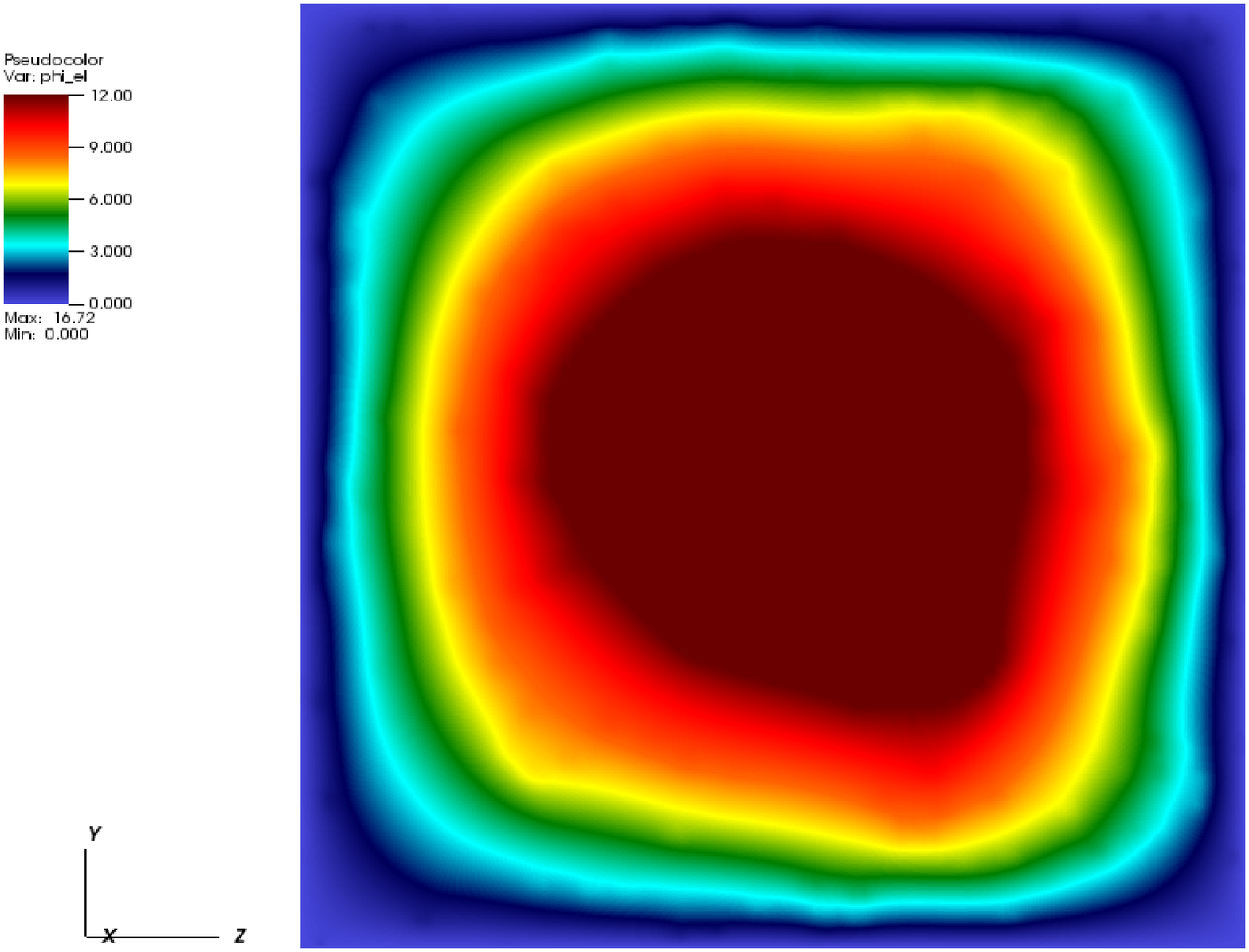}\label{fig:}}~
 \subfigure[$Re_\mathrm{\tau}=$\,600, $\rho_\mathrm{p} / \rho=$\,7500]{\includegraphics[trim=7.2cm 2cm 6.5cm 1cm,clip=true,width=0.29\textwidth]{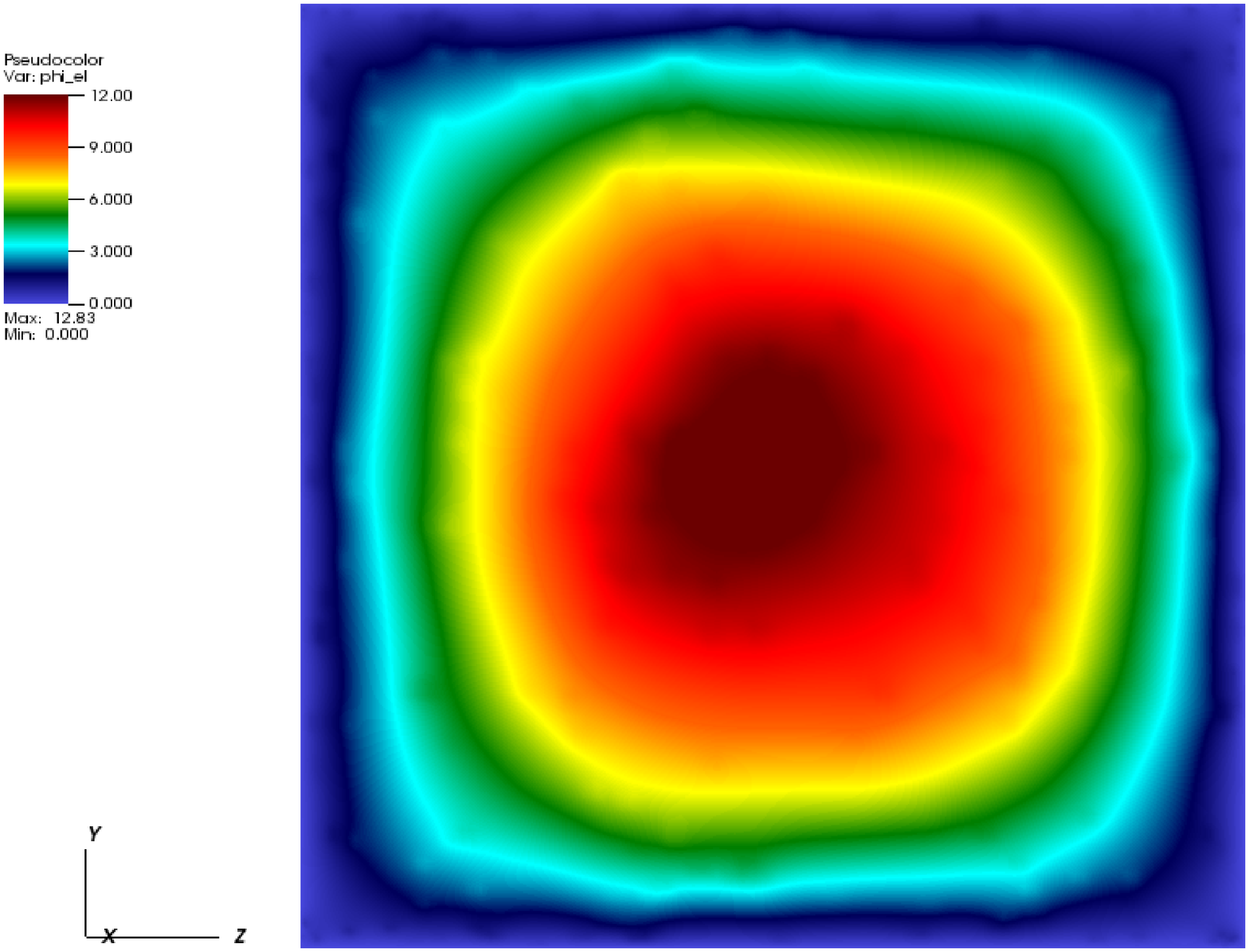}\label{fig:}}
\begin{tikzpicture}
\node[anchor=south west,inner sep=0] (Bild) at (0,0)
{\includegraphics[trim=0cm 0cm 0cm 0cm,clip=true,width=5mm,height=25mm]{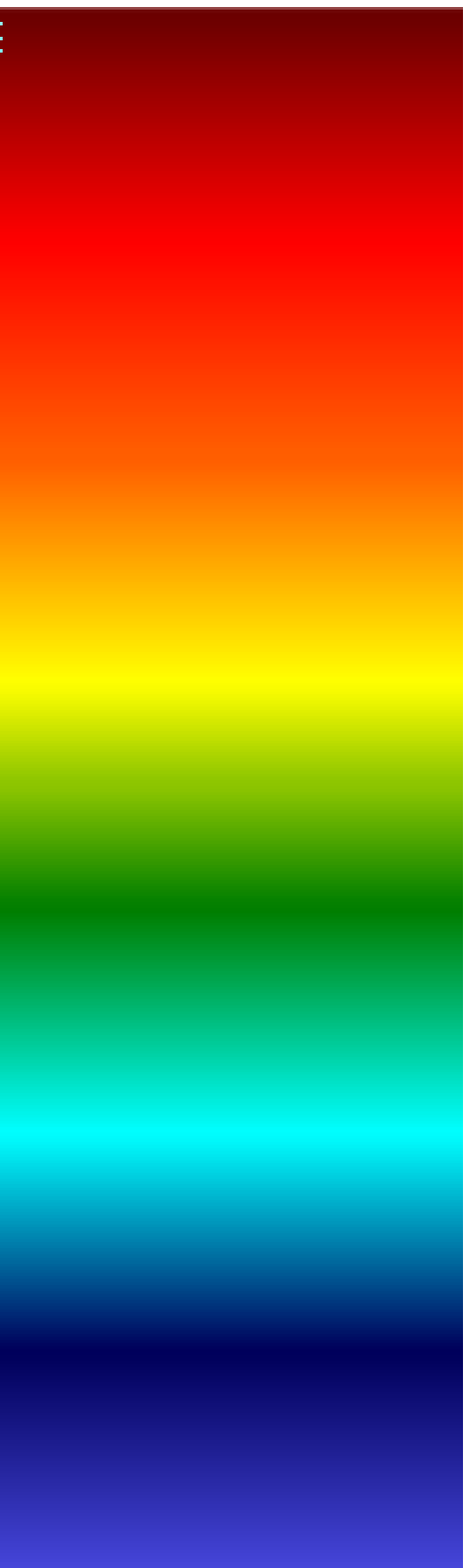}};
\begin{scope}[x=(Bild.south east),y=(Bild.north west)]
\draw [](0.6,1.0) node[anchor=south]{$\varphi_\mathrm{el}$/V};
\draw [](0.7,0.95) node[anchor=west,xshift=1mm]{12};
\draw [](0.7,0.02) node[anchor=west,xshift=1mm]{0};
\end{scope}
\end{tikzpicture}
\end{center}
\caption[]{Instantaneous snapshots of the electric potential across the cross-section of the duct,  induced by charges carried by the particles.}
\label{fig:phi}
\end{figure*}

To illustrate the direction of the arising electrostatic forces, instantaneous snapshots of the electric potential $\varphi_{\rm el}$ are presented in figure~\ref{fig:phi}.
We recall that the electric field strength ${\bm E}$ is equal to $-\nabla \varphi_{\rm el}$.
Accordingly, one can deduce from these plots that the particle encounters a higher electrostatic force when moving from the bisectors of the walls towards the centreline of the duct than when moving along the diagonal from the centreline to the corner.
This implies that, the emerging electric field significantly reduces particle circulation due to secondary flows of a second kind.

\begin{figure*}
\begin{center}
\subfigure[Slice: $2.5<z^+<5$ for $Re_\mathrm{\tau}=$~300 and  $5<z^+<10$ for $Re_\mathrm{\tau}=$~600.]{\includegraphics[trim=0cm 0cm 0cm 0cm,clip=true,width=0.47\textwidth]{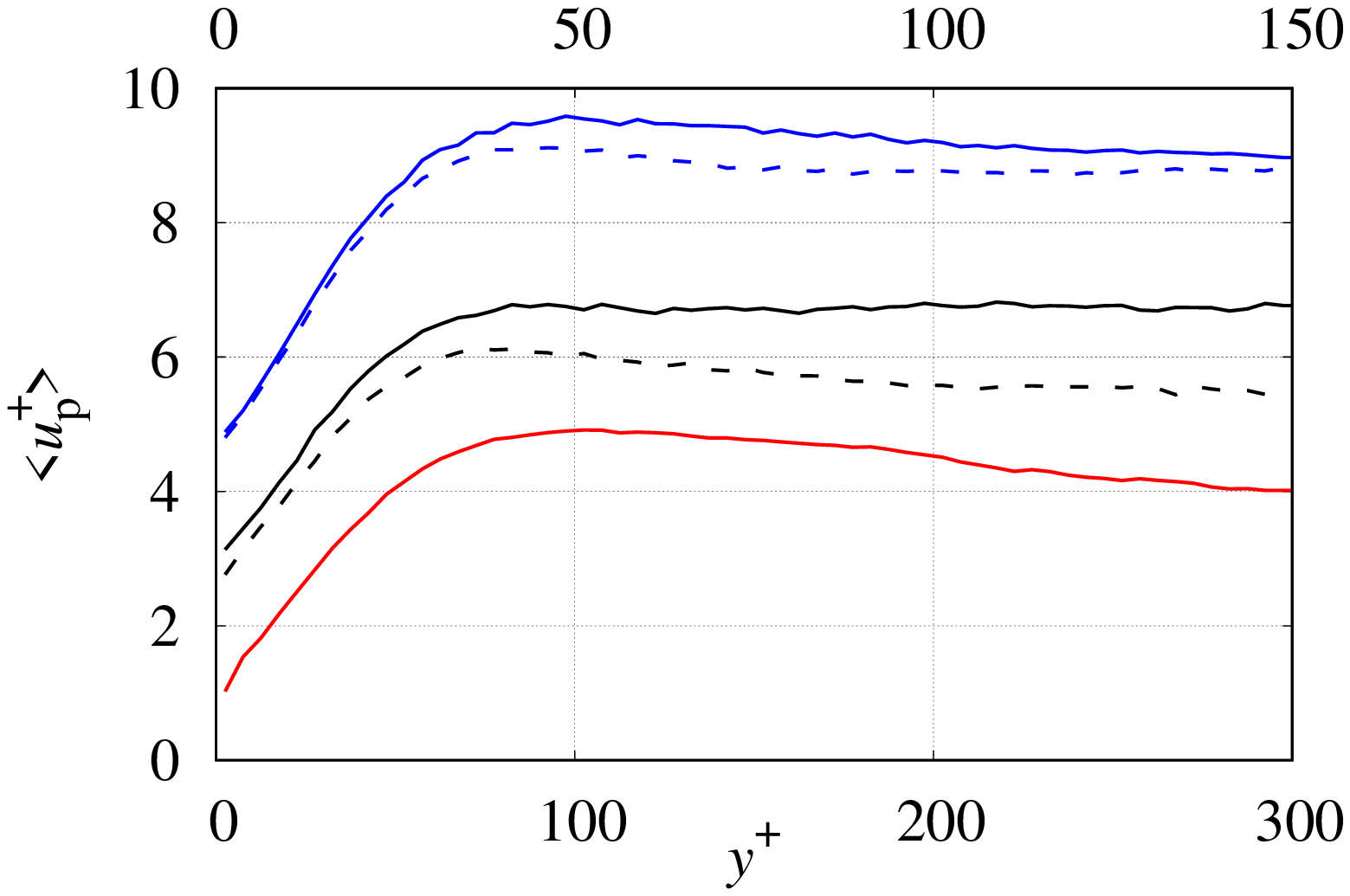}\label{fig:1du_1}}\quad
\subfigure[Slice: $22.5<z^+<25$ for $Re_\mathrm{\tau}=$~300 and $45<z^+<50$ for  $Re_\mathrm{\tau}=$~600.]{\includegraphics[trim=0cm 0cm 0cm 0cm,clip=true,width=0.47\textwidth]{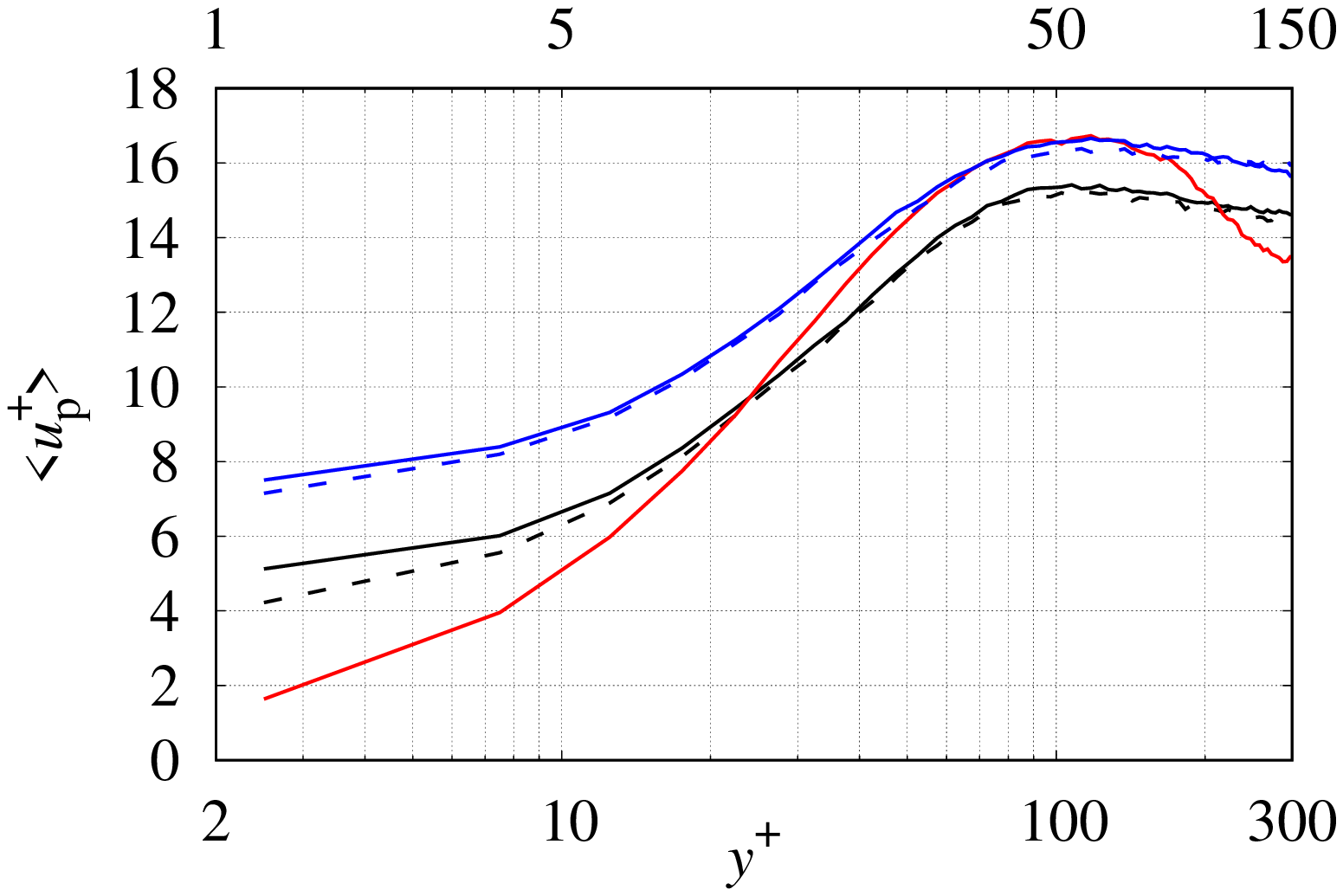}\label{fig:1du_2}}\\
\subfigure[Slice: $35<z^+<37.5$ for  $Re_\mathrm{\tau}=$~300 and $70<z^+<75$ for  $Re_\mathrm{\tau}=$~600.]{\includegraphics[trim=0cm 0cm 0cm 0cm,clip=true,width=0.47\textwidth]{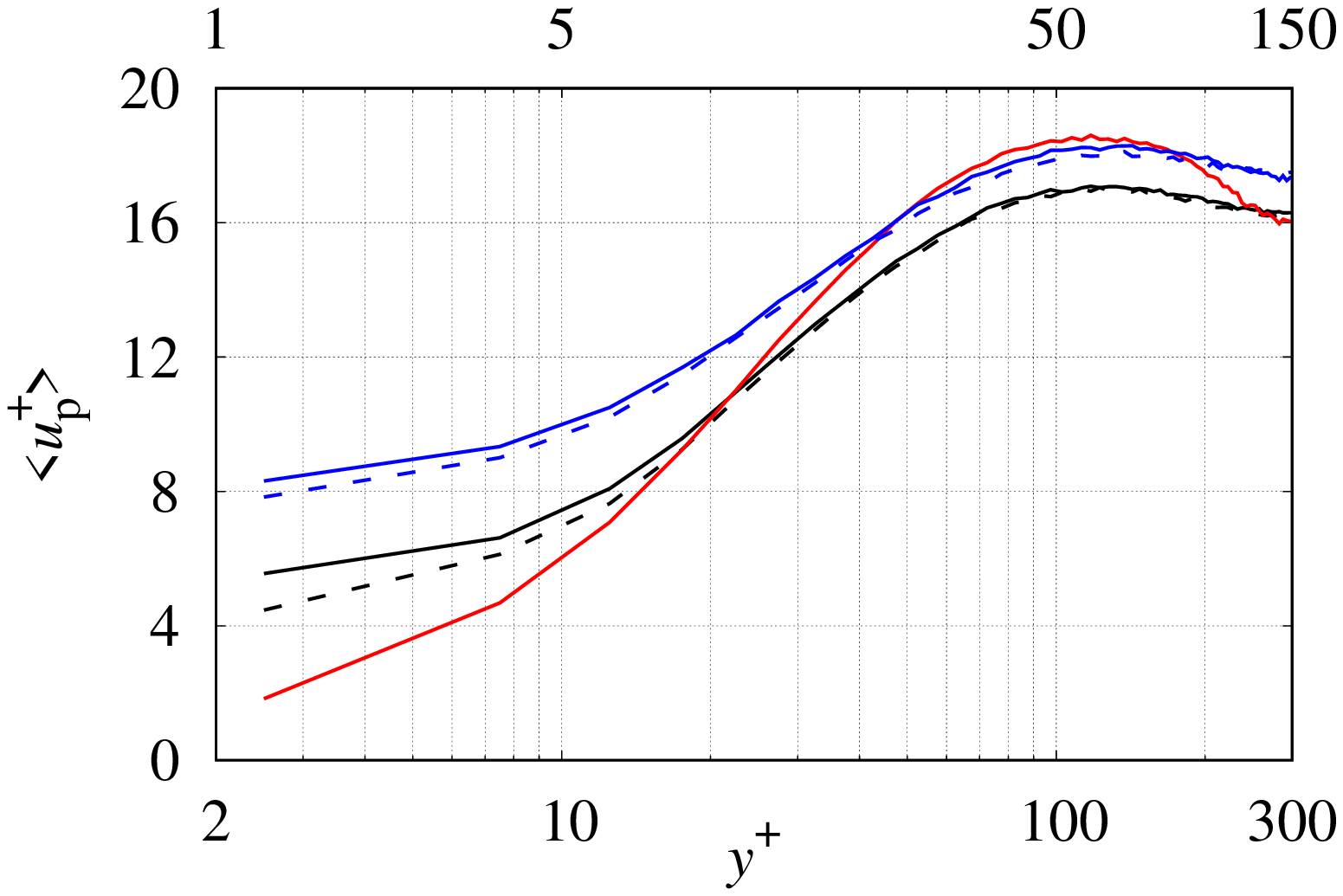}\label{fig:1du_3}}\quad
\subfigure[Slice: $147.5<z^+<150$ for $Re_\mathrm{\tau}=$~300 and $295<z^+<300$ for  $Re_\mathrm{\tau}=$~600.]{\includegraphics[trim=0cm 0cm 0cm 0cm,clip=true,width=0.47\textwidth]{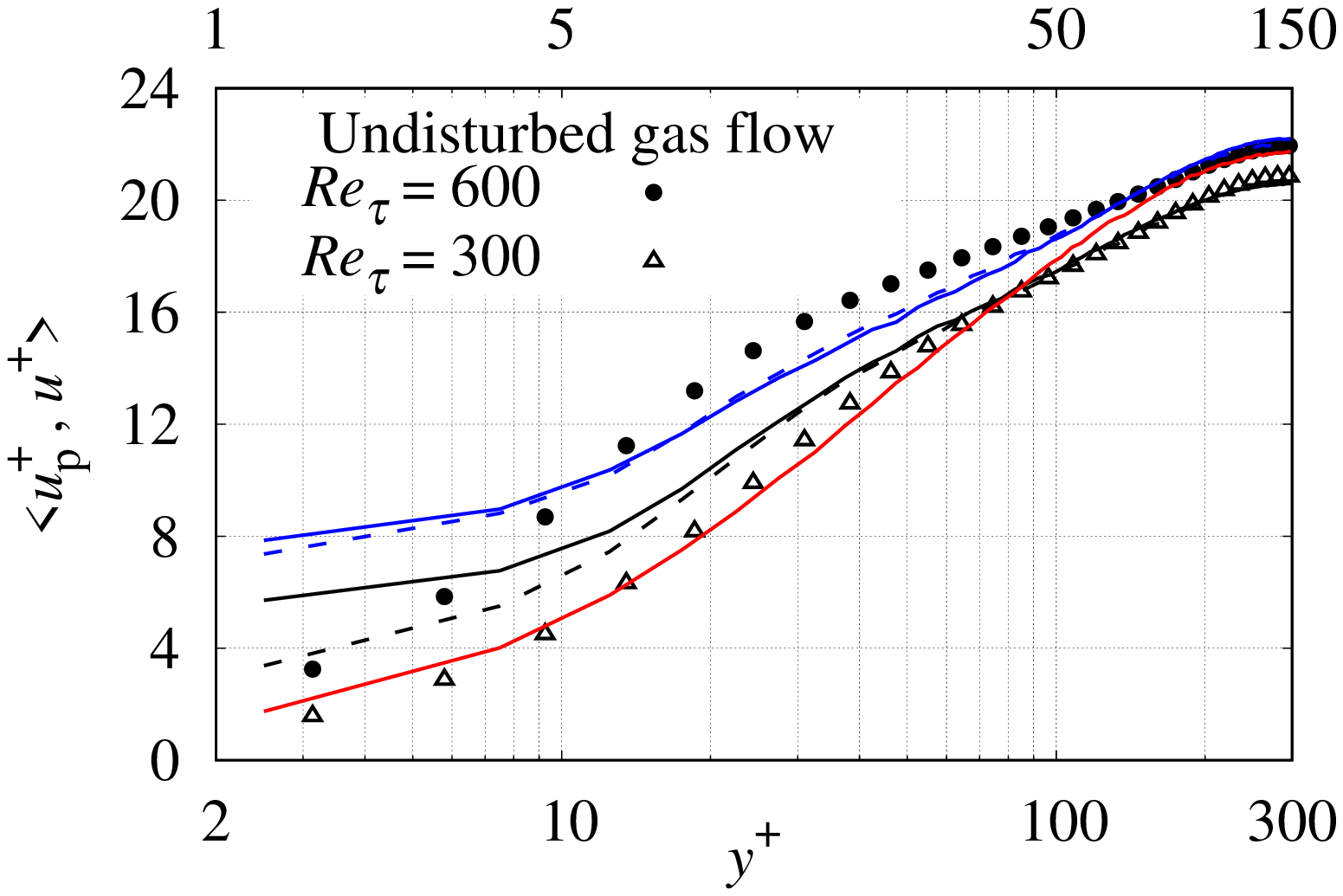}\label{fig:1du_4}}\quad
\end{center}
\caption[]{
Normalized mean particle velocity in the streamwise direction in four different slices as a function of the distance to the duct's wall.
For visibility purposes, in (a) the horizontal axes is linear, whereas in (b), (c), and (d) they are logarithmic.
The upper axis in each figure relates to the cases of $Re_\mathrm{\tau}=$~300 and the lower axis to the cases of $Re_\mathrm{\tau}=$~600.
Additionally, in (d) the mean velocity of the undisturbed gaseous phase is displayed.
\begin{tabular}{llll}
(\raisebox{1mm}{\tikz{\draw[thick] (0,0)--(.5,0);}}) &
$Re_\mathrm{\tau}=$~600, $\rho_\mathrm{p} / \rho=$~1000, $F_\mathrm{el}/F_\mathrm{g}=$~0 &
\qquad  (\raisebox{1mm}{\tikz{\draw[thick,dashed] (0,0)--(.5,0);}}) &
$Re_\mathrm{\tau}=$~600, $\rho_\mathrm{p} / \rho=$~1000, $F_\mathrm{el}/F_\mathrm{g}=$~0.026 \\
(\raisebox{1mm}{\tikz{\draw[thick,red] (0,0)--(.5,0);}}) &
$Re_\mathrm{\tau}=$~300, $\rho_\mathrm{p} / \rho=$~1000, $F_\mathrm{el}/F_\mathrm{g}=$~0 &
\qquad  (\raisebox{1mm}{\tikz{\draw[red,thick,dashed] (0,0)--(.5,0);}}) &
$Re_\mathrm{\tau}=$~300, $\rho_\mathrm{p} / \rho=$~1000, $F_\mathrm{el}/F_\mathrm{g}=$~0.026 \\
(\raisebox{1mm}{\tikz{\draw[thick,blue] (0,0)--(.5,0);}}) &
$Re_\mathrm{\tau}=$~600, $\rho_\mathrm{p} / \rho=$~7500, $F_\mathrm{el}/F_\mathrm{g}=$~0 &
\qquad (\raisebox{1mm}{\tikz{\draw[blue,thick,dashed] (0,0)--(.5,0);}}) &
$Re_\mathrm{\tau}=$~600, $\rho_\mathrm{p} / \rho=$~7500, $F_\mathrm{el}/F_\mathrm{g}=$~0.004
\end{tabular}
}
\label{fig:1d_u}
\end{figure*}

Next, we investigate the influence of electrostatic charges on the streamwise velocity of the particles.
Plots of the
particle mean streamwise velocity component are provided in figure~\ref{fig:1d_u}.
Since particles follow to a certain extent the gaseous phase, the particle streamwise velocities close to the walls are lower than those in the bulk of the duct.
Moreover, as mentioned above, in ducts with square cross section, secondary flows play an important role in the transport of particles in the wall-normal directions.
In turn, the streamwise momentum of particles is also transported with them along the $y$- or $z$-axis.
More specifically, the vortical flow structures transport fast-moving particles towards the near-wall region and slow-moving ones back to the bulk of the duct.
Once the slow-moving particles enter the bulk of the duct, they begin to accelerate.
This momentum transport contributes to a more uniform streamwise velocity distribution as can be inferred from figure~\ref{fig:1du_4}.
In the slice shown in this figure, which corresponds to particle displacement along the arrow~C of  figure~\ref{fig:secondary} the particles in the near-wall region ($y^+<5$ or $y^+<20$, depending on the case), are faster than their surrounding fluid owing to the momentum they received when they were in the bulk of the duct.
When moving towards the centreline, beyond $y^+<5$, respectively $y^+<20$, the surrounding gas is faster than the particles which, therefore, experience an acceleration in $x$-direction.
Then, once they reach the centreline, their streamwise velocity is maximized.
The fact that the maximum particle velocity differs from one case to another and is not equal to the maximum gas velocity is due to the interplay between gravity and the aerodynamic forces acting on the particles.

Finally, we examine the effect of the electrostatic field to the particles moving close to the wall and in the direction of the arrow~B of figure~\ref{fig:1du_1}.
Comparison of the computed streamwise velocities, corroborates the fact that when the particles carry no charge they move significantly faster.
Indeed,  as can be observed in figure~\ref{fig:1dv_1}, the charged particles are slowed down in the wall-normal direction by the electrostatic forces and remain close to the walls for a long time.
Due to longer residence times, more streamwise momentum is exchanged between particles and gas, which results in the slowing down of the particles.

On the contrary, the electrostatic forces do not influence the velocity of particles moving outward diagonally, in the direction of the arrow~A.
With regard to this type of motion, charged and uncharged particles move at the same wall-normal velocity, which in turn implies that they exchange a similar amount of streamwise momentum with their surrounding gas.
As a result, their streamwise velocities are also similar, as can be inferred from figures~\ref{fig:1du_2} and~\ref{fig:1du_3} beyond $y^+\approx 30$.
These figures also confirm that the particles in the region $y^+ < 30$, i.e., those moving in the direction of the arrow~B, are transported slower in the $x$ direction.

On the basis of these observations, we conclude that electric charges not only attenuate particle motion induced by secondary flows but they also substantially reduce the streamwise momentum transfer between particles and carrier gas over the cross-section of the duct.

\section{Conclusions}

We studied, by means of direct numerical simulations, the dynamics of particle-laden flows through a duct of a square-shaped cross-section.
According to our simulations, when the particles are charged, the electrostatic forces
dampen significantly the particles' vortical motions that are induced by the secondary flows of the carrier gas.
The charged particles still migrate in the diagonal direction from the centreline of the duct towards its corners.
But on their way back from the wall to the center their velocity is reduced.
This modification to the particle dynamics results in significantly different characteristics of the particle number density.
The reduction of the vortical motion of particles in the regions of  secondary flows leads to an increase of the particle concentration at the walls, especially at the bisectors of the walls and the corners of the duct.
Also, the streamwise momentum transfer in the cross-section, which relies on the wall-normal motion of particles, is significantly attenuated.
These results demonstrate the fundamental influence of electric forces on the emerging pattern of dispersed two-phase flows. 
Such an understanding of the underlying mechanisms opens new perspectives for the control of the flow dynamics of powders.
It would be interesting to test this influence also for rectangular ducts of different aspect ratios and to elaborate in detail on the dependence of the vortical particle motion on a successive increase of the powder charge.

\begin{acknowledgments}
The first and second authors gratefully acknowledge the financial support from the DECHEMA Max Buchner Research Foundation and the European Research Council (ERC) under the European Union’s Horizon 2020 research and innovation programme (grant agreement No. 947606 PowFEct).
The third and fourth authors gratefully acknowledge the financial support of the National
Research Fund of Belgium (FNRS) under the FLOW-CHARGE grant.
\begin{center}
\includegraphics[trim=3cm 5cm 2cm 5cm,clip=true,scale=0.15]{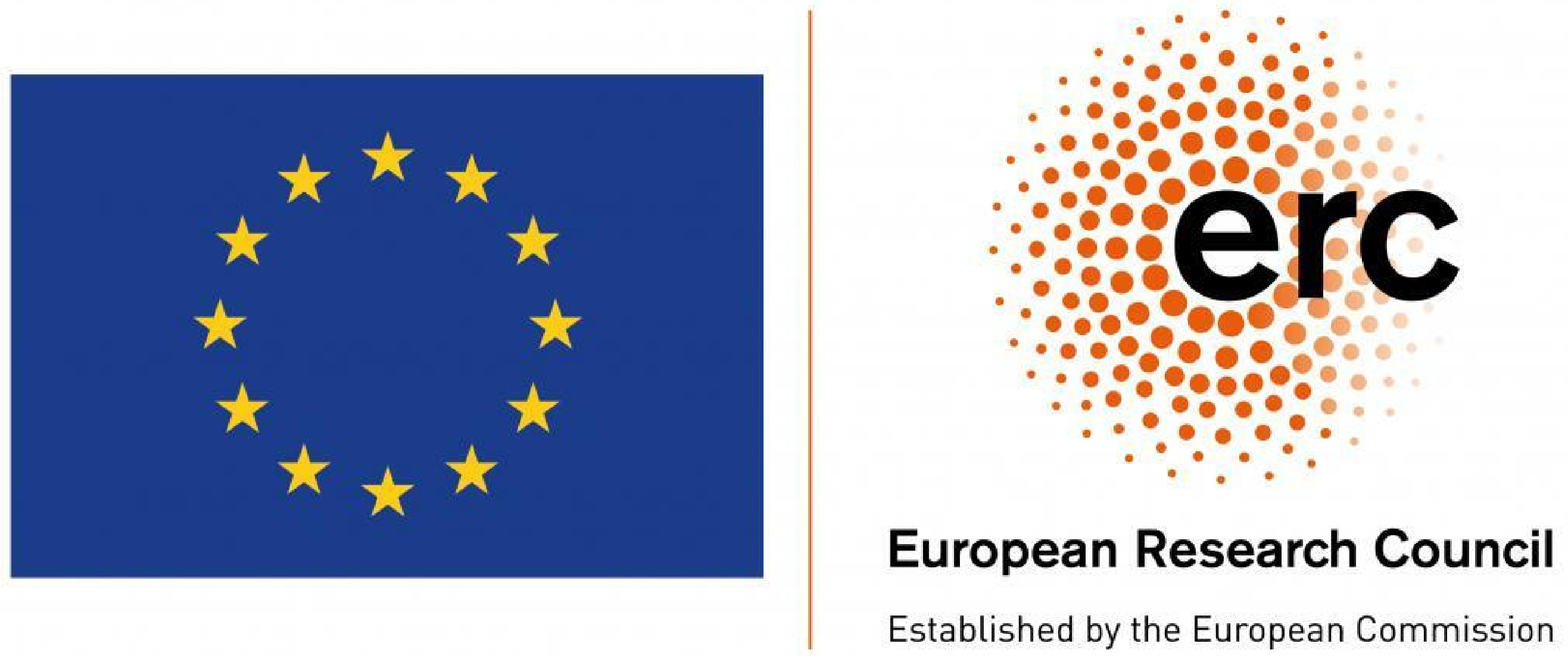}
\end{center}
\end{acknowledgments}

\medskip
Declaration of Interests. The authors report no conflict of interest.

\appendix

\section{Scalability}
\label{app:s}

\begin{figure*}
\begin{center}
\includegraphics[trim=0cm 0cm 0cm 0cm,clip=true,width=0.75\textwidth]{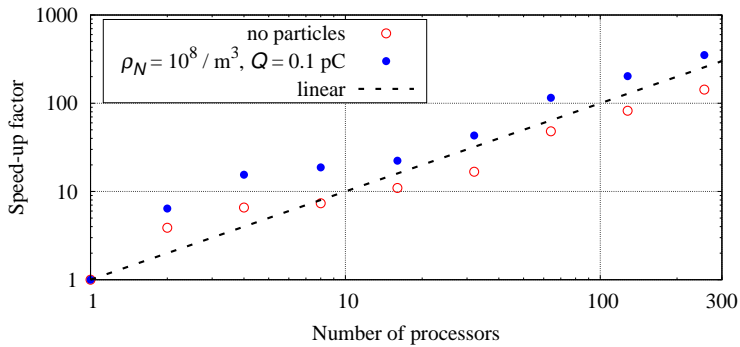}
\caption{Scalability of the fluid solver and the fluid plus the particle and electrostatic solver.
The straight dashed lone denotes the ideal linear scaling.}
\label{fig:speedup}
\end{center}
\end{figure*}

As a first test, we examined the scalability of pafiX.
More specifically, we simulated flows in a duct with square cross-section for which the computational domain, depicted in figure~\ref{fig:grida}, was discretized by 256\,$\times$\,120\,$\times$\,120 grid cells.
In our tests, two different cases were considered.
The first one is single-phase flow, i.e. without solid particles, at $Re_\mathrm{\tau}=$~600.
In this manner, only the scalability of the Navier-Stokes solved is accessed.
The second one is flow laden with $10^8$/m$^3$ charged particles of $Q=$~0.1~pC; this is also the second case given in table~\ref{tab:param}.

With regard to domain decomposition, the domain was decomposed in the $x$-direction.
For the geometry considered herein, this is the most efficient domain decomposition in terms of load balance between processors.
Evidently, this choice also poses an upper limit to the number of processors that can be used.
For the grid employed herein this limit is 256 processors.
The speed-up factor of the simulation, $S_k$, is defined as the ratio of the run-time of the parallel code running on a single processor, $T_1$, to the run-time of the same code running on $k$ processors, 
\begin{equation}
\label{eq:speedup}
S_k \;=\; \dfrac{T_1}{T_k}\,,
\end{equation}

The resulting speed-up for both cases and for $k=2^n$ ($n=0,1, \ldots, 8$) is plotted in figure~\ref{fig:speedup}.
According to figure~\ref{fig:speedup}, the code scales very well in the range of number of processors considered herein.
Our tests confirmed that, as expected, the fraction of the MPI communication time, i.e.~the time needed to send and receive data packages from one processor to the other, compared to the total computing time, increases with the degree of domain decomposition.
More specifically, for the case of single-phase flow, it increases from 1.3\% for $k=2$ to 54.5\% for $k=256$.
The fact that the speed-up is not significantly reduced can be attributed to the non-optimized memory access procedure of the code which results in a rather expensive calculation when $k$ is small and the variable matrices are large.

Moreover, a particularly long computing time was recorded for the simulation of the second test case (particle-laden flow with charged particles) on only one processor.
Upon detailed inspection, it was revealed that this is directly related to the criterion~(i) of the collision algorithm described in \S\ref{sec:math}.
As a matter of fact, the computational cost of this criterion scales with $\mathcal{O}(N_p^2)$ where $N_p$ is the number of points per subdomain and scales with $1/k$.
Accordingly, the computational expense of the solver for the particulate phase is significantly reduced if $k>1$.

It is also worth  mentioning that, for each case, the results of these simulations were identical, i.e.
independent of $k$.
In summary, the proposed algorithm scales very well for $k>1$.
The numerical simulations reported below were performed each on 32 processors.

\section{Code validation}
\label{app:v}

\begin{figure}
\begin{center}
\subfigure[$Re_\tau = 600$]{\includegraphics[trim=0cm 0cm 0cm 0cm,clip=true,width=0.47\textwidth]{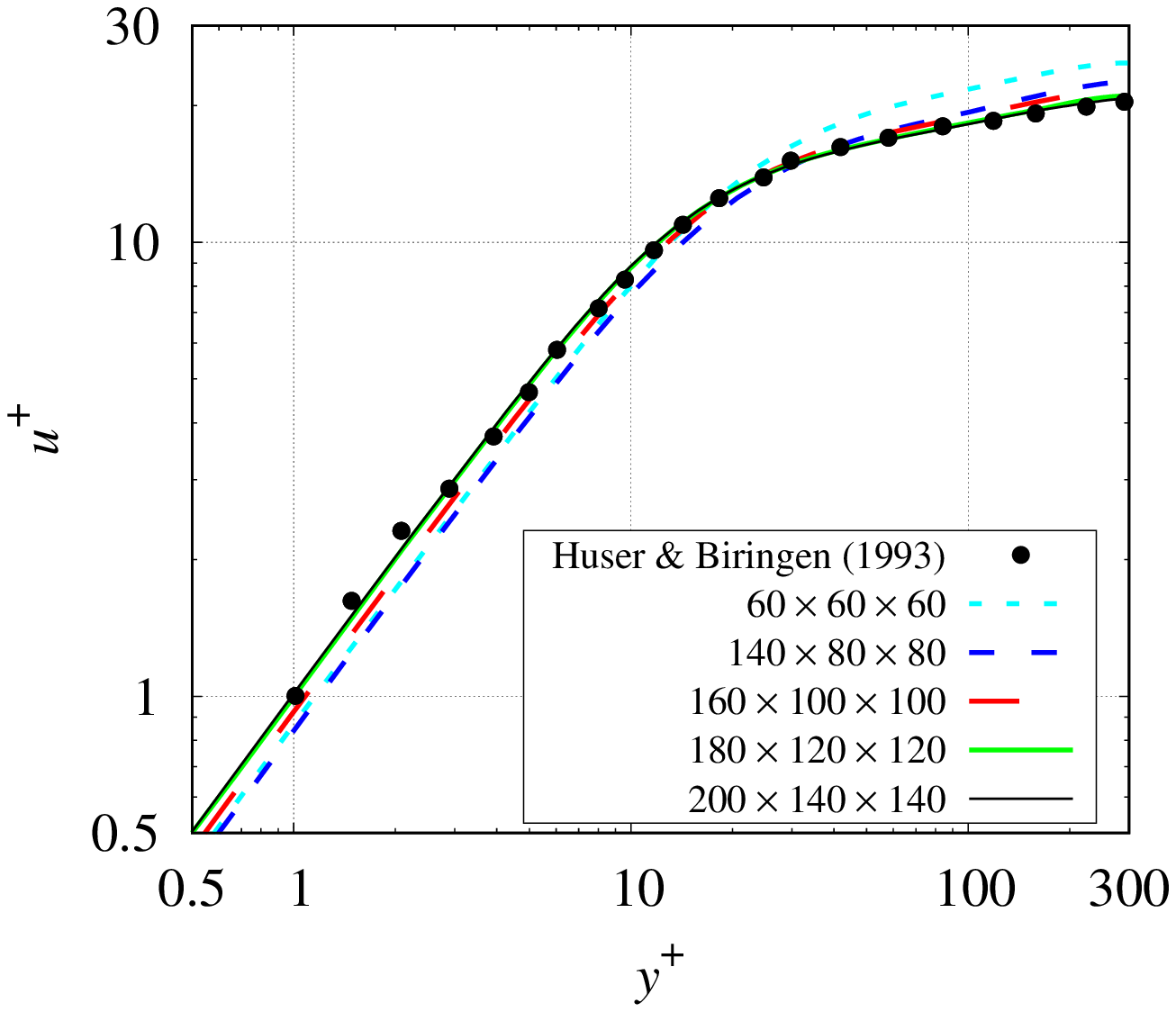}\label{fig:ugas600}}\quad
\subfigure[$Re_\tau = 300$]{\includegraphics[trim=0cm 0cm 0cm 0cm,clip=true,width=0.47\textwidth]{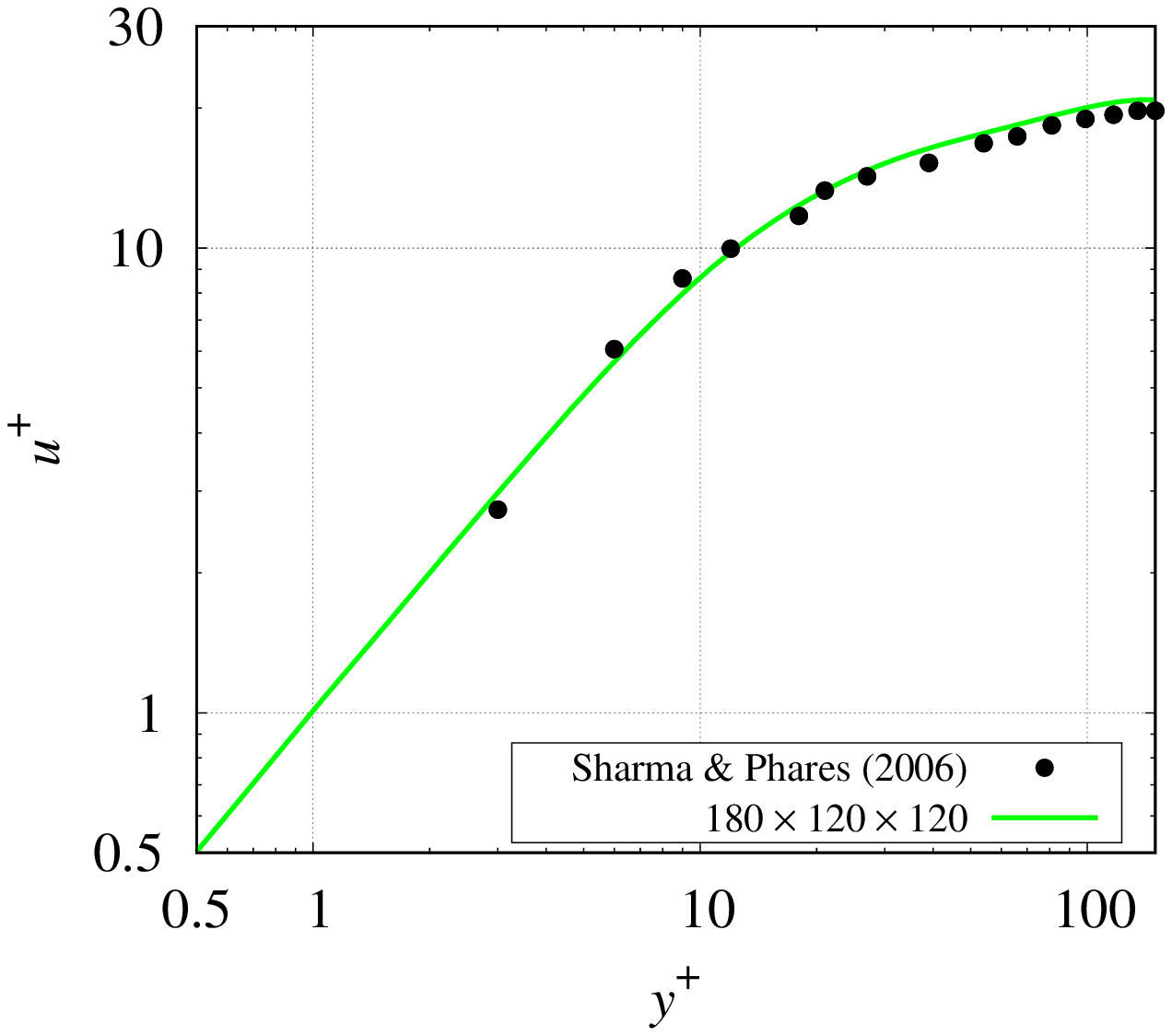}\label{fig:ugas300}}
\end{center}
\caption[]{DNS of flow in a duct without particles.
Comparison of the mean streamwise velocity at the centre plane of the duct with the DNS data of \citet{Hus93b} and \citet{Sha06}.
The profiles for $Re_{\tau}=$~600 are computed on five different grid resolutions, as shown in the legend of the figure.}
\label{fig:ugas}
\end{figure}

For validation purposes of the Navier-Stokes solver, we performed DNS of duct flows without particles and compared them
with earlier DNS that appeared in the literature.
For flow at $Re_{\tau}=$~600, we used five different meshes, ranging from 60\,$\times$\,60\,$\times$\,60 to 200\,$\times$\,140\,$\times$\,140 cells in the $x$-, $y$-, and $z$-directions, respectively.
In figure~\ref{fig:ugas600} we show plots of the mean streamwise velocity profile at the centre plane of the duct, $y=H$/2, as functions of the distance from the wall measured in wall units, $y^+$.
For comparison purposes, the DNS data of \citet{Hus93b} are also plotted in this figure.
It can be readily inferred that the velocity profiles converge as the grid resolution is increased.
In particular, the results obtained on the two finer grids, 180\,$\times$\,120\,$\times$\,120 and 200\,$\times$\,140\,$\times$\,140 cells respectively, are identical and match the DNS data of \citet{Hus93b}.

Moreover, we performed a simulation of a flow at $Re_{\tau}=$~300 on a grid consisting of 180\,$\times$\,120\,$\times$\,120 cells.
Our numerical predictions for the mean streamwise velocity at the centre plane are shown in figure~\ref{fig:ugas300}.
As can be seen, our predictions match those of \citet{Sha06}.

\begin{figure}
\begin{center}
\subfigure[$Re_\tau = 600$]{\includegraphics[trim=0cm 0cm 0cm 0cm,clip=true,width=0.32\textwidth]{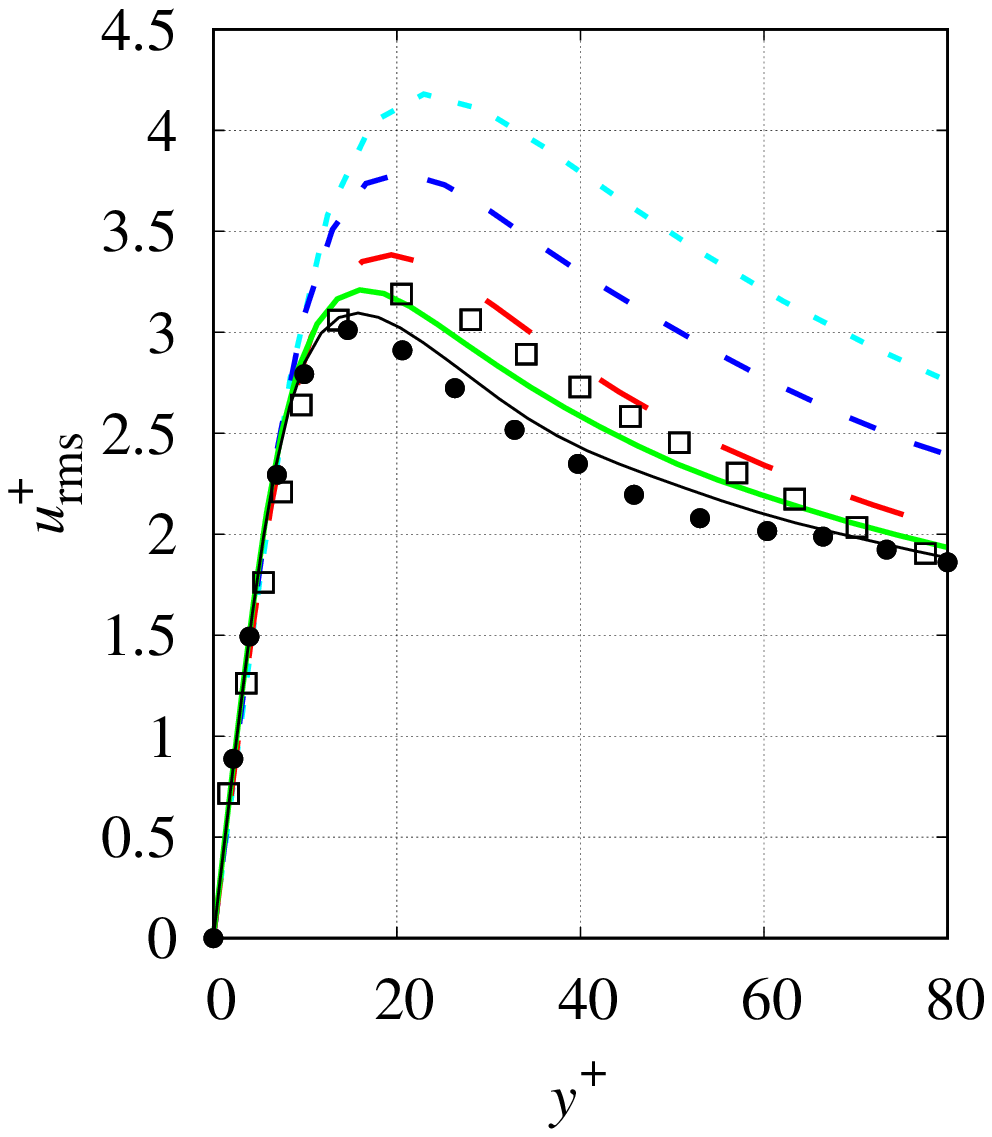}\label{fig:rms600u}}
\subfigure[$Re_\tau = 600$]{\includegraphics[trim=0cm 0cm 0cm 0cm,clip=true,width=0.32\textwidth]{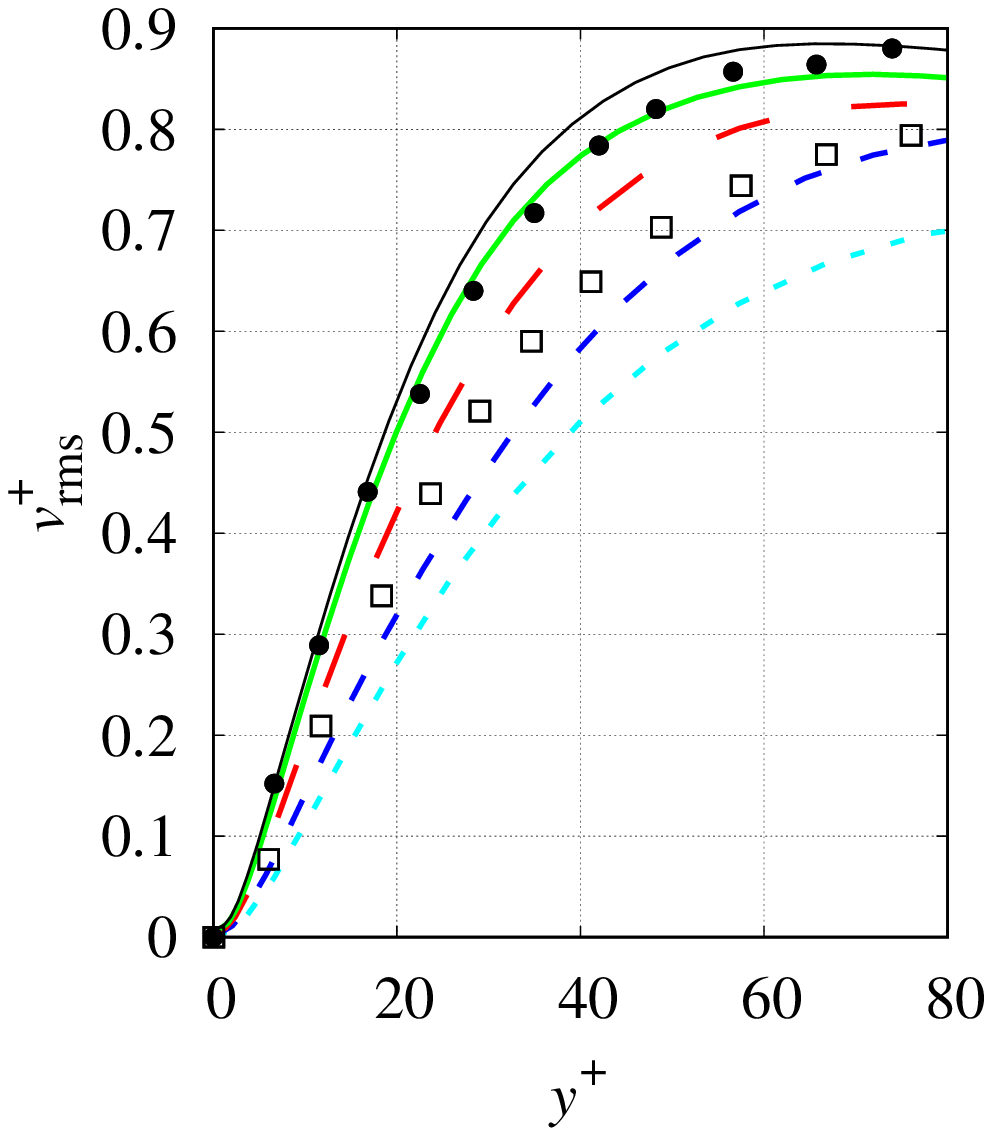}\label{fig:rms600v}}
\subfigure[$Re_\tau = 600$]{\includegraphics[trim=0cm 0cm 0cm 0cm,clip=true,width=0.32\textwidth]{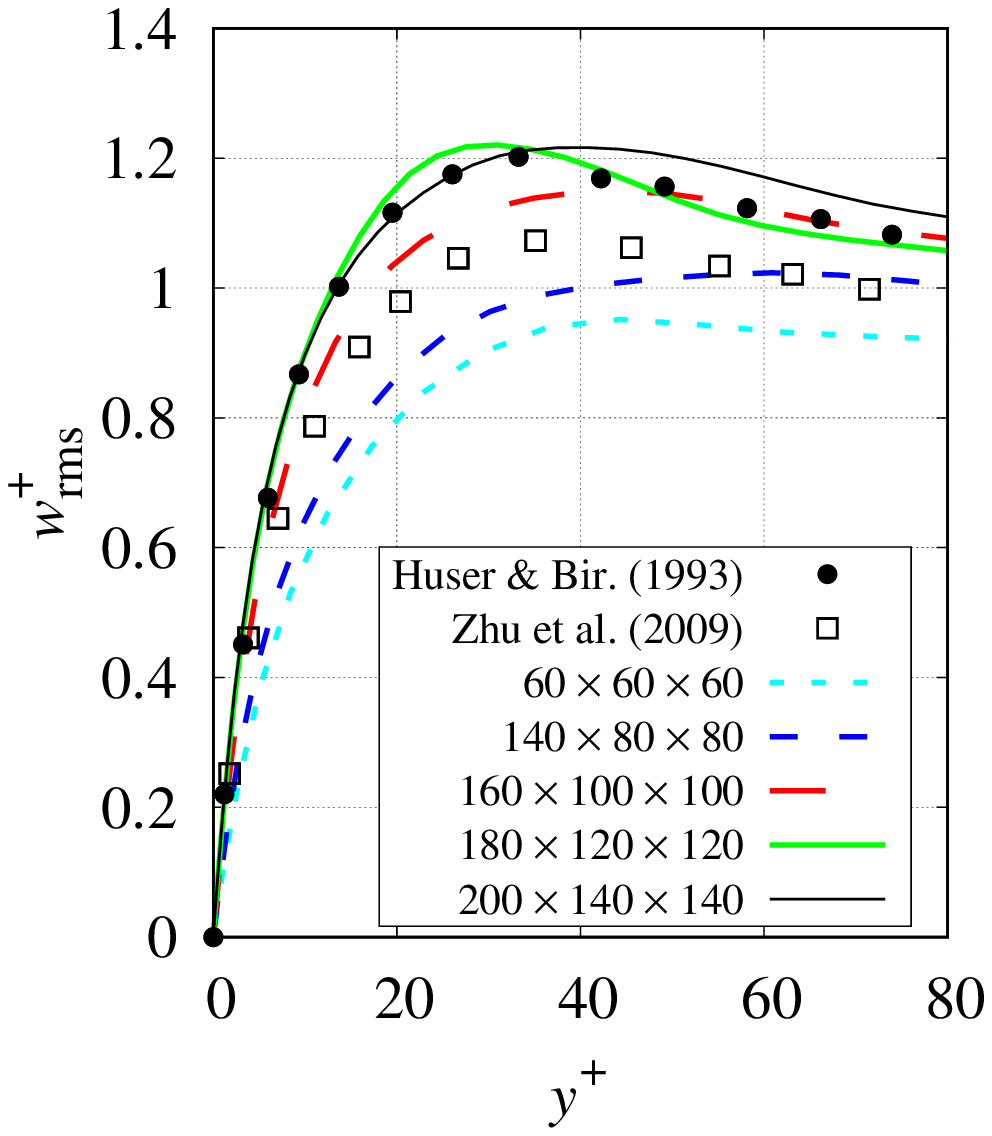}\label{fig:rms600w}}\\
\subfigure[$Re_\tau = 300$]{\includegraphics[trim=0cm 0cm 0cm 0cm,clip=true,width=0.32\textwidth]{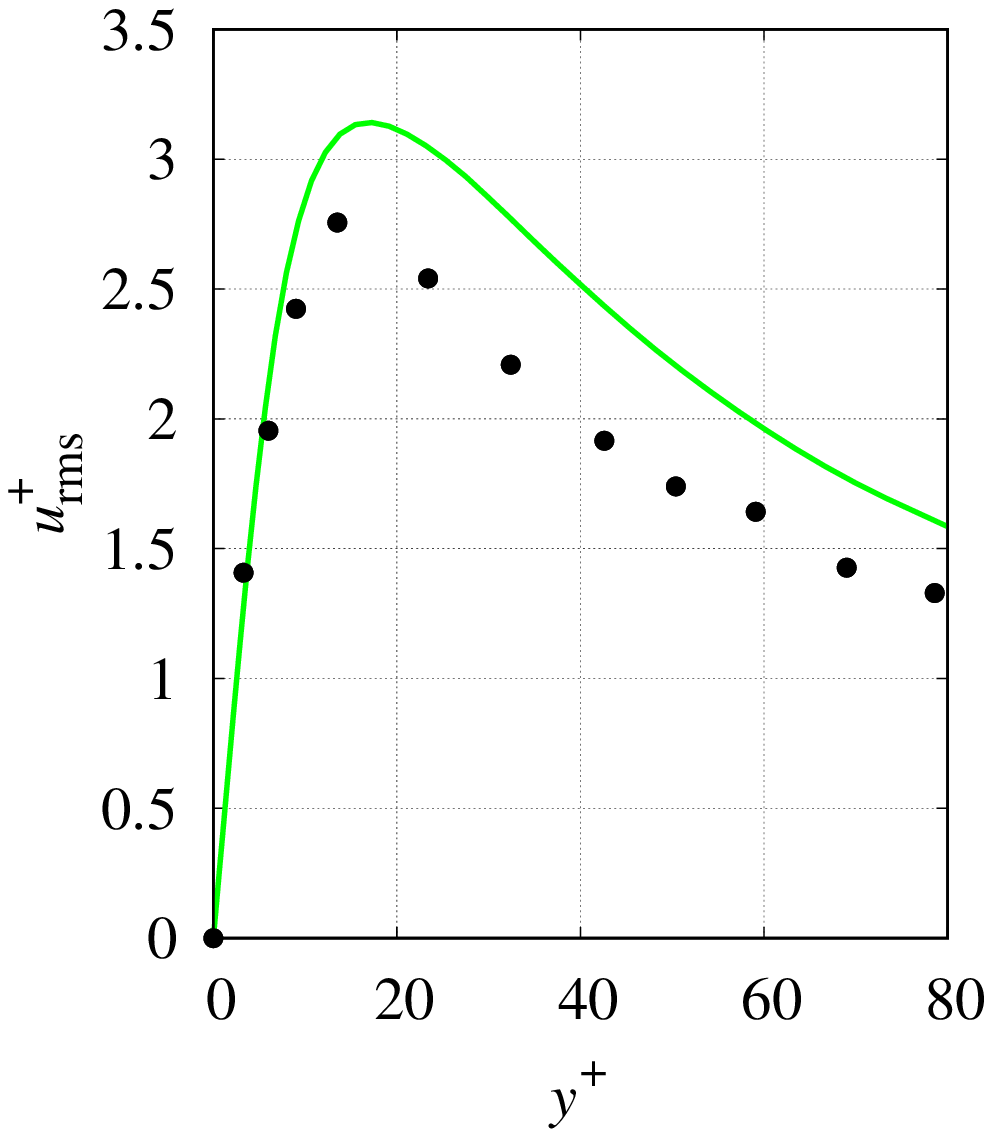}\label{fig:rms300u}}
\subfigure[$Re_\tau = 300$]{\includegraphics[trim=0cm 0cm 0cm 0cm,clip=true,width=0.32\textwidth]{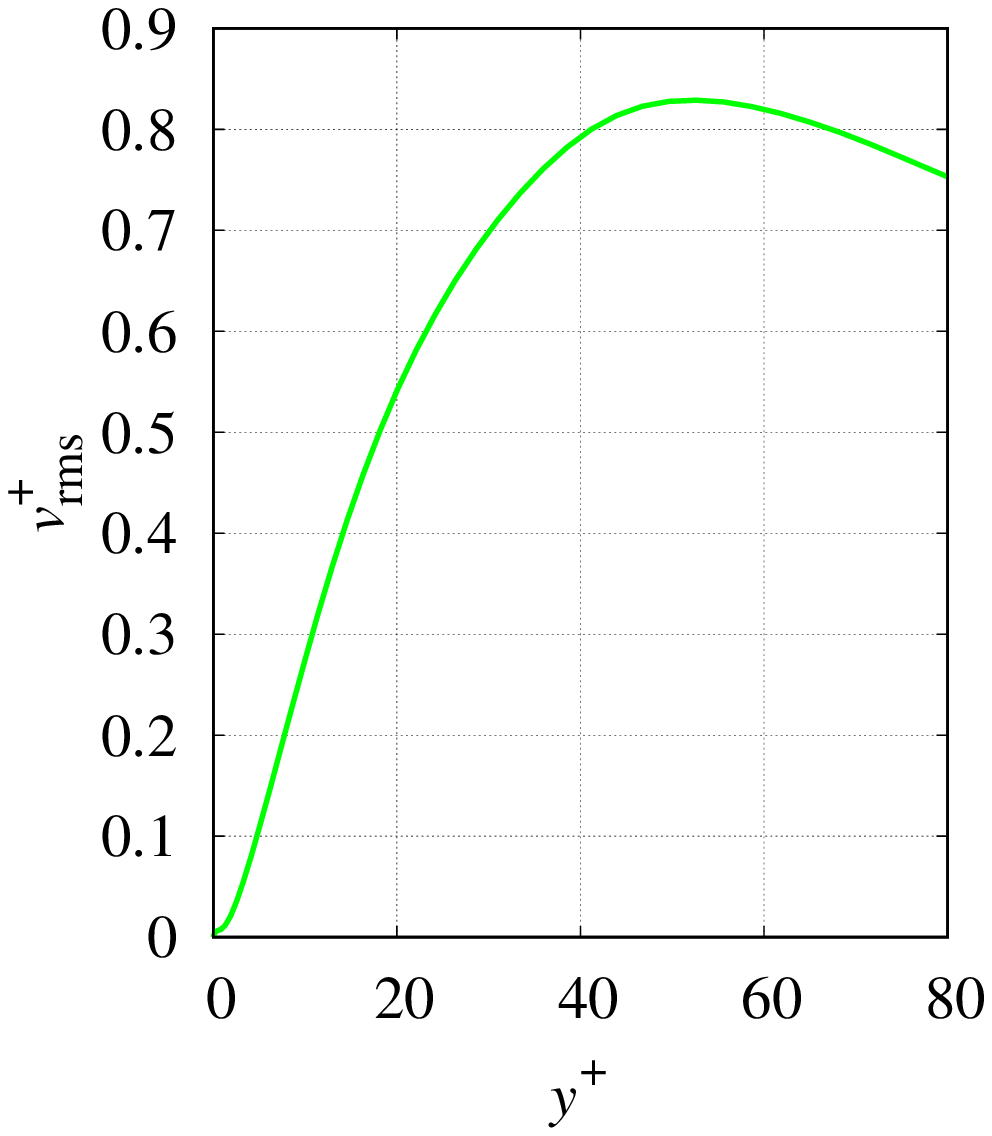}\label{fig:rms300v}}
\subfigure[$Re_\tau = 300$]{\includegraphics[trim=0cm 0cm 0cm 0cm,clip=true,width=0.32\textwidth]{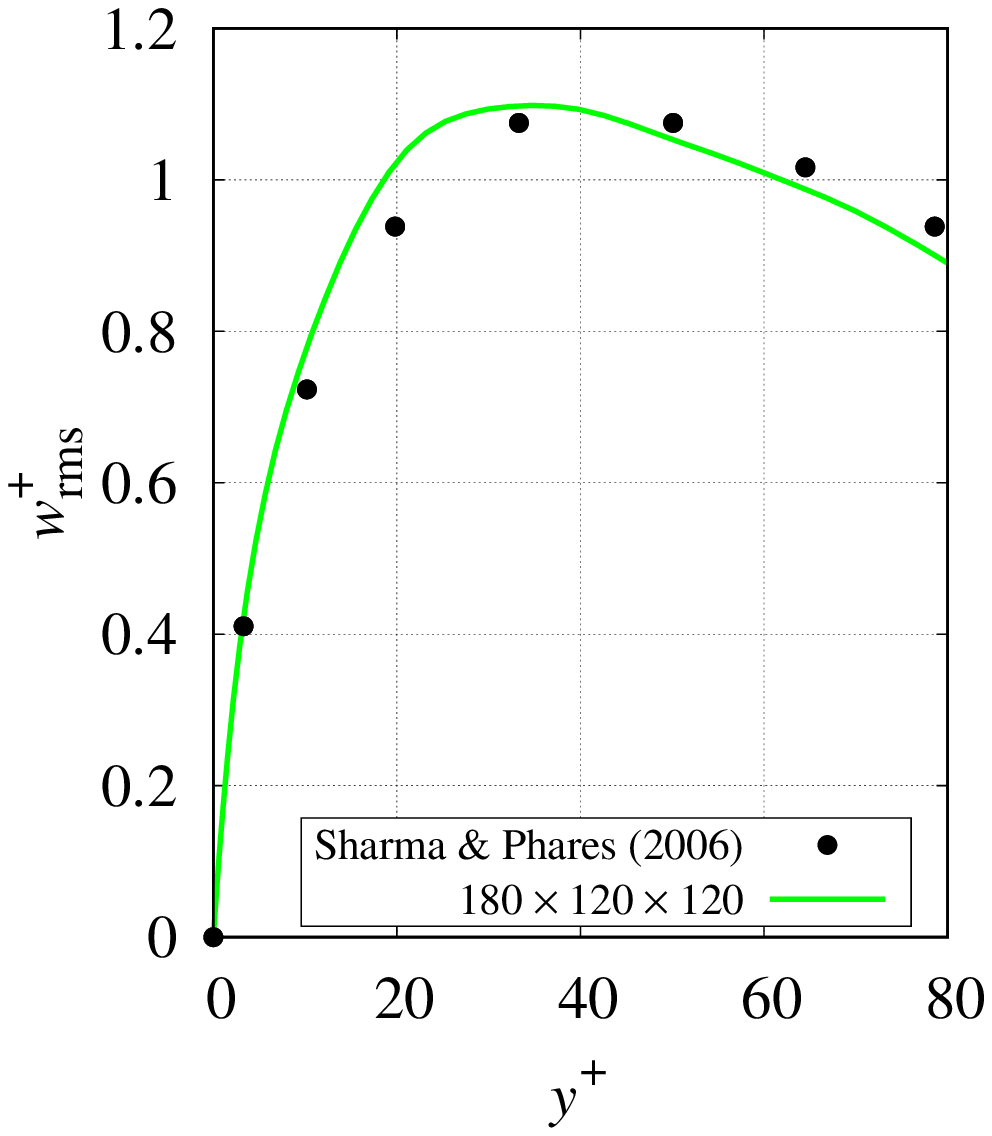}\label{fig:rms300w}}
\end{center}
\caption[]{DNS of flow in a duct without particles.
Comparison of the rms velocity components with the DNS data of \citet{Hus93b}, \citet{Zhu09}, and \citet{Sha06}.
The profiles for $Re_{\tau}=$~600 are computed on five different grid resolutions, as shown in the legend of the figure.
Note that \citet{Sha06} did not provide data for $v^+_\mathrm{rms}$.
}
\label{fig:rms}
\end{figure}

Further, the rms values of the velocity components are presented in figure~\ref{fig:rms}.
From the results for the flow at $Re_{\tau}=$~600 (figures~\ref{fig:rms600u}--\ref{fig:rms600w}) we can infer once again that numerical grid convergence is attained with grid refinement.
For the fine grid resolutions, only slight discrepancies are observed between our predictions for the rms of the wall-normal velocity components, $v_{\rm rms}$ and $w_{\rm rms}$ and those of \citet{Hus93b} and \citet{Zhu09}.

The rms of the velocity components for the flow at $Re_{\tau}=$~300 are plotted in figures~\ref{fig:rms300u}--\ref{fig:rms300w}.
In general, our results compare favourably with the DNS data \citet{Sha06} (note that \citet{Sha06} did not provide the fluctuations of the velocity component $v$).
The most noticeable difference is in the $u_{\rm rms}$ profile far from the wall for which the predictions of \citet{Sha06} are 10\%--20\% lower than ours.

On the basis of these tests and the favorable comparison with earlier results, the DNS presented below have been performed on the grid consisting of 180\,$\times$\,120\,$\times$\,120 cells.

\section{Higher charge levels}

\begin{figure*}
\begin{center}
\subfigure[Slice: $10<z^+<12.5$ for $Re_\mathrm{\tau}=$~300 and $20<z^+<25$ for $Re_\mathrm{\tau}=$~600.]{\includegraphics[trim=0cm 0cm 0cm 0cm,clip=true,width=0.47\textwidth]{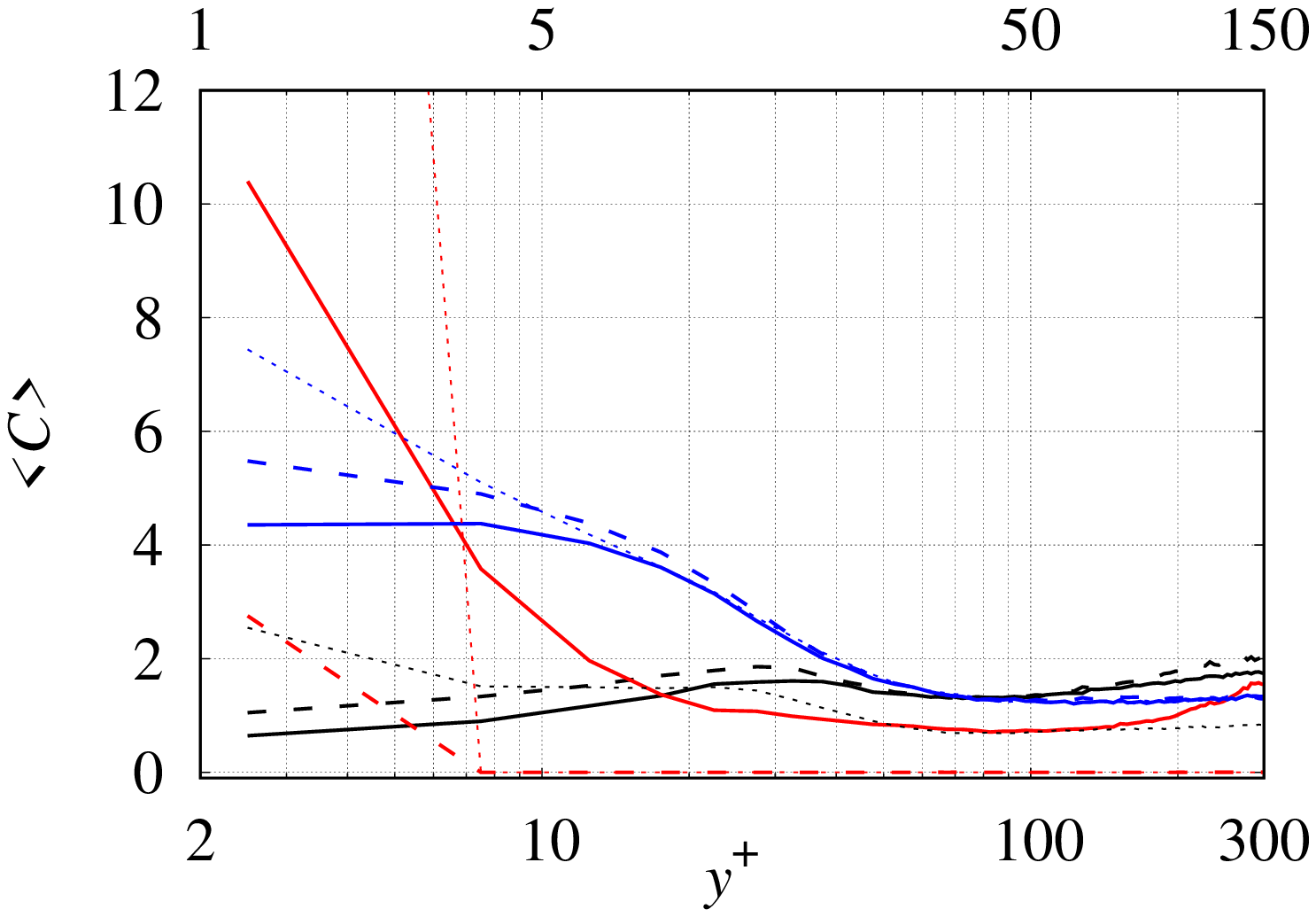}\label{fig:1d_2-r}}
\subfigure[Slice $147.5<z^+<150$ for $Re_\mathrm{\tau}=$~300 and $295<z^+<300$ for  $Re_\mathrm{\tau}=$~600.]{\includegraphics[trim=0cm 0cm 0cm 0cm,clip=true,width=0.47\textwidth]{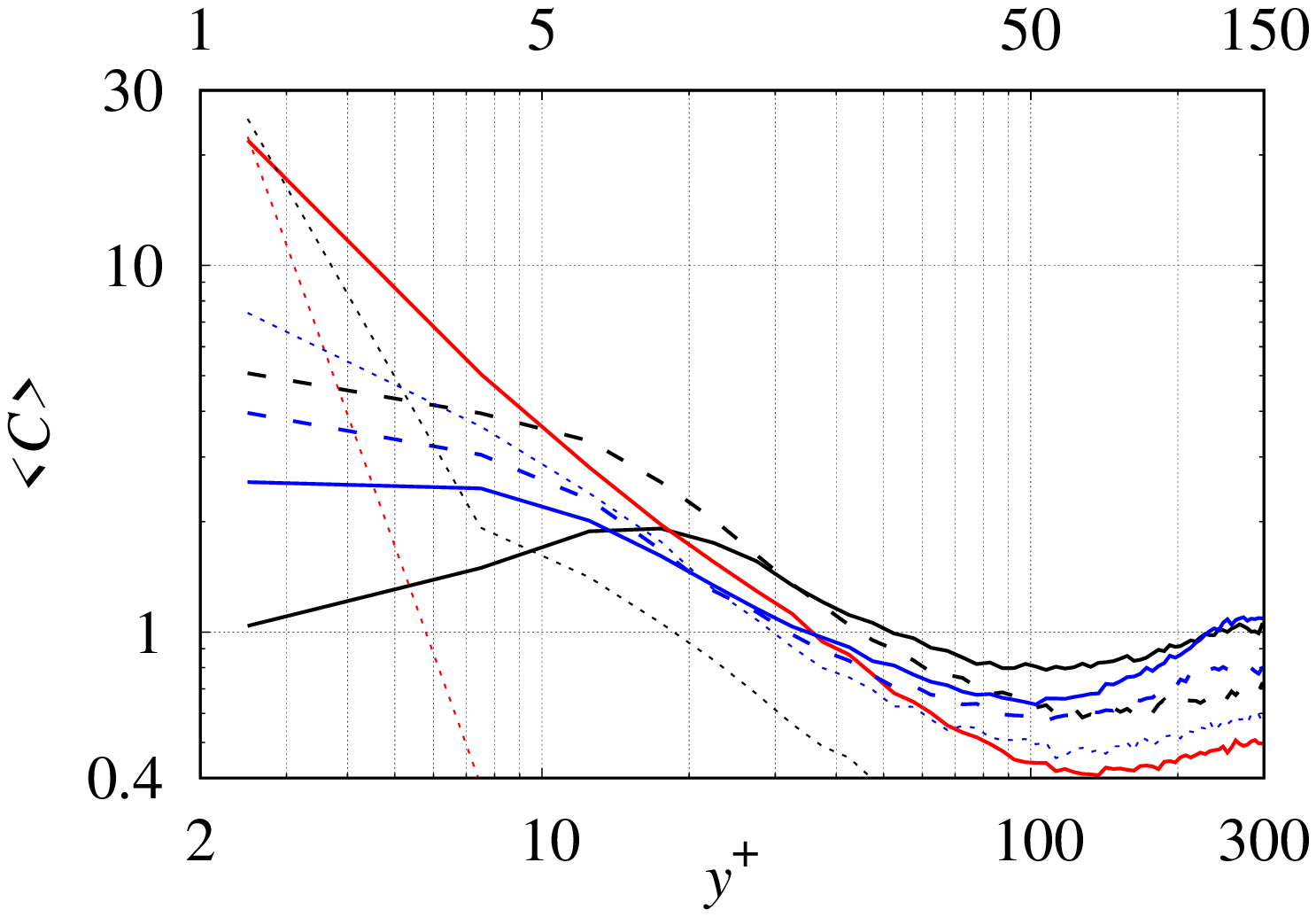}\label{fig:1d_4-r}}\quad
\end{center}
\caption[]{Mean normalized particle number density, $\langle C\rangle$, replotted from figure~\ref{fig:1d_2} and~\ref{fig:1d_4} with additional cases considering a higher charge level:\\
\begin{tabular}{ll}
(\raisebox{1mm}{\tikz{\draw[black,thick,dotted] (0,0)--(.5,0);}}) &
$Re_\mathrm{\tau}=$~600, $\rho_\mathrm{p} / \rho=$~1000, $F_\mathrm{el}/F_\mathrm{g}=$~0.104 \\
(\raisebox{1mm}{\tikz{\draw[red,thick,dotted] (0,0)--(.5,0);}}) &
$Re_\mathrm{\tau}=$~300, $\rho_\mathrm{p} / \rho=$~1000, $F_\mathrm{el}/F_\mathrm{g}=$~0.104 \\
(\raisebox{1mm}{\tikz{\draw[blue,thick,dotted] (0,0)--(.5,0);}}) &
$Re_\mathrm{\tau}=$~600, $\rho_\mathrm{p} / \rho=$~7500, $F_\mathrm{el}/F_\mathrm{g}=$~0.017
\end{tabular}
}
\label{fig:part_conc1d-rebuttal}
\end{figure*}

Exemplary results of an exploratory study considering a higher charge level of 0.2~pC on each particle are presented in figure~\ref{fig:part_conc1d-rebuttal}.
In this figure, we replotted figures~\ref{fig:1d_2} and~\ref{fig:1d_4} and added the mean particle concentration profiles of the additional cases.
For this new charge level qualitatively the same phenomena occur as for the lower particle charge.
The vortical motion of the particles in the cross-section is even more retarded by the arising electrostatic forces.
Consequently, even more particles accumulate close to the walls.

For clarity of the presentation, we focus in the main results section of this paper on one charge level per flow condition.

\bibliography{../../../publications/publications.bib}

\end{document}